\documentclass[a4paper,11pt]{article}
\usepackage{graphicx,amssymb,amstext,amsmath,mathabx}
\usepackage{color, yfonts, tcolorbox}
\usepackage{floatflt}
\usepackage{bm}
\usepackage{tikz}

\usepackage{float}
\usepackage{ulem}
\usepackage{esint}
\usepackage{subcaption}
\usepackage{cite}
\usepackage{mathtools}
\usepackage{commath}
\usepackage{hyperref}
\usepackage{multirow}

\definecolor{cbl}{rgb}{0,0,1}                

\newcommand{\bc}{\begin{center}}
\newcommand{\ec}{\end{center}}
\def\ba#1{\begin{array}{#1}\displaystyle}
\newcommand{\ea}{\end{array}}

\newcommand{\beq}{\begin{equation}}
\newcommand{\eeq}{\end{equation}}
\newcommand{\beqa}{\begin{eqnarray}}
\newcommand{\eeqa}{\end{eqnarray}}

\newcommand{\bi}{\begin{itemize}}
\newcommand{\ei}{\end{itemize}}
\global\long\def\ud{\mathrm{d}}%

\global\long\def\th{\theta}%
 
\global\long\def\r{\rangle}%
 
\global\long\def\oo{\mathcal{O}}%

\global\long\def\eps{\varepsilon}%

\newcommand{\bra}{\langle}
\newcommand{\ket}{\rangle}

\newcommand{\Tr}{{\rm Tr}}

\newcommand{\OCA}{\color{black}}
\newcommand{\DXH}{\color{black}}

\begin{document}
\begin{titlepage}
\vspace{0.2cm}
\begin{center}

{\large{\bf{Entanglement of Stationary States in the Presence of Unstable Quasiparticles}}}

\vspace{0.8cm} 
{\large D\'avid X. Horv\'ath{$^{\spadesuit}$}, Pasquale Calabrese$^\clubsuit$ and Olalla A. Castro-Alvaredo$^\heartsuit$}

\vspace{0.8cm}
{\small
{$^{\spadesuit,\clubsuit}$}  SISSA and INFN Sezione di Trieste, via Bonomea 265, 34136 Trieste, Italy\\
\medskip
{$^{\clubsuit}$}  International Centre for Theoretical Physics (ICTP), Strada Costiera 11, 34151 Trieste, Italy\\
\medskip

$^{\heartsuit}$ Department of Mathematics, City, University of London, 10 Northampton Square EC1V 0HB London, UK\\
\medskip
}
\end{center}

\medskip
\medskip
\medskip
\medskip

The effect of unstable quasiparticles in the out-of-equilibrium dynamics of certain integrable systems has been the subject of several recent studies. In this paper we  focus on the stationary value of the entanglement entropy density, its growth rate, and related functions, after a quantum quench. We consider several quenches, each of which is characterised by a corresponding squeezed coherent state. In the quench action approach, the coherent state amplitudes $K(\theta)$ become input data that fully characterise the large-time stationary state, thus also the corresponding Yang-Yang entropy.  We find that, as function of the mass of the unstable particle, the entropy growth rate has a global minimum signalling the depletion of entropy that accompanies a slowdown of stable quasiparticles at the threshold for the formation of an unstable excitation. We also observe a separation of scales governed by the interplay between the mass of the unstable particle and the quench parameter, separating a non-interacting regime described by free fermions from an interacting regime where the unstable particle is present.
This separation of scales leads to a double-plateau structure of many functions, where the relative height of the plateaux is related to the ratio of central charges of the UV fixed points associated with the two regimes, in full agreement with conformal field theory predictions. The properties of several other functions of the entropy and its growth rate are also studied in detail, both for fixed quench parameter and varying unstable particle mass and viceversa.  

\noindent 
\medskip
\medskip
\medskip
\medskip

\noindent {\bfseries Keywords:} Entanglement Dynamics, Integrable Quantum Field Theory, Unstable Quasiparticles, Thermodynamic Bethe Ansatz
\vfill
\noindent 
{$^{\spadesuit}$} esoxluciuslinne@gmail.com\\
{$^\clubsuit$} calabrese@sissa.it\\
{$^\heartsuit$} o.castro-alvaredo@city.ac.uk\\

\hfill \today

\end{titlepage}

\section{Introduction}
The dynamics of out-of-equilibrium many body quantum systems is an area of enormous current interest and scientific activity. Over the past few years there have been many developments, with experimental and theoretical work both influencing, inspiring and stimulating each other. In this context, a key quantity of interest has been the entanglement entropy and its time evolution following a quantum quench. The latter is understood as a sudden change of the evolution hamiltonian as introduced in \cite{quench1, quench2} (see also the reviews \cite{Eisert,EEDyn,CEM,pssv-11,EF16}).  
Since the famous Quantum Newton's Cradle experiment \cite{kinoshita} it has been known that integrability (i.e.~the presence of a large number of conservation laws) in one-dimensional systems leads to a distinct post-quench dynamics, which has later been understood in terms of a Generalized Gibbs Ensemble (GGE) \cite{Rigol,vr-16}. This means that the partition function of the generic stationary state involves all local and quasi-local conserved charges in the system
\beq
{\mathcal Z}= \Tr\left(e^{-\sum_{i} \beta_i Q_i}\right)\,.
\label{gge}
\eeq
Therefore integrable systems do not {\it{thermalize}} in the usual sense but they do {\it{relax}} towards a GGE.  In particular, the role of quasi-local and semi-local conserved quantities in the GGE has been the subject of a lot of investigation \cite{fcec-13,failure,Prosen1, Prosen2, Prosen3, Prosen4, doyon2017,ilinardo}.  A good summary of many results in this area can be found in the special issues \cite{CEM,specialrec}. The fact that, after sufficiently long times, every integrable system settles to a stationary state characterised by a GGE implies that such a state and, subsequently, all related observables, can be obtained from the knowledge of all expectation values of conserved quantities in the initial, pre-quench state. 
However, the full description of all conserved local and quasi-local charges is sometimes very challenging, as shown in the pioneering works \cite{failure,ilinardo}. 

Several alternative ways of obtaining the stationary values of  dynamical quantities have been proposed and employed successfully in recent  years. Particularly successful approaches are generalised hydrodynamics (GHD) \cite{ourhydro,theirhydro,benreview}, Quantum Transfer Matrix (QTM) \cite{p-13,IntegrableQuench,pvcp-18,pvcp-18b,ppv-18} and the quench action method (QAM) \cite{Cfab1,QuenchActionReview}. The latter will be employed in this paper.  The basis of this method is the intuitive idea that, no matter how complicated the post-quench states is, there will always be a ``dominant" pure state that will provide the leading large-time contribution to the stationary value of any local observable. Identifying and describing such a state allows us to then carry out explicit computations. This has been done very successfully for a variety of models: in Heisenberg
spin chains \cite{ilinardo,Ams1, Ams2, failure, Hun2, AC2,mbpc-17,pvcr-16,pvc-16}, interacting Bose gases \cite{NWB,PCE,Bucci,pce-16,NA1, NA2, NA3,bpc-18,pc-17}, the Hubbard model \cite{rbc-22,rcb-22b},  transport problems \cite{LMV}, and interacting quantum field theories \cite{Cfab2,BPC}. The latter are most closely related to the work presented in this paper. 

In many-body quantum integrable models, exact computations are facilitated by a mathematical description of the quasiparticle content. Such a formulation has been known for a long time and comes through the (thermodynamic) Bethe ansatz approach \cite{takahashi,tba1,tba2,Mossel}. This lies at the heart of both the GHD and QAM approaches. 

In particular, the QAM admits a manageable formulation when the pre-quench state can be expressed as a ``squeezed coherent state" of the eigenstates of the post-quench theory. 
When such an expression exists  \cite{FM,SFM,HST}, then the overlaps between pre- and post-quench states can act as input to generalised versions of the TBA equations \cite{Cfab1,QuenchActionReview}, from which the Yang-Yang entropy of the stationary state can be read off. More generally, such a representation of the state can be employed in conjunction with form factor techniques to obtain the full time-evolution following a quench, as done, e.g., for the  Ising model \cite{Cfab1,ES,TIsing}, for the interacting Bose gas \cite{NA1,NA2} and {\OCA for other interacting integrable models \cite{Cfab2,BPC,DS}}.

In this paper we carry out a study which is similar in spirit to the works {\OCA \cite{Cfab2, BPC,DS}} where the QAM approach was employed for two interacting integrable QFTs, the sine- and sinh-Gordon models, respectively. Unlike those papers however, we want to focus only on features of the stationary (entanglement) entropy. Our aim is to explore the imprints of instability that can be seen in the stationary values of the entropy and related functions, when unstable particles are present in the spectrum. 
We focus on a very simple integrable QFT known as the $SU(3)_2$-homogeneous sine-Gordon (HSG) model \cite{hsg, ntft, FernandezPousa:1997iu, smatrix}. This is a theory whose spectrum contains two stable particles $(\pm)$ of the same mass $m$. The two-particle scattering matrix $S_{+-}(\theta)$, characterised by the rapidity difference $\theta$ has a pole in the unphysical sheet at $\theta=\sigma-i\pi/2$. This can be interpreted as associated with the creation of an unstable particle, with a finite decay width and mass which are functions of the resonance parameter $\sigma$.  

The free parameter $\sigma$ introduces an energy scale related to the mass of the unstable particle so that below this energy, the unstable particle is too massive and cannot be formed. In this regime the theory is identical to two free fermions, with UV limit characterised by a conformal field theory of central charge $c=1$. For energies above the unstable particle mass, the latter is formed and the theory flows in the UV limit to a conformal field theory of central charge $c=6/5=1.2$. As a result, all quantities that can be analysed for the model, including the entropy, can be usually understood as exhibiting three regimes with respect to the parameter $\sigma$, corresponding to the free, interacting and transition regime. This transition regime corresponds to the threshold for the formation of the unstable particle.  

Some of the dynamical properties of the model have been studied in \cite{ourU, nextU, nnextU} employing the GHD approach. In those works the focus was on the particle densities, effective velocities and state densities associated to the stable particles. It was found that all these functions display new features in the presence of the unstable particle so that despite it not being part of the asymptotic spectrum, its formation and decay can still be dynamically observed. As we shall see in this work, also the stationary value and growth rate of the entropy display interesting new properties when unstable excitations come into play. The qualitative and quantitative understanding of these properties for different global quenches are the main focus of this paper.

Although all our computation are performed for the manageable ``squeezed coherent states", we do not believe that such a restriction generically implies qualitative difference for larger classes of initial states. 
Based on the results for this class of states, our main observations go as follows:
\begin{itemize}
\item The stationary state spectral particle densities and effective velocities of the stable excitations display, as expected, many features already found in previous work \cite{ourU,nextU,nnextU}. For the densities, that is an additional local maximum (whose area is related to a persistent density of unstable particles in the stationary state) and for the velocities an additional plateau (whose presence signals the slowdown of stable particles that precedes the formation of unstable ones), respectively.
\item The stationary entanglement entropy per unit length $S/L$ displays, as a function of the unstable particle's mass (or the parameter $\sigma$), a double-plateau structure, where the transition between plateaux is determined by the energy threshold for the formation of the unstable particle and the relative height of the plateaux is given by the ratio of UV central charges in the interacting and non-interacting regimes. 
\item This threshold can be determined exactly for each quench and corresponds to a precise relationship between the resonance parameter $\sigma$ and a quench parameter, which we call $\alpha$ (related to an effective inverse temperature $\beta_{\rm FF}$, a measure of the energy injected by the quench).  
\item Close to the threshold, we observe the emergence of a local minimum in the entanglement entropy production rate, a feature that has previously been argued to precede the formation of  stable excitations (this is described as the dynamical manifestation of  Gibbs' paradox in \cite{Gibbsparadox}). Our work suggests that a similar signature, albeit less pronounced, is found before the emergence of the unstable particle and that it is fundamentally linked to the increase of the number of effective degrees of freedom.
\item This minimum coincides with a plateau of the effective velocities of stable excitations around the value zero, indicating that the formation of the unstable particle requires stable particles to slow down first, which in turn leads to depleted entropy growth. 
\end{itemize}
  
This paper is organised as follows: In Section \ref{model} we introduce the model and the equilibrium TBA equations. In Section \ref{quenches} we review the theory of global quenches and squeezed coherent states. In Section \ref{action} we review the quench action method and the quasiparticle picture of entanglement evolution after a quench, and apply them to our model.  The core of our paper is Section \ref{results} in which we present analytic results and their numerical analysis for the entanglement entropy and other auxiliary quantities. Our calculations and analyses provide a qualitative and partially quantitative explanation of the entanglement dynamics, in line with the bullet points above. 
We conclude in Section  \ref{conclus}. A simple derivation of some scaling properties of the entropy per unit length and entropy growth rate of free fermions is presented in Appendix \ref{AppA}. In Appendix \ref{AppB} we analyse the properties of the entropy and related quantities for an additional quench not discussed in the main text.

\section{The Model at Thermal Equilibrium}
\label{model}
In this section we introduce the basic description of our model: the two-particle scattering matrix, first derived in \cite{smatrix}, and the thermal equilibrium properties of the theory as described by its TBA equations, solved and analysed in \cite{ourtba,CastroAlvaredo:2002nv,Dorey:2004qc}. 

\subsection{The Scattering Matrix}
The HSG models are a family of integrable quantum field theories possessing a diagonal scattering matrix and both stable and unstable bound states. 
 These models were first studied in a series of papers in the late 90s where their classical and quantum integrability were established \cite{hsg,ntft}, the particle spectrum determined \cite{FernandezPousa:1997iu}, and a scattering matrix proposed \cite{smatrix}. The scattering matrix was then extensively tested through the TBA \cite{ourtba,CastroAlvaredo:2002nv,Dorey:2004qc}  and form factor approaches \cite{KW,SmirnovBook, CastroAlvaredo:2000em,CastroAlvaredo:2000nk, CastroAlvaredo:2000ag,CastroAlvaredo:2000nr}. The mass-coupling relation for the $SU(3)_2$-homogeneous sine-Gordon model
was studied in \cite{MC1,MC2}. 

Each HSG model is associated with a simply-laced algebra $g$ and an integer $k$, called the level. The models may be seen as massive perturbations of a critical Wess-Zumino-Novikov-Witten model \cite{WZNW1,WZNW2,WZNW3,WZNW4,WZNW5} associated to the coset $G_k/U(1)^r$, where the level $k$ is a parameter of the model, $r$ is the rank of $g$ {\OCA and $G$ is the group associated to the algebra $g$}. The HSG models are part of the Toda family. Indeed, each HSG model can be seen as $r$ copies of $A_{k-1}$ minimal Toda field theory which interact with each other non-trivially, giving rise in the process to the formation of unstable excitations. The simplest model in this large family is the $SU(3)_2$-HSG model which we consider in this paper.  

The $SU(3)_2$-HSG model has a two-particle spectrum. It is convenient to label the particles as ($\pm$). The scattering matrices are:
\beq\label{Smatrix}
S_{\pm\pm}(\theta)=-1, \qquad S_{\pm\mp}(\theta)=\pm \tanh\frac{1}{2}\left(\theta\pm \sigma-\frac{i\pi}{2}\right)\,,
\eeq
where $\sigma$ is a free parameter of the theory. Note that 
$
\lim_{|\sigma| \rightarrow \infty} S_{\pm\mp}(\theta)=1
$,
 which means that in this limit parity symmetry is restored and a theory of two Majorana fermions is recovered.  An important consequence of this property is that the behaviour of any quantity we compute at or out of equilibrium  should reduce to the free  fermion case if $|\sigma|$ is large compared to the overall energy scale in the system (i.e. temperature, if at equilibrium or the quench parameter in the present work). 
  
For finite $\sigma$, the theory is interacting and the scattering amplitudes $S_{\pm\mp}(\theta)$ have a pole outside the physical sheet at $\theta=\mp\sigma-i\pi/2$, in the strip $-\pi \leq \mathrm{Im}(\theta)\leq 0$. As discussed in \cite{CastroAlvaredo:2000ag}, the mass $M$ and decay width $\Gamma$ of this unstable particle can be obtained from the Breit-Wigner formula. It is particularly useful to note that
\beq
{M}\sim \frac{1}{\sqrt{2}} m e^{\frac{|\sigma|}{2}} \quad \mathrm{and} \quad  \Gamma\sim \sqrt{2} m e^{\frac{|\sigma|}{2}}\, \quad \mathrm{for} \quad |\sigma|\gg 1\,,
\eeq
so that, the larger the value of $|\sigma|$, the more massive and short-lived the unstable excitation becomes. For general values of $\sigma$ (not necessarily large) the mass is given by
\beq
M=m\sqrt{1+\cosh\sigma}\,.
\label{exactM}
\eeq
A very clear picture then emerges, namely that {\OCA at thermal equilibrium in a Gibbs ensemble with temperature $T$ and for $\sigma$ sufficiently large}, the variable 
\beq
\kappa:=\log(2T)-\frac{\sigma}{2}\,,
\label{kappa}
\eeq 
characterises three distinct regimes: $\kappa<0$, is the free fermion regime, where the energy is not large enough for unstable particles to be formed, $\kappa=0$ is the threshold for the formation of unstable particles and $\kappa>0$ is the interacting regime, where the unstable particle is present.  
In this paper we will always choose $\sigma>0$.

 In accordance with this separation of energy scales, many TBA functions develop staircase patterns, where the position and size of the steps are related to the value of $\sigma$. For $\kappa\ll -1$ the theory reaches the UV limit of a two free fermion theory, with central charge $c=1$. In contrast, for $\kappa\gg 1$, the UV fixed point is determined by the coset $SU(3)_2/U(1)^2$ and corresponds to $c=6/5=1.2$. As we shall see, the role of these central charges as counting degrees of freedom at high energies will also become apparent when we study the entropy per unit length in the stationary state. 
 The natural role played by the variable $\kappa$ is also seen in the TBA equations, as we discuss below. 

\subsection{The thermodynamic Bethe ansatz equations}
Since there are two stable particles, there are also two TBA equations, which are related to each other by a parity transformation. Let $\varepsilon_\pm(\theta)$ be the pseudoenergies and $L_{\pm}(\theta)=\log(1+e^{-\varepsilon_{\pm}(\theta)})$ the $L$-functions, then the TBA equations can be written as
\beq
\varepsilon_{\pm}(\theta)=\omega(\theta)-(\varphi_{\pm\mp}\star L_\mp)(\theta)\, \quad \mathrm{with}\quad \varepsilon_+(\theta)=\varepsilon_-(-\theta)\,,
\label{tba}
\eeq
$\varphi_{\pm \mp}(\theta)={\rm sech}(\theta\pm \sigma)$ is the logarithmic derivative of the scattering matrix (scattering phase) and $\star$ indicates the convolution:
\beq
(a\star b)(x)=\frac{1}{2\pi} \int_{-\infty}^\infty a(x-y)b(y) dy\,.
\eeq
At equilibrium, the driving term, $\omega(\theta)$ is given in terms of the one-particle eigenvalue of the energy $E(\theta)=m\cosh\theta$ (the same for both particles, as they have the same mass) as $\omega(\theta)=\beta E(\theta)$, where $\beta$ is the inverse temperature. More generally, in a GGE, $\omega(\theta)$
is a sum over one-particle eigenvalues of any conserved charges involved in the GGE with coefficients which are generalised inverse temperatures. 

An important operation is the ``dressing" of a generic function $h(\theta)$. This is defined by differentiating the TBA equations above with respect to one of the generalised inverse temperatures involved in $\omega(\theta)$. This gives the equation
\beq
h_\pm^{\rm dr}(\theta)=h(\theta)+(\varphi_{\pm\mp}\star g_\mp)(\theta)\,,
\label{dressing}
\eeq
where $g_\pm (\theta)=n_\pm (\theta) h^{\rm dr}_\pm(\theta)$ and
\beq 
n_\pm(\theta)=\frac{1}{1+e^{\varepsilon_\pm(\theta)}}\,,
\label{filling}
\eeq 
are the occupation (or filling) functions. $h(\theta)$ represents a function that is independent of particle type, typically the single particle eigenvalue of a particular conserved charge. Many important thermodynamic quantities are defined through the dressing operation. For instance, the spectral particle densities $\rho_\pm(\theta)$ which represent the density of occupied states can be written as
\beq
\rho_\pm(\theta)=\frac{1}{2\pi} n_\pm(\theta) E_\pm^{\rm dr}(\theta)\,,
\label{rho}
\eeq 
and the effective velocity, which will play an important role later on is defined as 
\beq
v^{\rm eff}_\pm(\theta)=\frac{(E'_\pm)^{\rm dr}(\theta)}{(P'_\pm)^{\rm dr}(\theta)}\,,
\label{veff}
\eeq 
where the prime indicates differentiation with respect to $\theta$ and $P_\pm(\theta)=m\sinh\theta$ are the momenta.
An interesting feature of these equations is that if we define the shifted functions $\hat{\varepsilon}_{\pm}(\theta)=\varepsilon_{\pm}(\theta \mp \frac{\sigma}{2})$ and similarly for the $L$-functions, the TBA equations become
\beq
\hat{\varepsilon}_{\pm}(\theta)=\omega(\theta\pm\frac{\sigma}{2})-(\varphi\star {\hat{L}}_\mp)(\theta)\,,
\label{shifted}
\eeq
with the difference that $\varphi(\theta)={\rm sech}\theta$ no longer depends on $\sigma$ and the full $\sigma$ dependence is now in the driving term. Then if $\omega(\theta)=m\beta \cosh\theta$ and $\sigma$ is large, the driving term can be approximated by a function of $\kappa$ only (\ref{kappa}), so that all TBA functions are functions of this scale. In other words, irrespective of the values of $T$ and $\sigma$ any TBA functions will ``collapse" to a single curve, when plotted against the scale $\kappa$. 
A generalisation of this kind of collapse will also be seen later in our study of the entropy (especially in the figures for $S/S_{\rm max}$), where the role of temperature is instead played by a function of the quench parameter.

\section{Global Quenches and Squeezed Coherent States}
\label{quenches}

Let us start by recalling the definition of a global quench, as given in \cite{quench1,quench2}. A model described by a hamiltonian $H(\alpha)$ depending on some global parameter $\alpha$ is initially in its ground state $|\Psi_0\ket$. At time $t=0$ the parameter $\alpha$ is suddenly changed to a new value $\hat{\alpha}$. Henceforth, the model evolves in time with the hamiltonian $H(\hat{\alpha})$ whose eigenstates do not generically include $|\Psi_0\ket$. The question is then, what is the long-term stationary state this system will  reach? 

As mentioned in our introduction, for integrable models we expect the {\OCA long-term stationary values of  any local observables} to be described by some GGE. The challenge is then to characterise this GGE. In the QAM, the main assumption is that in the large-time limit one particular state will come to dominate the dynamics of the model. Identifying such a state is not easy in general, but it becomes easier when the initial state $|\Psi_0\ket$ has the structure of an squeezed coherent state. This means that the state $|\Psi_0\ket$ can be written in terms of eigenstates of the post-quench hamiltonian in a systematic fashion. 
In order to explain this in more detail we need to introduce some basic definitions first, and then specialise them to our model. 

\subsection{States and Charges}
In integrable quantum field theories we characterise quasiparticles by means of the Zamolodchikov-Faddeev algebra \cite{ZA,FA,Lechner1,Lechner2,Gutkin}. This algebra consists of particle creation and annihilation operators satisfying
\begin{eqnarray}
Z_{a}^{\dagger}(\theta_{1})Z_{a'}^{\dagger}(\theta_{2}) & = & S_{aa'}(\theta_{1}-\theta_{2})Z_{a'}^{\dagger}(\theta_{2})Z_{a}^{\dagger}(\theta_{1})\:,\nonumber \\
Z_{a}(\theta_{1})Z_{a'}(\theta_{2}) & = & S_{aa'}(\theta_{1}-\theta_{2})Z_{a'}(\theta_{2})Z_{a}(\theta_{1})\:,\nonumber \\
Z_{a}(\theta_{1})Z_{a'}^{\dagger}(\theta_{2}) & = & S_{a'a}(\theta_{2}-\theta_{1})Z_{a'}^{\dagger}(\theta_{2})Z_{a}(\theta_{1})+\delta_{aa'}2\pi\delta(\theta_{1}-\theta_{2})\boldsymbol{1}\:,\label{eq:}
\end{eqnarray}
where, for our model, $a,a'=\pm$ and the operator
$Z_{a}^{\dagger}(\theta)$ creates a ($a$) particle
excitation with rapidity $\theta$. The scattering matrices were given in (\ref{Smatrix}). Starting with these operators, asymptotic states can be constructed by their repeated
action on the vacuum $|0\rangle$. In particular, the incoming and outgoing states can be written as
\begin{eqnarray}
|\theta_{1},\theta_{2},...,\theta_{n}\rangle_{a_{1},...,a_{n}}^{\rm in}&=&Z_{a_{1}}^{\dagger}(\theta_{1})Z_{a_{2}}^{\dagger}(\theta_{2})...Z_{a_{n}}^{\dagger}(\theta_{n})|0\rangle,\nonumber
\\
|\theta_{n},\theta_{n-1},...\theta_{1}\rangle_{a_{1},...,a_{n}}^{\rm out}&=&Z_{a_{n}}^{\dagger}(\theta_{n})Z_{a_{n-1}}^{\dagger}(\theta_{n-1})...Z_{a_{1}}^{\dagger}(\theta_{1})|0\rangle,\qquad\theta_{1}>\theta_{2}>...>\theta_{n}
\label{eq:basis}
\end{eqnarray}
where the particular ordering in the rapidities ensures the normalisation
\beq
\phantom{}{}_{a'_{1},a'_{2}}^{\quad \rm out}\langle\theta'_{1},\theta'_{2}|\theta_{1},\theta_{2}\rangle_{a_{1},a_{2}}^{\rm in}=S_{a_{1}a_{2}}(\theta_{1}-\theta_{2})\,\delta_{a_{1}a'_{1}}\, \delta_{a_{2}a'_{2}}\, 2\pi\delta(\theta_{1}-\theta'_{1})\, 2\pi\delta(\theta_{2}-\theta'_{2})\,,
\eeq
for $\theta_{1}>\theta_{2}$ and $\theta'_{1}>\theta'_{2}$ and similarly for higher particle states. 

In massive relativistic IQFT, the action of conserved charges on asymptotic states is the sum of the one-particle eigenvalues introduced earlier, that is
\begin{equation}
Q_{s}^{\rm e}|\th_{1},...\th_{n}\ket_{a_{1},\ldots,a_{n}}=\sum_{i=1}^{n}q_{a_{i}}^{s}\cosh\left(s\th_{i}\right)|\th_{1},...\th_{n}\ket_{a_{1},\ldots,a_{n}}\label{eq:ChargeEven}
\end{equation}
for even charges and
\begin{equation}
Q_{s}^{\rm o}|\th_{1},...\th_{n}\ket_{a_{1},\ldots,a_{n}}=\sum_{i=1}^{n}q_{a_{i}}^{s}\sinh\left(s\th_{i}\right)|\th_{1},...\th_{n}\ket_{a_{1},\ldots,a_{n}}\label{eq:ChargeOdd}
\end{equation}
for odd charges, where $s$ is the Lorentz-spin and $q_{a}^s \cosh(s \theta), q_a^s \sinh(s\theta)$ are the
one-particle eigenvalues of the charges. This means that the charges themselves admit representations of the form
\begin{eqnarray}
Q_{s}^{\rm e}&=&\sum_{i}\int\frac{d\th}{2\pi}\,q_{a_{i}}^{s}\cosh\left(s\th_{i}\right)Z_{a_{i}}^{\dagger}(\th)Z_{a_{i}}(\th)\,,\label{eq:ChargeEvenOp}\\
Q_{s}^{\rm o}&=&\sum_{i}\int\frac{d\th}{2\pi}\,q_{a_{i}}^{s}\sinh\left(s\th_{i}\right)Z_{a_{i}}^{\dagger}(\th)Z_{a_{i}}(\th)\,.
\end{eqnarray}

\subsection{Integrable Quenches}
\label{Ks}
Integrable quenches are defined as quenches in integrable models from initial states where {\OCA the expectation values of} all the parity-odd charges vanish \cite{IntegrableQuench}. They can generally be written as squeezed coherent
states, such as those explicitly constructed in \cite{FM,SFM, HST}, which means that they either take the form,
\begin{equation}
|\Psi_0\rangle=\mathcal{N}\exp\left(\frac{1}{2}{\OCA \sum_{a,b}}\int\frac{d\theta}{2\pi}K_{ab}(\theta)Z_{a}^{\dagger}(-\theta)Z_{b}^{\dagger}(\theta)\right)|0\rangle\,,\label{IntegrableState}
\end{equation}
or
\begin{equation}
|\Psi_0\rangle=\mathcal{N}\exp\left({\OCA \sum_{a}}\frac{g_{a}}{2}Z_{a}^{\dagger}(0)+\frac{1}{2}{\OCA \sum_{a,b}}\int\frac{d\theta}{2\pi}K_{ab}(\theta)Z_{a}^{\dagger}(-\theta)Z_{b}^{\dagger}(\theta)\right)|0\rangle\;.\label{IntegrableState1pt}
\end{equation}
These types of states have been previously studied in the context of boundary integrable quantum field theory, where the functions $K_{ab}(\theta)$ are related to boundary reflection amplitudes \cite{GZ}. 

For the HSG model we will only consider integrable initial states in which
$K_{\pm\pm}(0)=0$ and no one-particle state is present in the
exponential. Such states can be specifically written in terms of $K$-functions $K_{\pm\mp}(\theta)$ which satisfy the consistency equations (boundary crossing equations)
\begin{equation}
K_{\pm\mp}(\th) =S_{\pm\mp}(2\th)K_{\mp\pm}(-\th)\,,\label{eq:BoundaryCrossing}
\end{equation}
giving
\beqa 
|\Psi_0\rangle & = & \mathcal{N}\exp\left(\frac{1}{2}\int\frac{d\theta}{2\pi}K_{+-}(\theta)Z_{+}^{\dagger}(-\theta)Z_{-}^{\dagger}(\theta)+\frac{1}{2}\int\frac{d\theta}{2\pi}K_{-+}(\theta)Z_{-}^{\dagger}(-\theta)Z_{+}^{\dagger}(\theta)\right)|0\rangle \nonumber \\
 & = &\mathcal{N}\exp\left(\int\frac{d\theta}{2\pi}K_{+-}(\theta)Z_{+}^{\dagger}(-\theta)Z_{-}^{\dagger}(\theta)\right)|0\rangle
\label{IntegrableStateIHSG}
\eeqa 
The $S$-matrices $S_{\pm\mp}(\theta)$ can be
written in terms of the scattering phases $\delta_{\pm\mp}(\theta)$ as $S_{\pm\mp}(\th) =e^{i\delta_{\pm \mp}(\th)}$ with (note that the kernels $\varphi_{\pm\mp}(\theta)=\delta'_{\pm\mp}(\theta)$, where the prime means differentiation w.r.t. $\theta$) 
\begin{equation}
\delta_{\pm\mp}(\th) =\pm \frac{\pi}{2}\mp 2\arctan\left(\tanh\frac{\th\pm\sigma}{2}\right)\,.
\end{equation}
From this structure, it is easy to find a solution of the boundary crossing equation (\ref{eq:BoundaryCrossing}), namely
\begin{equation}
K_{\pm\mp}(\th) =\sqrt{S_{\pm\mp}(2\theta)}=e^{\frac{i\delta_{\pm\mp}(2\th)}{2}}\,.
\end{equation}
Other solutions can be obtained from the above, by multiplying with CDD factors $f(\theta)$. Explicitly, we then have
\begin{equation}
K_{\pm\mp}(\th)  =f(\th) \exp\left[\pm \frac{i\pi}{4} \mp i\arctan\left(\tanh\frac{2\th\pm\sigma}{2}\right)\right]
\,,
\end{equation}
where $f(\th)=f(-\th)$ is an arbitrary even function. Except for this function $f(\theta)$, the solution above is a pure phase. As it turns out, only the function $f(\theta)$ plays a role in our analysis as all quantities of interest are functions of $|K_{+-}(\theta)|$ only. The task is then to find real functions $f(\theta)$ that give rise to a sensible dynamics. This can be done very explicitly for free models \cite{SFM} but is much more difficult for interacting theories \cite{HST}. 

One natural way to further constrain the set of functions $f(\theta)$ is to require that they decay rapidly for large $\theta$. In fact, many known solutions such as those for the free boson \cite{SFM} and sinh-Gordon \cite{BPC,DS} model decay as
 $f(\theta) \sim e^{-2\theta}$ for $\theta \gg 1$ in order to ensure that the injected
energy density (due to the quench) is finite. 

A property that all these solutions also generally share is that $|K(\theta)|\leq 1$ 
{\DXH as long as they have no singularity at zero rapidity. Nevertheless as was shown in \cite{SingularOverlaps}, the singularity of the $K$-functions implies the presence of a 1-particle overlap, which we chose to be vanishing for simplicity.} This requirement seems natural from several points of view, including the fact that the $K(\theta)$ functions are solutions of equations for boundary reflection amplitudes, which must clearly satisfy this property. In this paper we consider three solutions which are compatible with all the above properties. 

{\OCA Although the properties just discussed do constrain the set of possible solutions $K_{+-}(\theta)$ they are not sufficient to entirely fix them. As mentioned earlier finding the functions $K(\theta)$ that correspond to a particular global quench is generally difficult for interacting theories. A general structure for mass quenches in diagonal theories is known (i.e. sinh-Gordon) but this takes as input a solution of the boundary reflection equation. Such a solution is still outstanding for the present model. In addition, for our model the mass is not the only parameter that might be changed in a global quench. We also have $\sigma$ and here we know even less about what the amplitude $K_{+-}(\theta)$  should look like. 

For these reasons, our strategy in the present paper is to consider the three $K$-functions listed below. These are all functions that have appeared previously in the literature but which cannot be linked explicitly to an specific global quench in the current model. Instead, they should be seen as representing types of solutions, which have all the main functional properties required and are expected to lead to results whose main features will be common to a large family of global quenches. In other words, we expect that we will still be able to find some universal properties for the entanglement density and related quantities (which are the focus of this paper) starting from the functions below.}

\begin{enumerate}
    \item ``Free Boson Solution":
\begin{equation}
| K_{ \rm B}(\theta)|=\left|\frac{\sqrt{\alpha^2 \sinh^2\theta+1}-|\alpha \cosh\theta|}{\sqrt{\alpha^2 \sinh^2\theta+1}+|\alpha \cosh\theta|}\right|\,,
\label{freeB}
\end{equation}
which is the $K$-function for a mass quench in the massive free boson QFT (see \cite{SFM} for a derivation and alternative representation) and $\alpha=m_{0}/m$ in this case, i.e., the ratio of the pre-quench mass $m_0$ and the post-quench mass $m$. A useful property is
\beq
|K_{\rm B}(\theta)|\sim C_{\rm B}(\alpha) e^{-2|\theta|}\qquad \mathrm{with}\qquad C_{\rm B}(\alpha):=\frac{|1-\alpha^2|}{\alpha^2} \quad {\rm for} \quad |\theta|\gg 1\,.
\label{asiB}
\eeq 
\item ``Squared Solution":
\begin{equation}
|K_{\rm S}(\theta)|=\frac{\sinh(2|\log\alpha|)}{2 \cosh(\theta+|\log\alpha|)\cosh(\theta-|\log\alpha|)}\,,
\label{asi}
\end{equation}
which, for $\alpha\ll 1$ is very similar in shape to the free boson solution and is characterised by a large plateau of width $2|\log\alpha|$ centered around $\theta=0$.
The asymptotics of this function is:
\beq
|K_{\rm S}(\theta)|\sim C_{\rm S}(\alpha) e^{-2|\theta|}\qquad \mathrm{with}\qquad C_{\rm S}(\alpha):=\frac{|1-\alpha^4|}{\alpha^2} \quad {\rm for} \quad |\theta|\gg 1\,.
\eeq 

\begin{figure}[t]
\begin{center}
	\includegraphics[width=9cm]{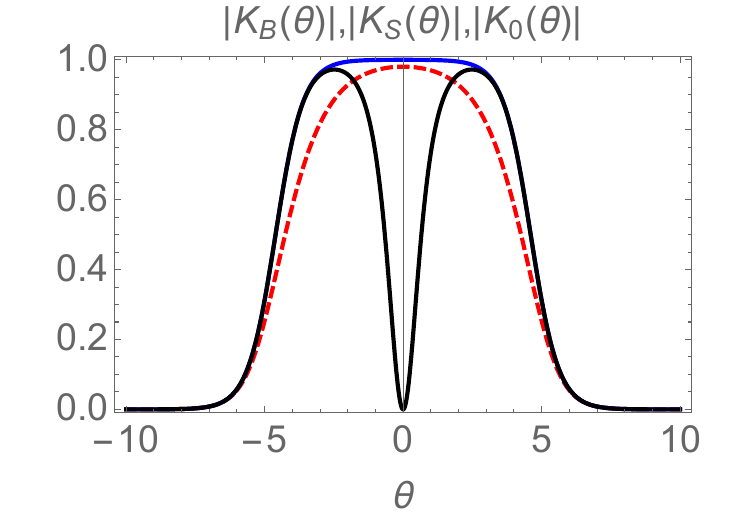}
    \caption{The functions $|K_{\rm 0}(\theta)|$ (black), $|K_{\rm S}(\theta)|$ (blue) and $|K_{\rm B}(\theta)|$ (red, dashed) for $\alpha=0.01$. In this case $|\log\alpha|\approx 4.6$ which is half the size of the plateau of the second function and also gives the location of the maxima of $|K_0(\theta)|$ at $\theta=\pm \log\alpha$. Reducing the value of $\alpha$ all functions become increasingly square-shaped, with $K_{\rm 0}(\theta)$ preserving its zero at the origin.}
    \label{Kfun}
    \end{center}
\end{figure}
\item ``Modified Squared Solution":
\begin{equation}
| K_{\rm 0}(\theta)|=| K_{\rm S}(\theta)|-\frac{\tanh|\log\alpha|}{\cosh(2\theta)}\,,
\end{equation}
which is very similar to $| K_{\rm S}(\theta)|$ but incorporates a zero at zero rapidity. The asymptotics in this case reads:
\beqa
|K_{\rm 0}(\theta)|&\sim & 2(\sinh(2|\log\alpha|)-\tanh|\log\alpha|) e^{-2|\theta|}\nonumber\\
&=& C_{\rm 0}(\alpha) e^{-2|\theta|}\qquad \mathrm{with}\qquad C_{\rm 0}(\alpha):=\frac{(1+\alpha^4)|1-\alpha^2|}{\alpha^2(1+\alpha^2)} \quad {\rm for} \quad |\theta|\gg 1\,.
\eeqa 
\end{enumerate}
In all examples $\alpha$ is a quench parameter related to pre- and post-quench features of the model. However, because the $K$-functions above are not derived from first principles for our model, this parameter $\alpha$ can not be easily linked to any of the known parameters in the theory. Instead, we will think of $\alpha$ as an effective variable whose value gives a measure of the energy that is injected in the system by the quench. {\OCA Note also that our amplitudes $K_B(\theta), K_S(\theta)$ and $K_0(\theta)$ can not be interpreted as reflection amplitudes, even if they characterise a state that has the form of a boundary state. As we have seen, all three amplitudes decay for large $|\theta|$ whereas reflection amplitudes usually tend to a finite constant for large energies (see for instance the integrable boundary reflection matrices which were studied in \cite{GZ} for the Ising model).}

In all our numerical and analytical results we have taken $\alpha<1$.
The smaller $\alpha$ is, the larger the quench, that is the larger the amount of energy that is injected in the system at $t=0$. 
A plot of the three $K$-functions for $\alpha=0.01$ is presented in Fig.~\ref{Kfun}. 
Another feature of all $K$-functions is that they vanish identically for $\alpha=1$, that is, when there is no quench and the pre- and post-quench ground states coincide $|\Psi_0\ket=|0\ket$.

\section{The Quench Action Method}
\label{action}

The quench action method (QAM) or overlap TBA method \cite{Cfab1, Cfab2, IlievskiCauxQPPicture,QuenchActionReview,ExcitedBetheStates} was primarily developed to characterise the stationary expectation
values of local operators after (integrable) quantum quenches. Under certain, yet often natural assumptions, the QAM can be used to compute the von Neumann \cite{QuasiAgain,ac-17c} and R\'enyi \cite{ac-17a,ac-17b,mac-18,bka-22,pvcc-22} entanglement entropies, both in the stationary state and their time evolution.  In particular, the computation of the time evolution of von Neumann entropies requires an additional ingredient, namely the applicability of the quasiparticle picture \cite{QuasiParticlePicture} which we review at the end of this section. For certain initial states the time evolution of R\'enyi entropies can be computed by space-time duality techniques \cite{bka-22} that, for integrable models, generalise the QAM \cite{bka-22}.

The central quantity in this method is an effective free energy or quench action functional (QAF) from which various quantities can be computed. Following the logic of the original literature \cite{Cfab1}, the starting point is  the post-quench time-dependent expectation value of a local operator. This expectation value can be
written as

\begin{equation}
\langle\Psi_{0}|\oo(t)|\Psi_{0}\rangle=\frac{1}{\langle\Psi_{0}|\Psi_{0}\rangle}\sum_{\Phi,\Phi'}e^{-\eps_{\Phi}^{\text{*}}-\eps_{\Phi'}}e^{i\left(\omega_{\Phi}-\omega_{\Phi'}\right)t}\langle\Phi|\oo(t)|\Phi'\rangle\,,
\end{equation}
where $|\Psi_0\ket$ is the pre-quench state, $\oo(t)$ is the time-evolved operator in the Heisenberg picture, $\Phi$ and $\Phi'$ label eigenstates of the post-quench Hamiltonian
with energies $\omega_{\Phi}$ and $\omega_{\Phi'}$ respectively, $\eps_{\Phi}$ is the logarithmic overlap
\begin{equation}
\eps_{\Phi}=-\log\langle\Phi|\Psi_{0}\rangle\,,
\end{equation}
and $\eps^\star_\Phi$ its complex conjugate.
The main idea is to take a continuum or thermodynamic limit in which summation over eigenstates is replaced by a functional
integral over the root densities $\rho$ giving

\begin{equation}
\langle \Psi_0| \oo(t)|\Psi_0\rangle=\frac{1}{\langle\Psi_{0}|\Psi_{0}\rangle}\int\mathcal{D}[\rho]e^{S[\rho]}\sum_{\Phi}\left[e^{-\eps_{\Phi}^{\text{*}}-\eps[\rho]}e^{i\left(\omega_{\Phi}-\omega[\rho]\right)t}\langle\Phi|\oo(t)|\rho\rangle+\Phi\leftrightarrow\rho\right]\,,\label{eq:QAOp}
\end{equation}
where $S[\rho]$ is the Yang-Yang entropy, which is equal to the logarithm
of the number of micro-states corresponding to a given macro-state.
If $\mathcal{O}$ is a local operator, the matrix element $\langle\Phi|\oo(t)|\Phi'\rangle$
is non zero if $\Phi$ and $\Phi'$ correspond to the same macro-state
in the thermodynamic limit up to microscopic differences or in other words a finite
number of excitations \cite{ExcitedBetheStates}. A key assumption
is that the functional integral is dominated by
a single saddle-point root density $\rho_{sp}$. This saddle-point
density can be determined by minimising an effective free energy functional $\mathcal{F}[\rho]$ which is defined through
\begin{equation}
\langle\Psi_{0}|\Psi_{0}\rangle=\int\mathcal{D}[\rho]e^{\mathcal{F}[\rho]}\, \quad {\rm with} \quad \mathcal{F}[\rho]=-2{\rm Re}(\eps[\rho])+S[\rho]\label{eq:QAOverlap}
\end{equation}
which ensures the normalisation of $\langle \Psi_0| \oo(t)|\Psi_0\rangle$ by the condition
\begin{equation}
\frac{\delta\mathcal{F}[\rho]}{\delta\rho}=0\,.\label{eq:AQSaddlePoint}
\end{equation}
An important observation is that the characterisation of the saddle point by
\eqref{eq:QAOverlap} is also valid for the expansion \eqref{eq:QAOp} as long as the matrix elements of $\oo$ do not grow exponentially with system size. 
Since matrix elements of local operators are
usually of order 1 in system size, the insertion of a local operator does not change
the saddle point \eqref{eq:QAOverlap} (a remarkable example for which this is not true consists of the R\'enyi entropies with index different from $1$ \cite{ac-17a,ac-17b,mac-18}, 
a fact that also reflects into a very unusual time evolution \cite{bka-22}).
To calculate the saddle-point density $\rho_{\rm sp}$ by Eq. (\ref{eq:AQSaddlePoint}), we then just need to know 
the overlaps, or more
precisely, the extensive part of the logarithm of the overlaps. Once
the overlaps are known, the construction of the root density is relatively
straightforward. 

Having computed the saddle-point density $\rho_{\rm sp}$, operator expectation
values can be computed, and in particular,
the von Neumann entropy of the stationary state can be immediately
obtained as well, since, as we shall see, the Yang-Yang entropy of the representative
state is the thermodynamic entropy which corresponds to the stationary entanglement entropy \cite{QuasiAgain}. 

\subsection{Finite Volume Initial State and Logarithmic Overlap}

In order to compute the saddle-point density we need to express
the logarithmic overlaps as  functions of the density $\rho$. To
do so, we follow the logic of {\OCA \cite{BPC,AC2,Cfab2,OnePointFunctions,108}} and consider the theory and the initial state in finite volume. In
finite volume, the eigenstates of the theory are characterised by solutions of
the Bethe-Yang equations, which for a set if integers $\{I_{k}\}$
specify the allowed rapidites of the stable particles. In our case
having two different particle species and a non-trivial $S$-matrix
only between the two species, the equations can be written as 
\begin{equation}
\begin{split}Q_{k}^{+}= & ML\sinh\theta_{k}+\sum_{j=1}^{m}\delta_{+-}(\theta_{k}-\theta'_{j})=2\pi I_{k}\;,\quad\quad k=1,\dots,n\;\\
Q_{l}^{-}= & ML\sinh\theta'_{l}+\sum_{j=1}^{n}\delta_{-+}(\theta'_{l}-\theta_{j})=2\pi J_{l}\;,\quad\quad l=1,\dots,m\;
\end{split}
,\label{eq:BY}
\end{equation}
where the set of rapidities $\{\th_{k}\}$ and $\{\th'_{l}\}$ as
well as the set of quantum numbers $\{I_{k}\}$ and $\{J_{l}\}$ correspond
to the $(+)$ and $(-)$ particles respectively and we also defined the
functions $Q_{k}^{\pm}$ which can be general functions of rapidites. $\varphi_{\pm\mp}(\theta)$ are the scattering phases introduced in Section \ref{model}. 
In contrast to the Bethe Ansatz solution on spin chains \cite{BetheWaveFunction},
this scattering state is just an approximate solution of the model
in which finite size effects that decay exponentially in the volume are neglected
\cite{Luscher}. 

The state corresponding to quantum numbers $\{I_{1},\dots,I_{n},J_{1},\dots,J_{m}\}$
is denoted by 
\beq 
|\{I_{1},\dots,I_{n},J_{1},\dots,J_{m}\}\rangle_{L}^{(n,m)}\;,
\eeq 
and is independent (up to a possible phase ambiguity) of the ordering
of $I$-s and $J$-s. Such states are normalised so that their
scalar products are 
\beq
_{\quad\,L}^{\;(n,m)}\langle\{I_{1},\dots,I_{n},J_{1},\dots,J_{m}\}|\{I'_{1},\dots,I'_{n'},J'_{1},\dots,J'_{m'}\}\rangle_{L}^{(n',m')}=\delta_{nn'}\delta_{mm'}\prod_{k=1}^n\delta_{I_{j}I'_{j}}\prod_{l=1}^m\delta_{J_{l}J'_{l}}\;,
\eeq
with the quantum numbers ordered by convention as $I_{1}<\dots<I_{n}$,
$J_{1}<\dots<J_{m}$ and similarly for the primed indices. The total energy and momentum can be expressed as 
\beq 
E=\sum_{k=1}^{n}M\cosh\theta_{k}+\sum_{l=1}^{m}M\cosh\theta'_{l}+O(e^{-\mu L})\,,\quad P=\sum_{k=1}^{n}M\sinh\theta_{k}+\sum_{l=1}^{m}M\sinh\theta'_{l}+O(e^{-\mu L})
\eeq 
up to exponential corrections governed by some mass scale $\mu$. A systematic treatment of exponential corrections to excitation
energies can be found in \cite{Luscher,KlassenM,BajnokJanik,Hatsuda}.

It is useful to introduce the rapidity space density of $n$-particle
states, which is given by the determinant of the Jacobian 
\begin{equation}
\rho_{n+m}(\theta_{1},\dots,\theta_{n},\theta'_{1},\dots,\theta'_{m}):=\det \mathcal{J}\;,\qquad \mathcal{J}_{kl}=\frac{\partial\tilde{Q}_{k}}{\partial\tilde{\theta}_{l}},
\end{equation}
with
\begin{equation}
\tilde{Q}_{\kappa}=  \begin{cases}
Q_{k}^{+} & \text{if}\quad k \in[1,n]\\
Q_{l}^{-} & \text{if}\quad k\in [n+1,n+m]
\end{cases}\,,\qquad 
\text{\ensuremath{\tilde{\theta}}}_{l}=  \begin{cases}
\th_{l} & \text{if}\quad l \in [1,n]\\
\th'_{l} & \text{if}\quad l \in [n+1,n+m] \,.
\end{cases}
\end{equation}
Let us further characterise the integrable initial state \eqref{IntegrableStateIHSG}
in finite volume. In infinite volume, the state consists only of  pairs of $(+-)$
particles with opposite momentum. This feature is present
in the finite volume state as well, if properly defined. We can consider a general set of integers $\{I_{k}\}$ for the $(+)$ particles:
then the quantum numbers for the $(-)$ particles are fixed, and the
corresponding set is

\begin{equation}
\{J_{l}\}_{1}^{m}=\{-I_{1},\ldots,-I_{n}\}\,,\quad m=n\,.
\end{equation}
It is easy to check, that with this choice the solutions of the Bethe-Yang
equations \eqref{eq:BY} are such that for all $k=1,\ldots,m=n$
\begin{equation}
\th_{k}=-\th'_{k}\,,
\end{equation}
in other words, all parity-odd conserved charges annihilate such finite
volume states, that is
\begin{equation}
\begin{split} & Q_{s}^{\rm o}|\th_{1},...,\th_{n},-\th_{1},...,-\th_{n}\r_{L}^{(n,n)}=\\
 & \left(\sum_{i=1}^{n}q_{s}\sinh\left(s\th_{i}\right)+\sum_{i=1}^{n}q_{s}\sinh\left(-s\th_{i}\right)\right)|\th_{1},...,\th_{n},-\th_{1},...,-\th_{n}\r_{L}^{(n,n)}=0\,.
\end{split}
\end{equation}
This is in fact a natural definition of
integrable states in finite volume given the fact that in infinite
volume the squeezed coherent form of the state and annihilation by
all odd charges are completely equivalent.
For such a state we define the restricted $Q$-functions and the restricted
density of states, which incorporate the above constraint
\begin{equation}
\bar{Q}_{k}^{(n)}=ML\sinh\theta_{k}+\sum_{j=1}^{m}\delta_{+-}(\theta_{k}-\theta'_{j})=2\pi I_{k}\;,\quad\quad k=1,\dots,n\;\label{barQ}
\end{equation}
and
\begin{equation}
\bar{\rho}_{n}(\theta_{1},\dots,\theta_{n})=\det \mathcal{J}\;,\qquad \mathcal{J}_{kl}=\frac{\partial\bar{Q}_{j}}{\partial\th_{k}}\,.
\end{equation}
Then the complete initial state in finite volume reads \begin{equation}
\begin{split}|\Psi_0\rangle_{L}= & \mathcal{N}_{L}\sum_{n}\sum_{\{I_{k}\}_{1}^{n}}N_{n}(\th_{1,\ldots,}\th_{n};L)\left(\prod_{k=1}^{n}K(\th_{k})\right)|\{I{}_{1},\dots,I_{n},-I{}_{1},\dots,-I_{n}\}\rangle_{L}^{(n,n)}\,,\end{split}
\label{IntegrableStateIHSGFinVol}
\end{equation}
where the $1/n!$ factor from the expansion of the exponential function
is absent, since the summation is over the set $\{I_{k}\}_{1}^{n}$,
in which all possible quantum numbers appear but in an ordered way.
The normalisation factors $N_{n}$ were determined in \cite{OnePointFunctions}
up to finite size effects with exponential decay. Their general expression
reads 
\beq 
N_{n}(\th_{1,\ldots,}\th_{n};L)=\frac{\sqrt{\rho_{n+n}(\theta_{1},\ldots,\th_{n},-\th_{1},\ldots,-\th_{n})}}{\bar{\rho}_{n}(\theta_{1},\dots,\theta_{n})}=1+{O}(L^{-1})\,.
\eeq 
The functions $K(\theta):=K_{+-}(\theta)$ are any of the three $K$-functions discussed earlier. 
In order to describe the initial state as a representative state in terms of
continuous densities, we can first exploit the parity relation
\begin{equation}
\rho_{+}(\th)=\rho_{-}(-\th):=\rho(\theta)\,,
\end{equation}
so that the logarithmic overlap $-2{\rm Re}(\log{}_L \bra \rho|\Psi_0\ket_L )$ can be computed as
\begin{equation}
-2{\rm Re}(\log {}_L\bra  \rho|\Psi_0\ket_L ) =-2{\rm Re}\left[ \log\prod_{k=1}^{n}K(\th_{k})\right]+C =-\sum_{k=1}^{n}\log| K(\th_{k})|^{2}+C\,,
\end{equation}
where $|\rho\ket_{L}$ is one particular and appropriate
realisation of the macro-state dictated by the density $\rho(\theta)$ and $C$ is an unknown constant coming
from the normalisation of the state. In the thermodynamic limit, this formula becomes
\begin{equation}
\begin{split}
\lim_{L\rightarrow\infty}-2{\rm Re}(\log {}_L\bra  \rho|\Psi_0\ket_L )  & =-L\int_{-\infty}^{\infty}\text{d}\th\,\rho(\th)\log| K(\th)|^{2}\end{split}
+C.
\end{equation}

\subsection{Quench Action Functional}

We have now constructed all the ingredients to write the QAF of interest that reads 
\beqa 
\mathcal{F}[\rho]&=&-L\int_{-\infty}^{\infty}\text{d}\th\,\rho(\th)\log| K(\th)|^{2}+C+\nonumber\\
&& -L\int_{-\infty}^{\infty}\ud\th\left[\rho_{t}(\th)\log\rho_{t}(\th)-\rho(\th)\log\rho(\th)-\rho_{h}(\th)\log\rho_{h}(\th)\right]\,,\label{eq:QAFunctionalFinal}
\eeqa 
where we have explicitly written the Yang-Yang entropy of the macro-state (second line). 
An important fact is, as already mentioned, that due the pair structure,
the densities $\rho_{\pm}(\theta)$ are not independent and so the Yang-Yang entropy can be written in terms of just one density. As defined earlier, $\rho(\th)=\rho_+(\theta)$ is the density of occupied states, whereas $\rho_{\rm h}(\theta)$ represents the density of holes or unoccupied states and $\rho_{\rm t}(\theta)$ is the total density, that is their sum. 

The functional \eqref{eq:QAFunctionalFinal} is completely equivalent
to that of a Gibbs ensemble if $-\log| K(\th)|^{2}$
is replaced by the usual energy term $m\beta \cosh(\th)$. This means
that the saddle-point equations for $\mathcal{F}[\rho]$ are exactly given by (\ref{tba}) with driving term
\beq 
\omega(\th)=-\log|K(\theta)|^2\,.
\eeq 
It follows that the densities $\rho(\theta),\rho_{\rm h}(\theta)$ and $\rho_{\rm t}(\theta)$ are related in the usual way, namely
\beq 
\frac{\rho(\theta)}{\rho_{\rm t}(\theta)}=n_+(\th)\, \quad \mathrm{and}\quad  \frac{\rho_{\rm h}(\theta)}{\rho_{\rm t}(\theta)}=1-n_+(\th)\,,
\eeq 
where $n_+(\theta)$ that is the occupation function defined in (\ref{filling}).

\subsection{Quasiparticle Picture for Entanglement Evolution}
\label{quasi}

The Yang-Yang entropy of the representative state equals, by definition, the thermodynamic entropy. The latter is also the long-time limit of the extensive part of the stationary entanglement entropy associated with the equilibrated state \cite{EEDyn,QuasiAgain,ac-17c}.
However, a first principle analytic computation of the out-of-equilibrium time evolution of the entanglement entropy and related quantities is a notoriously difficult problem even for cases tractable by methods like QAM
(see, e.g., Refs. \cite{mck-22,ik-22} for some recently proposed truncated conformal space approaches). {\DXH From a phenomenological viewpoint, indeed a great variety of different behaviours have been detected ranging from a linear growth of entanglement and Rényi entropies to its suppression and the onset of oscillatory patterns associated with particle confinement \cite{confi} or the presence of 1-particle overlaps in the post-quench expansion of the initial state \cite{DelfinoOscillations, clsv-20,ch-sG-REOscillations}.} Despite the enormous difficulties, when one is interested in the evolution of the von Neumann entropy associated with pure initial states an intuitive and extremely accurate technique has been proposed in Ref. \cite{QuasiParticlePicture} and further justified in \cite{QuasiAgain}. This is based on the quasiparticle picture, which is applicable when quantum systems admit stable quasiparticles with purely elastic scattering and when the initial state is a low-entangled state (for instance the ground state of a gapped system) whose expansion in the post-quench basis is made up of pairs of quasiparticles with opposite momentum. These criteria are naturally fulfilled for the quenches we study in this work (obviously many of these assumptions can be relaxed \cite{btc-18,bfpc-18,abf-19,bc-18,a-19,lcp-22}, but this is not of interest here). Given these considerations the evolution of the entanglement entropy for an interval of length $L$ in an infinite system can be written as \cite{QuasiParticlePicture,QuasiAgain}
\begin{equation}
    S(L,t)=2t\int_{p>0}\text{d}p\, s_{\text{pair}}(p) 2v(p)\Theta(L-2 v(p)t)+2L\int_{p>0}\text{d}p\, s_{\text{pair}}(p) \Theta(2 v(p)t-L)
\label{QPPEntanglement}
\end{equation}
in a model consisting of one particle species. In the above formula, $v(p)$ is the velocity of the quasiparticles and is assumed to be a parity odd function of $p$ with $v(p)>0$  if $p>0$; and $s_{\text{pair}}(\theta)$ is a spectral entropy density, which accounts for the amount of entanglement carried by one pair of particles. The interpretation of Eq. \eqref{QPPEntanglement} is very natural: the entanglement between the subsystem and its complement is generated by particle pairs where one particle is in the subsystem and the other in its complementary region. In this reasoning we think of the particles with fixed trajectories and velocities $\pm v(p)$ created at single points equally distributed over space. Eq. \eqref{QPPEntanglement} predicts linear growth in time followed by saturation proportional to the subsystem's length $L$. To give quantitative predictive power to Eq. \eqref{QPPEntanglement}, it has been pointed out that (i) $v(p)$ must be identified with the effective velocity of particles, as defined in (\ref{veff}); (ii) $s_{\text{pair}}(p)$ is the spectral entropy density $s(p)$ of the stationary state. 

More precisely, in the long time and large subsystem limit, we end up with 
\begin{equation}
    \lim_{L\rightarrow \infty} \frac{S(L,t)}{L}=2\int_{p>0}\text{d}p\, s_{\text{pair}}(p)=\int \text{d}p\, s_{\text{pair}}(p)=\frac{1}{2}\int\text{d}p\, s(p)\, 
\label{QPPEntanglementTInf}
\end{equation}
where we exploited that the spectral entropy associated with a pair satisfies $s_{\text{pair}}(p)=s_{\text{pair}}(-p)$ and that due to the pair structure, the single particle spectral entropy density $s(p)$ eventually characterises the entire entropy contribution of a pair with opposite velocities. That is, $s_{\text{pair}}(p)=\frac{1}{2}s(p)$ as long as one intends to keep the whole real axis as the range of integration. In fact, it is often possible to obtain $s(p)$ via methods like QAM which gives information about the long time stationary state.

{\DXH It is important to stress that the validity and applicability of this set of ideas have been checked for the case of the XXZ chain as well in \cite{QuasiAgain}, that is also in an interacting integrable system for quenches that fulfil the requirements discussed above. A surprising observation of \cite{QuasiAgain} was that the predictions of the QPP regarding the initial time evolution are remarkably accurate already at very short times. Nevertheless, one generally expects that QPP becomes applicable at intermediate times and it may not capture the physics of very short times. This expectation is partially confirmed by \cite{TIsing} in which besides an initial growth additional power-law corrections with negative exponents were found in a free system, although for Rényi entropies.
For this reason we can rephrase \eqref{QPPEntanglement} in a slightly more precise way (already suggested in \cite{QuasiAgain}) and use an appropriate scaling of our space-time variables, i.e., the subsystem size $L$ and the time $t$ elapsed after the quench, yielding
\begin{equation}
   \lim_{L \rightarrow \infty} \frac{1}{L} S(L, t=\tau L)=2\tau\int_{p>0}\text{d}p\, s_{\text{pair}}(p) 2v(p)\Theta(1-2 v(p)\tau)+2 \int_{p>0}\text{d}p\, s_{\text{pair}}(p) \Theta(2 v(p) \tau-1)\,.
\label{QPPEntanglementScaled}
\end{equation}
}

The quasiparticle picture reviewed above can be easily applied to our particular model and to the quench protocols we consider, due to the integrability of the theory and the imposed pair structure between the ($+$) and ($-$) particles. To do so, we first  replace the integration variable with the rapidity $\theta$, which is more convenient for interacting integrable models. Nevertheless, some little care  has to be taken since in our case particles of different species form a pair. Additionally, the effective velocities satisfy $v^{\text{eff}}_+(\theta)=-v^{\text{eff}}_-(-\theta)$. For simplicity, we set $v^{\text{eff}}(\theta):= v^{\text{eff}}_+(\theta)$, which, therefore, can be unambiguously attributed to a pair. {\DXH For the sake of brevity, in the equations below, we do not yet make the scaling limit explicit, nevertheless it is implicitly assumed.} Given these considerations, Eq.~\eqref{QPPEntanglement} can be rewritten as
\begin{equation}
\begin{split}
    S(L,t)=&t\int_{v^{\text{eff}}(\theta)>0}\text{d}\theta\, s_{+-}(\theta) 2v^{\text{eff}}(\theta)\Theta(L-2 v^{\text{eff}}(\theta)t)\\
    +&t\int_{v^{\text{eff}}(\theta)<0}\text{d}\theta\, s_{-+}(\theta) 2|v^{\text{eff}}(\theta)|\Theta(L-2 |v^{\text{eff}}(\theta)|t)\\
    +&L\int_{v^{\text{eff}}(\theta)>0}\text{d}\theta\, s_{+-}(\theta) \Theta(2 v^{\text{eff}}(\theta)t-L)\\
    +&L\int_{v^{\text{eff}}(\theta)<0}\text{d}\theta\, s_{-+}(\theta) \Theta(2 |v^{\text{eff}}(\theta)|t-L)
\end{split}
\label{QPPEntanglementIGHS}
\end{equation}
where $s_{\pm\mp}$ is the associated spectral entropy density of a pair with a right-moving $\pm$ particle and a left-moving $\mp$ particle.
Based on the infinite time behaviour of the entropy in the thermodynamic limit, we can similarly relate the spectral entropy density of a pair with the single particle entropy densities of the stationary state. In particular, we have $s(\theta):=s_{+}(\theta)=s_{-}(-\theta)$, {\OCA where $s_{\pm}(\theta)$ are the spectral densities associated to each individual particle}. Due to the pair structure, these two contributions are not independent. The long time entropy of the infinite system reads
\begin{equation}
\begin{split}
   \lim_{L\rightarrow \infty} \frac{S(L,\infty)}{L}=\frac{1}{2}\int \text{d}\theta\, s_{+}(\theta)+\frac{1}{2}\int \text{d}\theta\, s_{-}(\theta)=\int \text{d}\theta\, s(\theta)\,,
\end{split}   
\label{QPPEntanglementIHSGTInf}
\end{equation}
where $s(\theta)$ is the same spectral entropy function that characterises the Yang-Yang entropy in Eq.~(\ref{eq:QAFunctionalFinal}).
This means that Eq. \eqref{QPPEntanglementIGHS} can be rewritten as

\begin{equation}
    S(L,t)=2t\int \text{d}\theta\, s(\theta) |v^{\text{eff}}(\theta)|\Theta(L-2 |v^{\text{eff}}(\theta)|t)
    +L\int \text{d}\theta\, s(\theta) \Theta(2 |v^{\text{eff}}(\theta)|t-L)\,,
\label{QPPEntanglementIGHS2}
\end{equation}
{\DXH or
\begin{equation}
    \lim_{L \rightarrow \infty} \frac{1}{L} S(L,t=\tau L)=2\tau \int \text{d}\theta\, s(\theta) |v^{\text{eff}}(\theta)|\Theta(1-2 |v^{\text{eff}}(\theta)|\tau)
    +\int \text{d}\theta\, s(\theta) \Theta(2 |v^{\text{eff}}(\theta)|\tau-1)\,.
\label{QPPEntanglementIGHS2Scaled}
\end{equation}
when the restriction of excluding short times is explicitly imposed.}
The initial linear growth of the entanglement entropy is implied by the above formula. It is therefore useful to characterise the initial entropy growth rate by the entropy production rate
{\DXH
\begin{equation}
\lim_{\tau\rightarrow 0}\lim_{L\rightarrow \infty}\frac{\text{d}S(L,\tau L)}{\text{d}\tau}=:\frac{\text{d}S}{\text{d}t}=2\int \text{d}\theta\, s(\theta)|v^{\text{eff}}(\theta)|\,.
\label{QPPEProdRateIGHS}
\end{equation}
}

\section{Quench Dynamics with Unstable Quasiparticles: Results}
\label{results}

In this section we present the core part of our paper, namely the explicit solution of the quench action TBA equations and the characterisation of the entanglement dynamics. We stress that although the solutions of TBA equations are numerical, all our results are exact. 

As detailed in Section \ref{model}, all quantities of interest can be obtained by solving the basic equations 
\beq 
\varepsilon_{\pm}(\theta)=-\log|K_i(\theta)|^2
-(\varphi_{\pm\mp}\star L_\mp)(\theta)\,\quad {\rm with} \quad \varepsilon_+(\theta)=\varepsilon_-(-\theta).
\label{main}
\eeq
The different functions $K_i(\theta)$, cf. Section \ref{Ks}, with $i=\rm B,S,0$ provide a useful way of parametrising certain families of GGEs with desirable properties and, in our case, dependent on a single quench parameter $\alpha$. Let us start by discussing briefly how the energy injected in the system depends on $\alpha$.
\subsection{Three Energy Regimes}
\label{regimes}
Following on from the discussion at the end of Section \ref{model}, we can rewrite the equations (\ref{main}) as in Eq.~(\ref{shifted}) so that all the $\sigma$ dependence is incorporated into a shifted driving term. Then, given the large $\theta$ asymptotics of the $K$-functions (see Subsection \ref{Ks}), we have that for large $\sigma$
\beq 
-\log|K(\theta+\frac{\sigma}{2})| \sim  2\theta+\sigma-\log C_i(\alpha)\,, \quad \mathrm{with} \quad i=\rm B,S,0\,,
\label{64}
\eeq 
where $C_i(\alpha)$ are the coefficients of $e^{-2|\theta|}$ in the large $\theta$ asymptotics of the $K$-functions (see again Subsection \ref{Ks}). In other words, quantities such as the entropy density and growth rate, that are obtained from the solutions of the TBA equations, will be functions of the universal scale 
\beq 
\kappa_i(\sigma,\alpha):=\log C_i(\alpha)-\sigma\,, \quad \mathrm{with} \quad i=\rm B,S,0,
\label{kappa2}
\eeq 
which now plays a role similar to an RG parameter. In particular, for $\alpha\ll 1$ which is the regime we are considering in most of our numerics, all functions $C_i(\alpha)\approx -2\log\alpha$ so that the energy injected by the quench grows logarithmically with $\alpha$. This implies the existence of three regimes: 
\begin{itemize}
    \item For { $\kappa_i(\sigma,\alpha)\gg 1$},  the energy injected by the quench is sufficient to form unstable particles.
    \item For $\kappa_i(\sigma,\alpha) \approx 0$, we are exactly at the threshold for the formation of unstable particles.
    \item For {$\kappa_i(\sigma,\alpha)\ll -1$}, the energy injected by the quench is not sufficient to excite unstable bound states and so the particles ($\pm$) do not interact. In fact, they behave as a system of two free fermions.
\end{itemize}

We should therefore expect to observe clear changes in all quantities of interest as the energy is varied, either by tuning the quench parameter $\alpha$ or the resonance parameter $\sigma$. We will study these changes for several physical quantities and for different choices of $K$-function. The quantities of interest are listed below:
\begin{enumerate}
    \item The stationary entanglement entropy per unity length $\frac{S}{L}$.
    \item The entanglement entropy production rate $\frac{{\rm d} S}{{\rm d}t}$.
    \item The total particle density $\frac{N_{\rm st}}{L}$ where
    \beq 
    N_{\rm st}=N_++N_- \qquad {\rm and} \qquad  N_{\pm} = \int_{-\infty}^\infty \rho_\pm(\theta) d\theta\,,
    \eeq 
    with $\rho_\pm(\theta)$  defined in Eq. (\ref{rho}).
    \item The entropy normalised by the total particle number $\frac{S}{N_{\rm st}}$.
    \item The entropy production rate normalised by the total particle density $\frac{L}{N_{\rm st}}\frac{{\rm d} S}{{\rm d}t}$.
    \item  The spectral densities $\rho_\pm(\theta)$ and effective velocities $v^{\rm eff}_\pm (\theta)$.
\end{enumerate}
Our aim is to highlight any features of these functions  that signal the presence of unstable particles in the steady state. 

{\OCA Although many properties such as (\ref{64}) and (\ref{kappa2}) are common to our three examples $K_{S,B,0}(\theta)$, for the rest of the paper we will focus on the (simpler) quenches $K_{S,B}(\theta)$ only. As we shall see, these quenches encapsulate the main new properties of the model, whilst the additional zero of $K_0(\theta)$ at the origin, somewhat complicates the analysis. Therefore, to improve readability, we leave the analysis of the $K_0(\theta)$ quench to Appendix \ref{AppB}}.

\medskip

\subsection{Changing the Quench Magnitude}

\begin{figure}[t]
\includegraphics[width=0.48\textwidth]{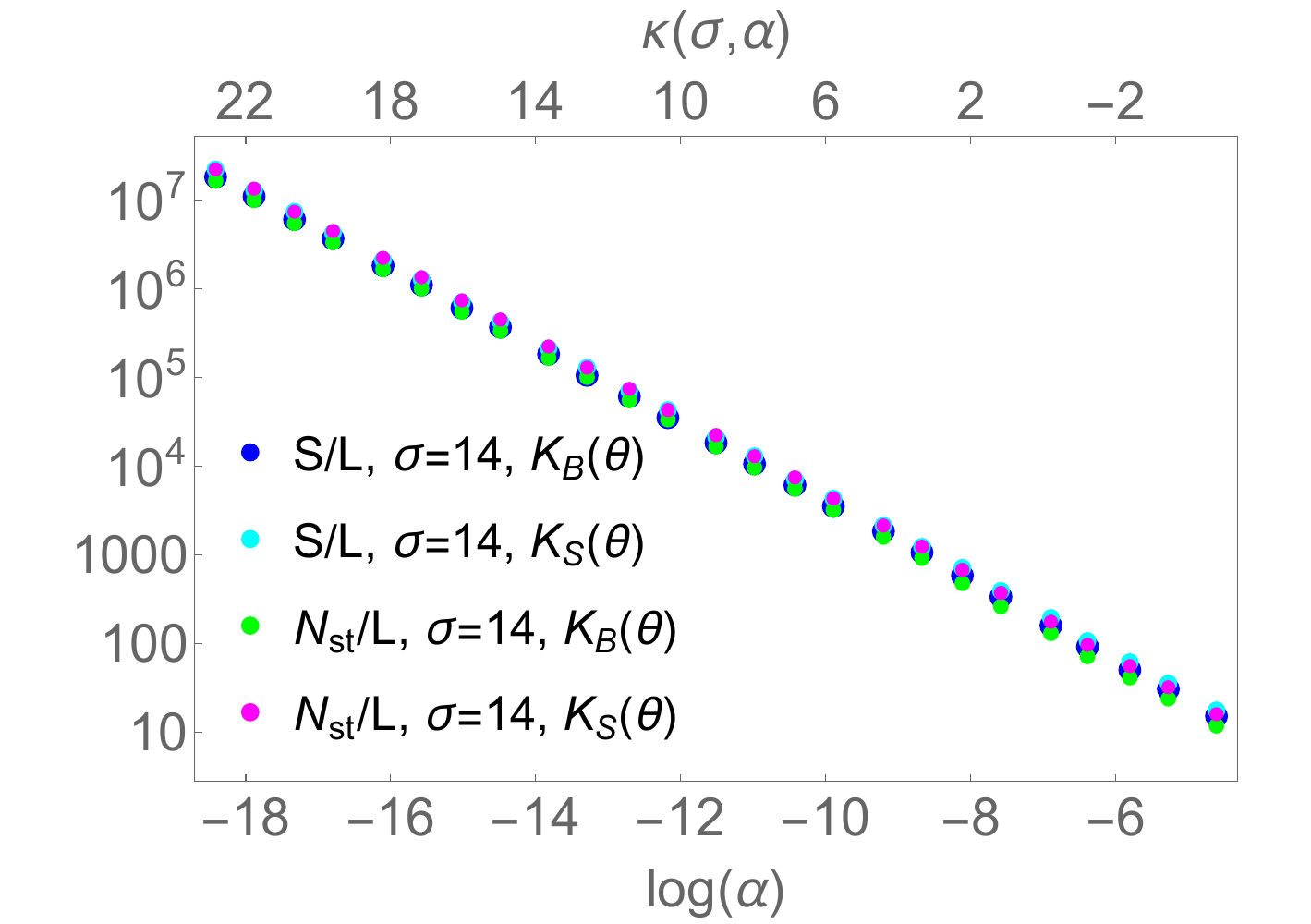}
\includegraphics[width=0.48\textwidth]{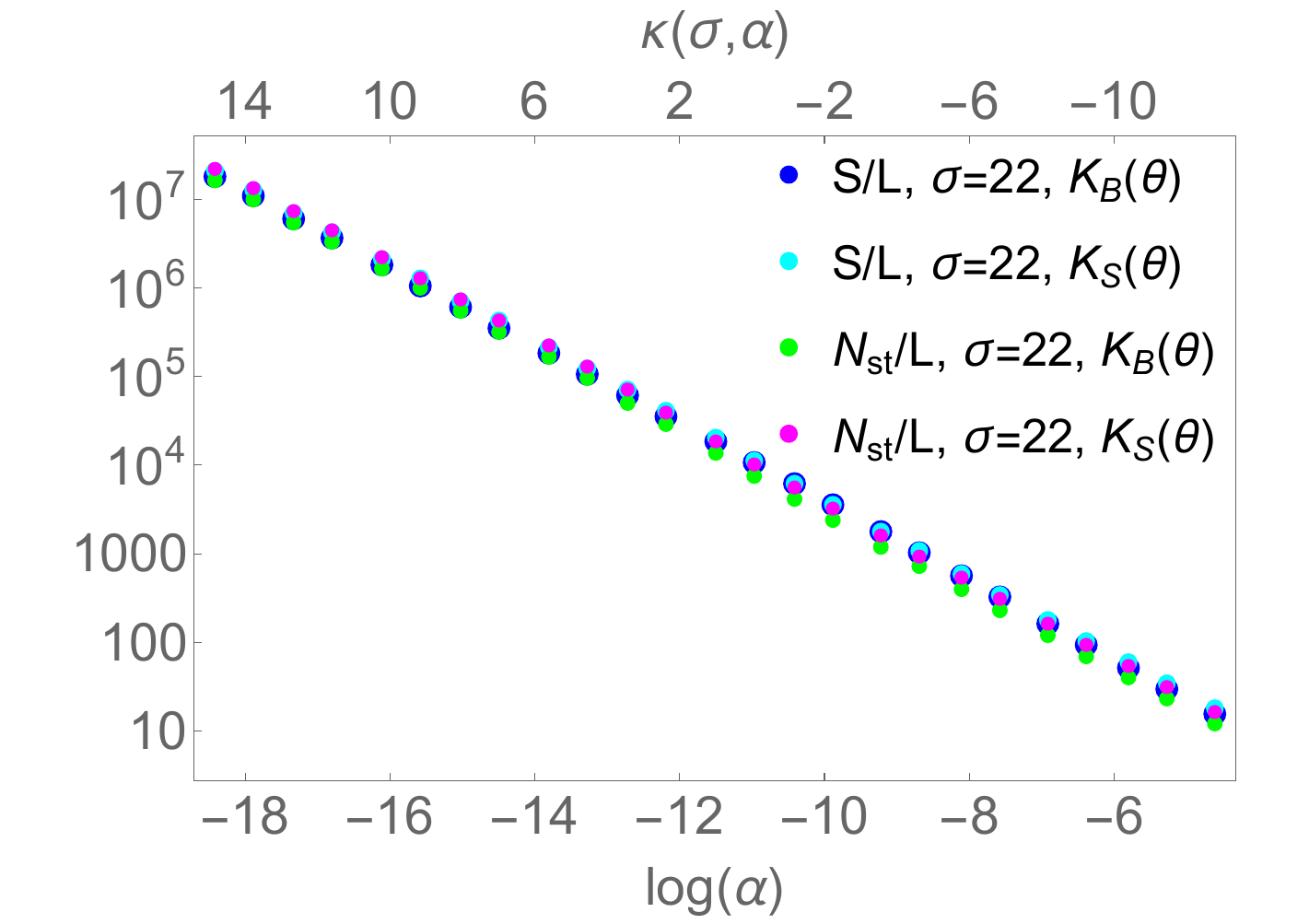}
\includegraphics[width=0.48\textwidth]{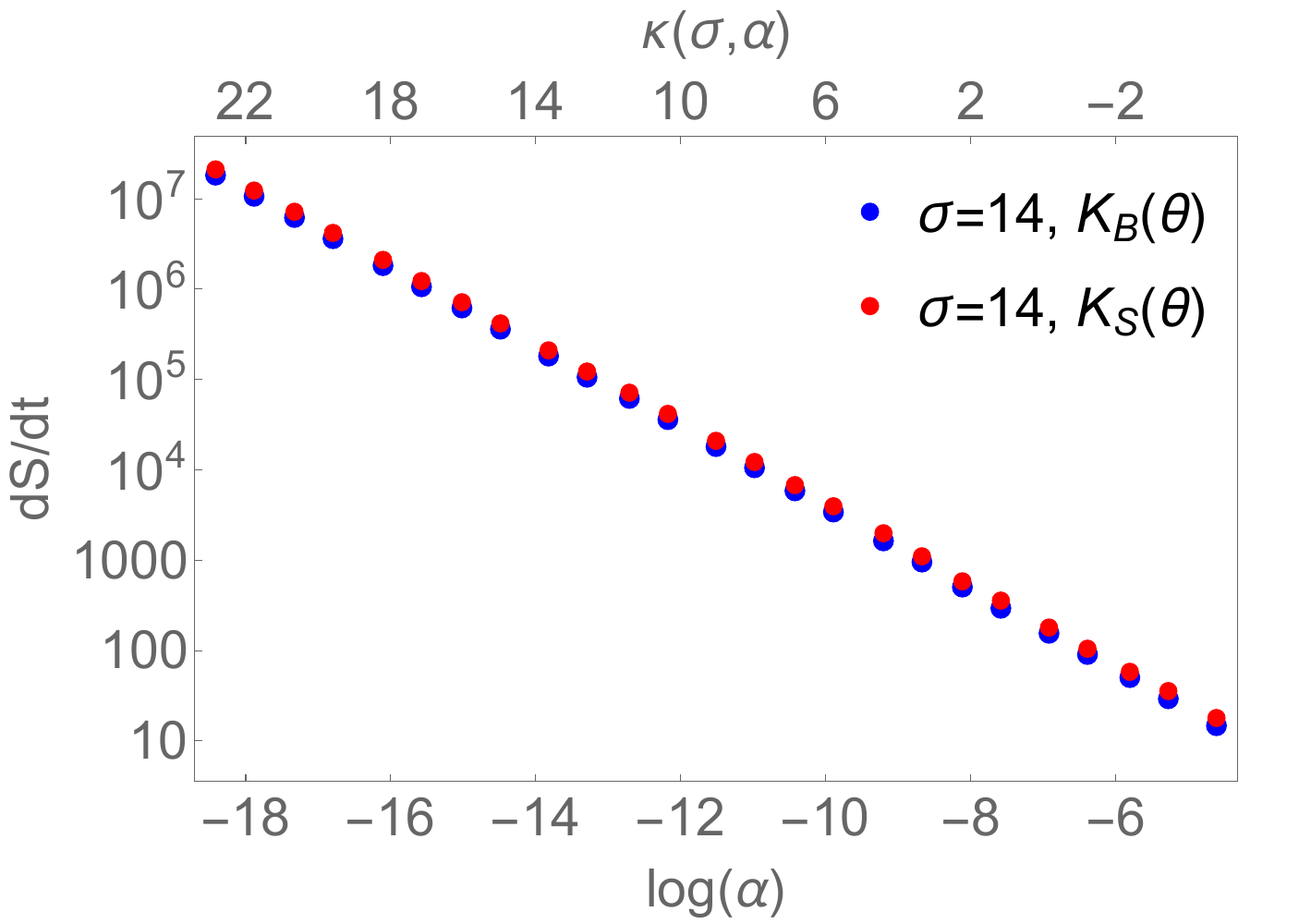}
\hspace{3mm}
\includegraphics[width=0.48\textwidth]{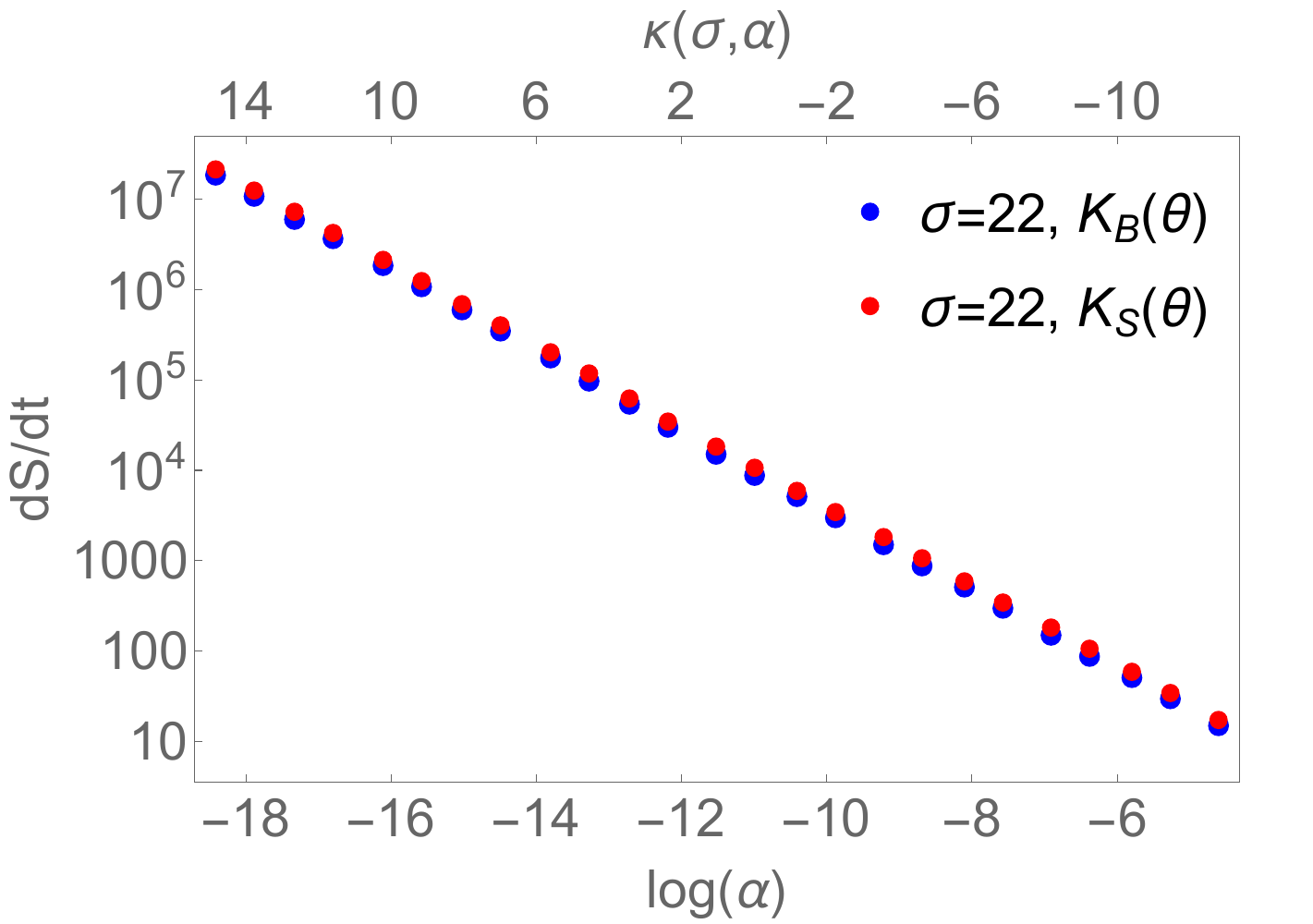}
\caption{$S/L$, $N_\text{st}/L$ and  $\text{d}S/\text{d}t$ as functions of $\log\alpha$ after quenches with different $K$-functions and fixed  resonance parameter $\sigma$ (reported in the legend). 
Note the $\log$ scale on the vertical axis.
The bottom and top of the labels are respectively the values of $\log\alpha$ and of $\kappa_{\rm S,B}(\sigma,\alpha)$ (we dropped the indices $\{\rm S,B\}$ because   $\kappa(\sigma,\alpha)\simeq-2\log\alpha-\sigma$ for $\alpha\ll 1$). 
}
\label{SNPerLFixedSigma}
\end{figure}

In the following we analyse the quantities listed above for  fixed $\sigma$ and varying the quench parameter $\alpha$. 
We focus on two particular values of the resonance parameter, $\sigma=14$ and $22$, but our results are qualitatively generic. 
As we will see below, there is a marked difference in the behaviour of quantities 1,2 and 3 in our bullet points above and quantities 4, 5, i.e. those that involve normalisation by $N_{\rm st}$. We split our discussion accordingly. 

\subsubsection{Linear Scaling} 
We start our analysis considering the functions $S/L$, $\text{d}S/\text{d}t$, and $N_{\text{st}}/L$. 
They can be computed from the QAM as explained above and the resulting curves as a function of $\alpha$ are shown in Fig.~\ref{SNPerLFixedSigma}. It is striking that all these functions (in $\log$ scale) have an approximately linear dependence on $\log \alpha$ which holds for a wide range of values of $\alpha$.
Furthermore there is only a little dependence on $\sigma$ and on the initial state $i=\rm B,S$.  
To elucidate this property, let us first focus on the quench with $K_{\rm B}(\theta)$ whose asymptotics is determined by (\ref{asiB}) so that, for any $\sigma$, we have
\beq
\log C_{\rm B}(\alpha)=\log(1-\alpha^2)-\log\alpha^2\approx -2\log\alpha \quad \mathrm{for} \quad \alpha\ll 1.
\label{asym}
\eeq 
Exactly the same expansion holds for $\log C_{\rm S}(\alpha)$, cf. Eq. (\ref{asi}). Therefore we explained why the dependence on $\sigma$ and $i=\rm B,S$ disappears for $\alpha\ll1$. 
However, as Fig.~\ref{SNPerLFixedSigma} shows,   all functions depend algebraically on $\alpha$ (linear behaviour in log-log scale), with a power (slope in the plot) that is compatible with $1$ and is  independent of the value of $\sigma$ and the choice of $K$-function. 
This algebraic scaling can be derived analytically in the free fermion regime, where it follows from the fact that  $K_{\rm S,B}(\theta)$ are piecewise constant functions equal to 1 for $-|\log\alpha|<\theta< |\log\alpha|$ and 0 otherwise. 
A simple derivation is presented in Appendix \ref{AppA} that however is difficult to extend to the interacting regime.

Although we cannot prove the algebraic behaviour with $\alpha$, we can strengthen this conjecture by more extensive numerical analysis that we present in Tables~\ref{TableCRatiosSPerLExtended} and \ref{TableCRatiosdSdtExtended}. 
In these table we report the results of the numerical fits of the  $\alpha$-dependent steady state entropy density $S/L$ and the entropy production rate $\text{d}S/\text{d}t$. 
The fitting function is 
\begin{equation}
\mathcal{C}(\kappa_i) \alpha^{-1}\,\qquad \mathrm{for} \qquad i=\rm S,B,
\label{Fit1}
\end{equation}
where $\mathcal{C}(\kappa_i)$ is 
the only fitting parameter. It turns out that $\mathcal{C}(\kappa_i)$  depends significantly only on the sign of  $\kappa_i$. 
Hence, we denote with $\kappa_i^\pm$ values of $\kappa_i(\sigma,\alpha)$ that are either positive $(+)$ or negative $(-)$, associated with the interacting and non-interacting regimes, respectively.
In the tables, we report the value of $\mathcal{C}(\kappa_i)$ and also the ratio $\mathcal{C}(\kappa_i^-)/\mathcal{C}(\kappa_i^+)$.
This ratio reveals a {\it universal} property of our quench protocol:  
in all cases we find that it is very close to $5/6=0.8333...$, that is the ratio of the two central charges associated with the UV physics of the theory depending on the presence ($c=6/5$) or absence ($c=1$) of the resonance. 
The value $5/6$ is better achieved for $\sigma$ large (i.e.  $\alpha \ll 1$ in the interacting regime), as this is the limit where the dependence of TBA quantities on the universal scale (\ref{kappa2}) works more precisely.

\begin{table*}[t]
\begin{center}
\begin{tabular}{|c||c|c|c|c|c|c|}
  \hline
    $\sigma$ & $12$ & $14$ & $16$ & $18$ & $20$ & $22$  \\
           \hline \hline 
            $\mathcal{C}(\kappa_{\rm B}^-)$  & 0.1574 & 0.1547 & 0.1530 & 0.1536 & 0.1526 & 0.1530\\\hline
           
            $\mathcal{C}(\kappa_{\rm B}^+)$  & 0.1825 &  0.1825 & 0.1825 & 0.1825 & 0.1825 & 0.1825 \\\hline
            
            $\mathcal{C}(\kappa_{\rm B}^-)/\mathcal{C}(\kappa_{\rm B}^+)$  & 0.863 & 0.848 & 0.839 &  0.842 & 0.836 & 0.838\\\hline
            $\mathcal{C}(\kappa_{\rm S}^-)$  & 0.1828 & 0.1834 & 0.1795 & 0.1818 & 0.1792 & 0.1807\\\hline
           
            $\mathcal{C}(\kappa_{\rm S}^+)$  & 0.2143 & 0.2143 & 0.2143 &  0.2143 & 0.2143 & 0.2143 \\\hline
            $\mathcal{C}(\kappa_{\rm S}^-)/\mathcal{C}(\kappa_{\rm S}^+)$ & 0.853 &  0.856 & 0.837 & 0.848 & 0.836 & 0.843 \\\hline
\end{tabular}
\caption{The values of $\mathcal{C}(\kappa_{i}^\pm)$ and the ratios $\mathcal{C}(\kappa_{i}^-)/\mathcal{C}(\kappa_{i}^+)$ for $i= \rm B,S$ for the steady state entropy density $S/L$. We note that the relative error of the fitted parameter  $\mathcal{C}(\kappa_i^{\pm})$ is typically of order $10^{-3}-10^{-4}.$}
\label{TableCRatiosSPerLExtended}
\end{center}
\end{table*}

\begin{table*}[t]
\begin{center}
\begin{tabular}{|c||c|c|c|c|c|c|}
  \hline
    $\sigma$ & $12$ & $14$ & $16$  &  $18$ & $20$ & $22$  \\
           \hline \hline
           $\mathcal{C}(\kappa_{\rm B}^-)$  & 0.1514 & 0.1515 & 0.1516 &  0.1519 & 0.1518 & 0.1519\\\hline
           
            $\mathcal{C}(\kappa_{\rm B}^+)$  & 0.1824 &  0.1824 & 0.1824 & 0.1824 & 0.1823 & 0.1820 \\\hline
            $\mathcal{C}(\kappa_{\rm B}^-)/\mathcal{C}(\kappa_{\rm B}^+)$  & 0.830 &  0.830 & 0.831 & 0.833 & 0.833 & 0.834\\\hline
            $\mathcal{C}(\kappa_{\rm S}^-)$  & 0.1774 & 0.1785 & 0.1783 &  0.1788 & 0.1787 & 0.1789\\\hline
           
            $\mathcal{C}(\kappa_{\rm S}^+)$  & 0.2143 &  0.2143 & 0.2143 & 0.2142 & 0.2141 & 0.2138 \\\hline
            $\mathcal{C}(\kappa_{\rm S}^-)/\mathcal{C}(\kappa_{\rm S}^+)$  & 0.828 &  0.833 & 0.832 &  0.835 & 0.835 & 0.836 \\\hline
\end{tabular}
\caption
{The values of $\mathcal{C}(\kappa_{i}^\pm)$ and the ratios $\mathcal{C}(\kappa_{i}^-)/\mathcal{C}(\kappa_{i}^+)$ for $i=\rm B,S$ for the entropy growth rate $\text{d}S/\text{d}t$. We note that the relative error of the fitted parameter  $\mathcal{C}(\kappa_i^{\pm})$ is typically of order $10^{-3}-10^{-4}$.}
\label{TableCRatiosdSdtExtended}
\end{center}
\end{table*}

This very interesting finding can be explained as follows. 
For the quenches considered here, energies are high enough as for the theory to reach the two UV fixed points. 
Hence the entropy and its growth rate are both well described by the conformal formula  \cite{QuasiParticlePicture} 
\beq
S= \frac{\pi c}{3 \beta_{\rm eff}} \min(2 v t, L)
\eeq
which indeed show proportionality to the central charge (here $\beta_{\rm eff}$ is the inverse effective temperature related to the energy of the quench and $v$ the sound velocity). 
In our case, the result is particularly interesting because it is the presence of the unstable particle and the additional tunable energy scale $\sigma$ it introduces, which allows us to access {\it  two conformal regimes in a single theory}.
Another consequence of the ratio $5/6$ is that the presence of  unstable particles increases both the stationary state entropy per unit length $S/L$ and the entropy growth rate. This is consistent with the fundamental interpretation of the steady state entropy as counting degrees of freedom in the theory.

\begin{figure}[t]
\includegraphics[width=0.48\textwidth]{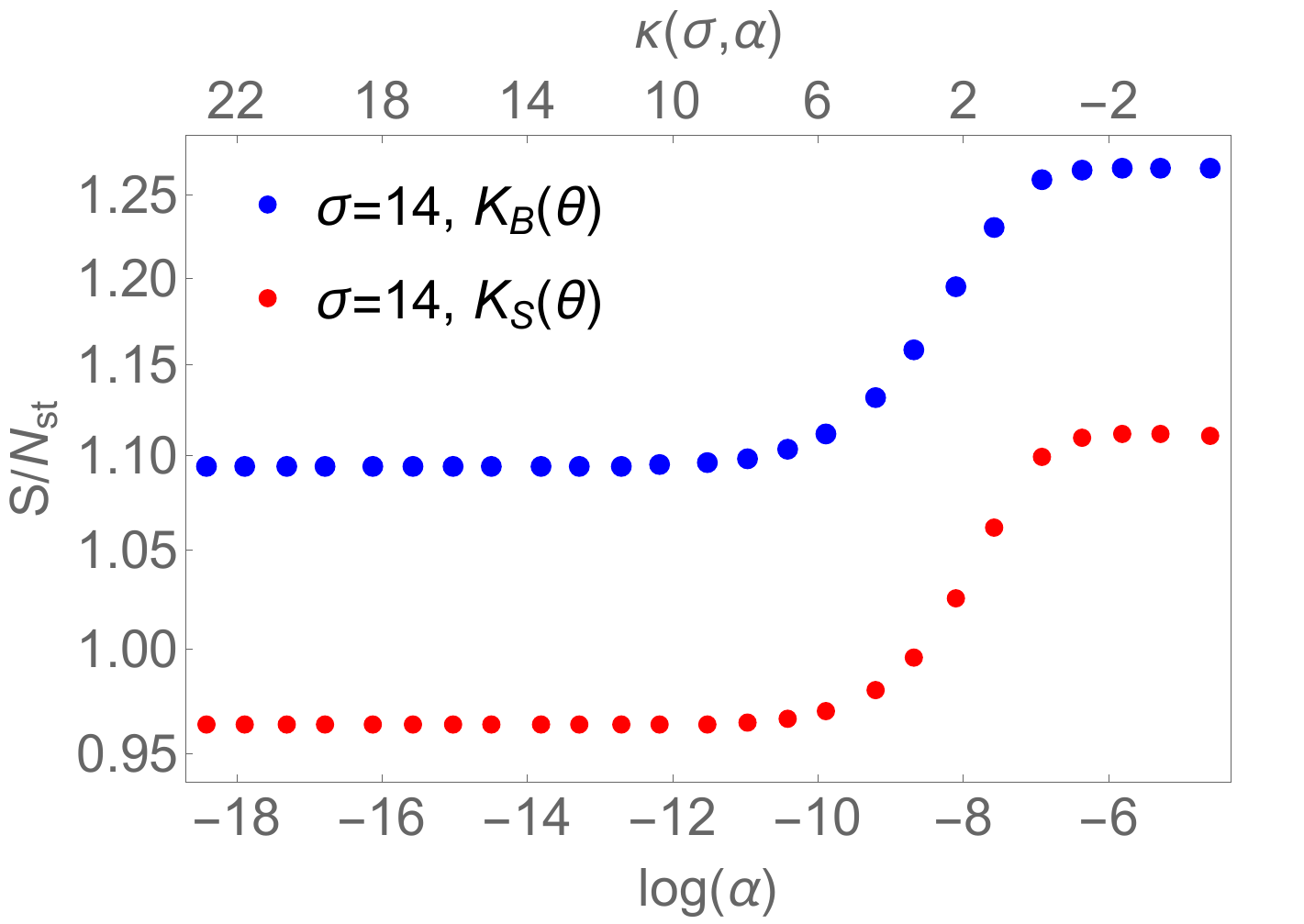}
\includegraphics[width=0.48\textwidth]{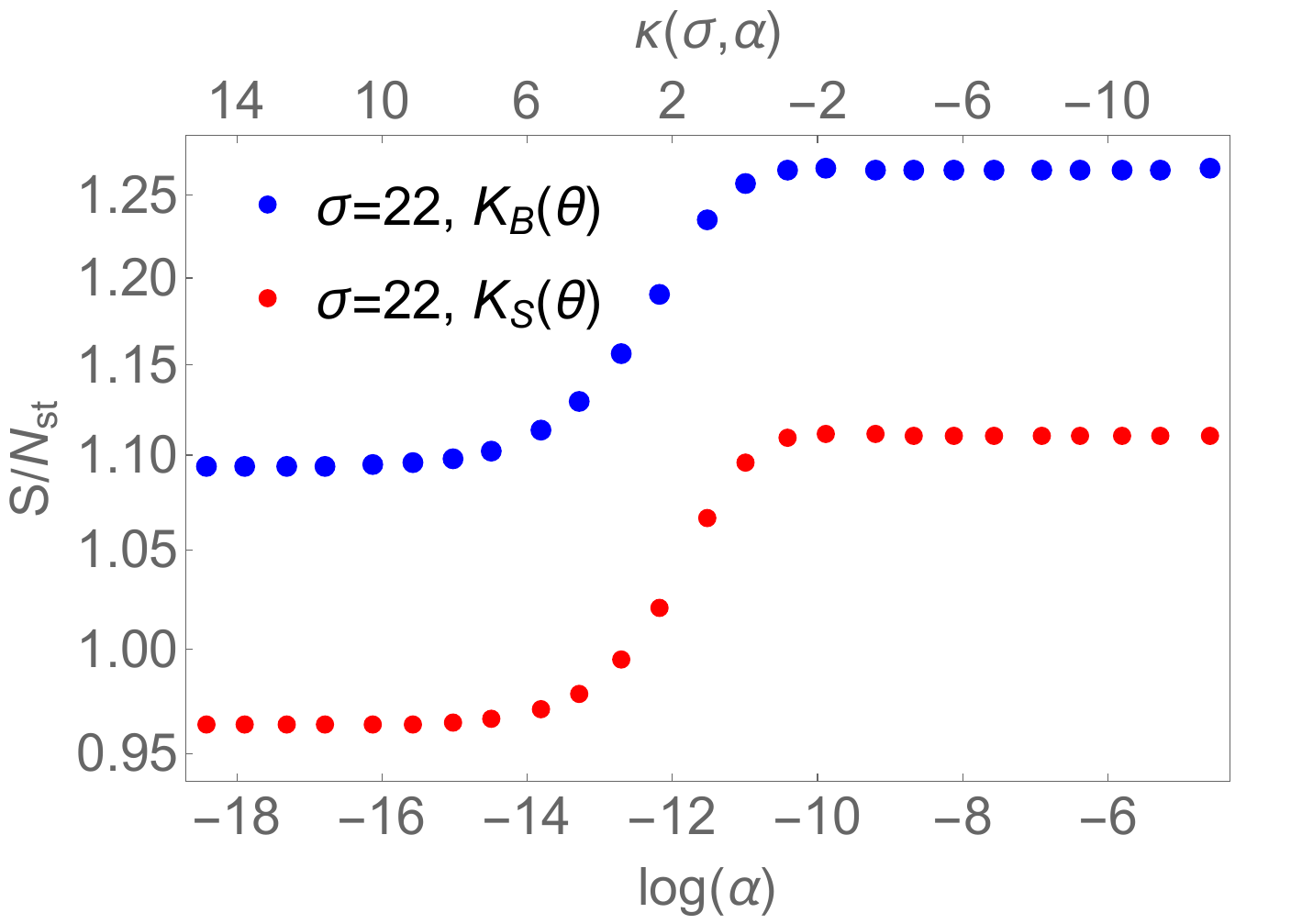}
\includegraphics[width=0.48\textwidth]{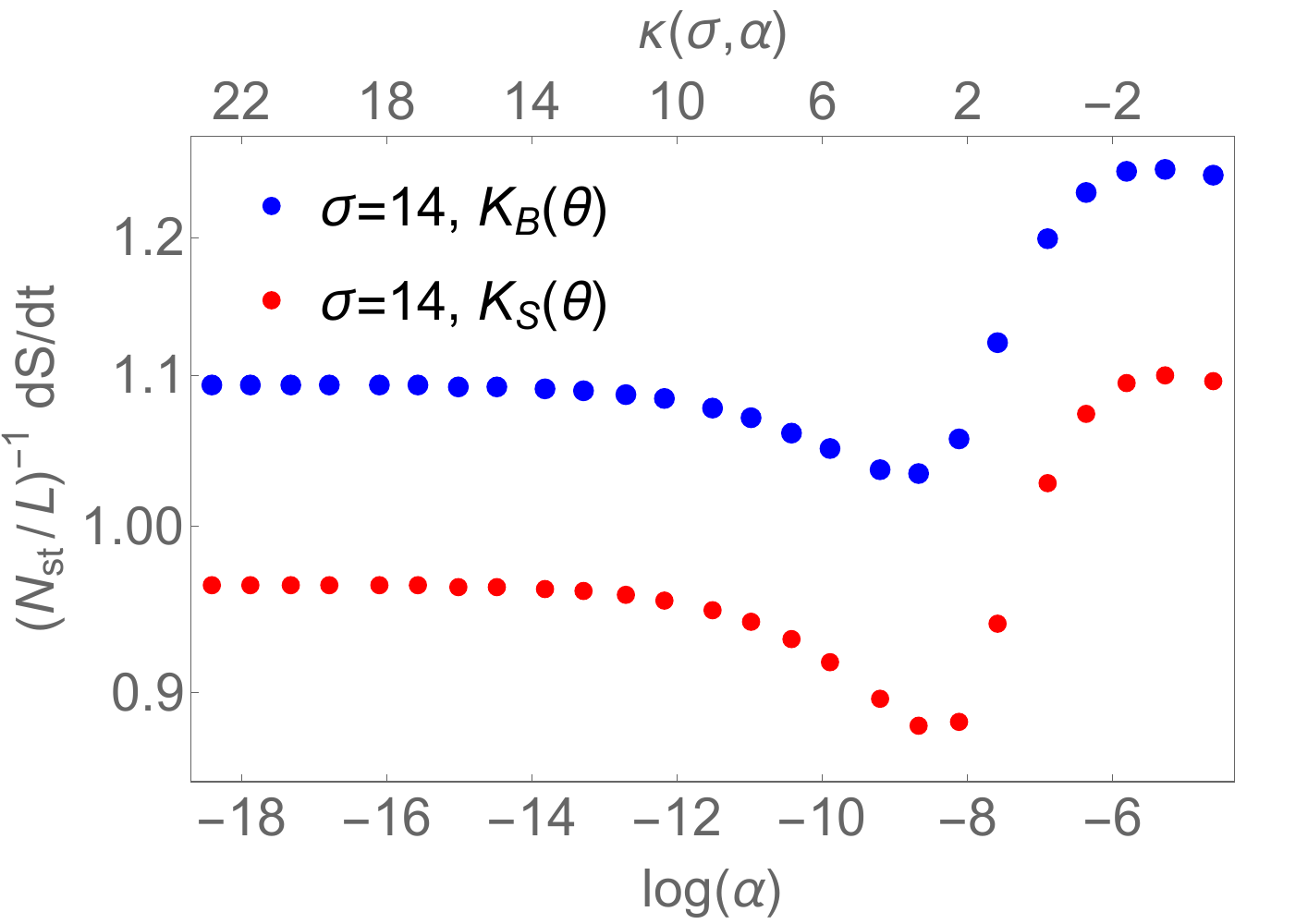}
\hspace{3mm}
\includegraphics[width=0.48\textwidth]{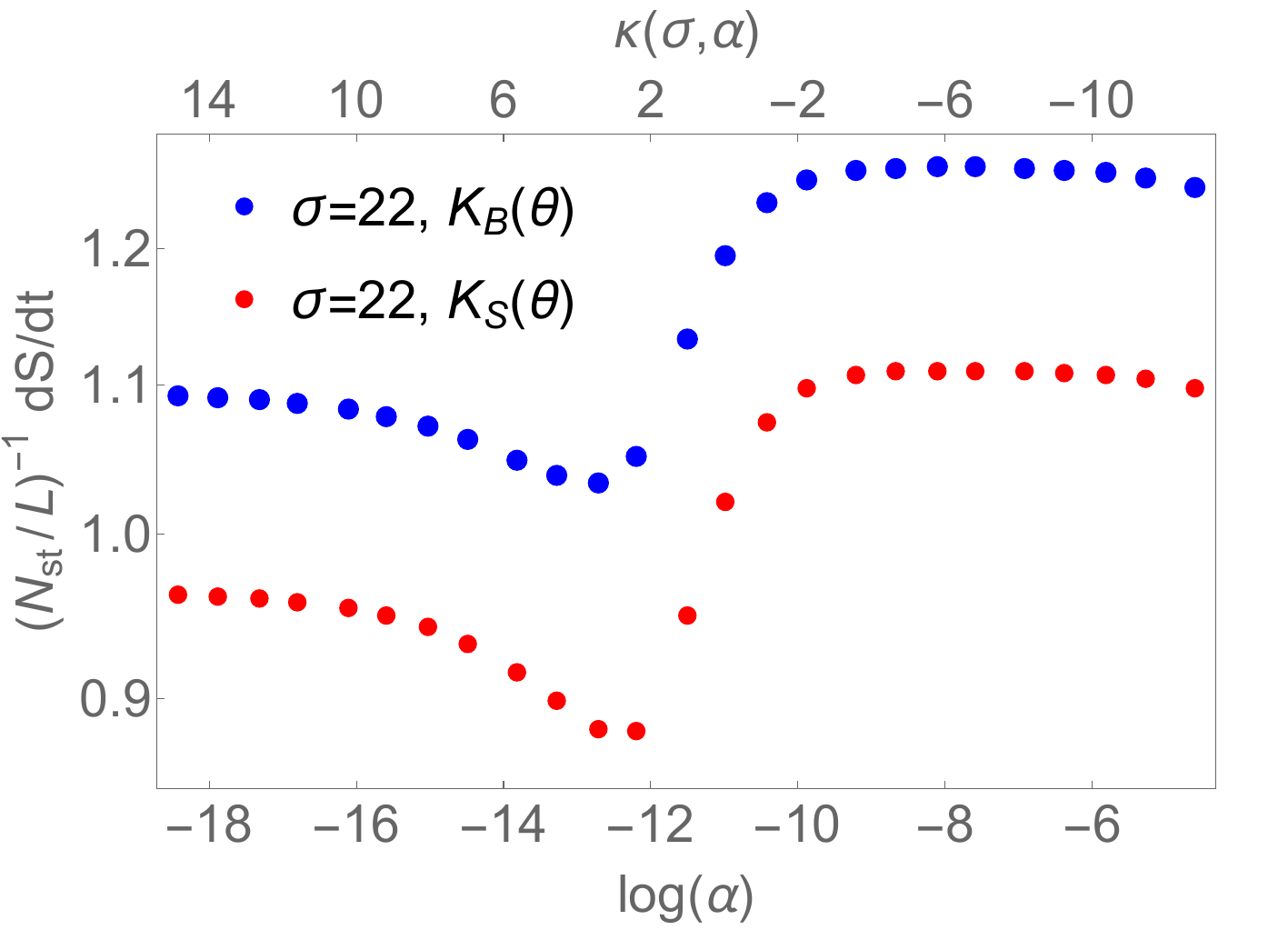}
\caption{
 $S/N_{\text{st}}$, $L/N_{\text{st}}$ and $\text{d}S/\text{d}t$ against $\alpha$ after quenches with different $K$-functions and fixed resonance parameter $\sigma$ (reported in the legend).
 }
\label{SPerNFixedSigma}
\end{figure}

\subsubsection{Emergence of two Plateaux and one Local Minimum}

We now consider the normalised quantities $S/N_\text{st}$ and $(N_{\text{st}}/L)^{-1}\times\text{d}S/\text{d}t$.
We report our exact numerical results in Fig.~\ref{SPerNFixedSigma} as a function of $\log\alpha$. 
These ratios are sensitive to the fine details of the quench because the ratio of two linear functions in $\log \alpha$ is
no longer a linear function.
Furthermore, the scaling with $\log\alpha$ is just the leading order for $\alpha\ll 1$ and there are corrections that play a role, especially when considering the ratios. Indeed, both functions $S/N_\text{st}$ and $(N_{\text{st}}/L)^{-1}\times\text{d}S/\text{d}t$ display a double plateau structure, where the transition between plateaux happens monotonically for $S/N_\text{st}$ and through the formation of a local minimum  for $(N_{\text{st}}/L)^{-1}\times\text{d}S/\text{d}t$. 

The arrows in Fig.~\ref{SPerNFixedSigma} marks the energy scale corresponding to the onset of unstable particles that, according to the discussion in Subsection \ref{regimes},  occurs for
 \beq 
 \log\alpha \approx -\frac{\sigma}{2}\,.
 \label{crita}
 \eeq 
It is evident that this scale
approximately marks also the midpoint between plateaux. 
In addition, the left plateau, corresponding to the interacting regime is lower than the right plateau, which corresponds to the free regime. 
At a superficial look, this may seem inconsistent with our earlier discussion in terms of degrees of freedom; this is not the case because we are dividing by particle density. Although the steady state entropy, the entropy growth rate and the particle density are all larger in the presence of interactions, the particle density grows faster so that the ratio is smaller in the presence of interaction. 
Notice that, however, the ratio between plateau heights (left/right) is still well approximated by the ratio of central charges $5/6$. 
We also observe that the results corresponding to different values of $\sigma$ (like the two blue-dotted curves on the top row) look very similar up to a shift. This feature is once more a consequence of the general dependence of our functions on the universal scale 
$\kappa(\sigma,\alpha)$ (we will discuss this further in Subsection \ref{uni}). 
 
Turning our attention to the normalised production rate $(N_{\text{st}}/L)^{-1}\times\text{d}S/\text{d}t$ as a function of the quench size (the bottom panels in  Fig. \ref{SPerNFixedSigma}), we observe a local minimum in the production rates which can be linked to the formation of unstable excitations. This local minimum is correlated with the emergence of an additional local maximum of the TBA density of particle $\rho(\theta)$ and simultaneous reduction of the effective velocities of stable quasiparticles at the threshold for the formation of unstable excitations. We will analyse this further in Subsection \ref{SpectralDensities}.

\subsubsection{Depletion as a Function of the Resonance Parameter }
\label{depli}

\begin{figure}[t]
\begin{center}
\includegraphics[width=7.5cm]{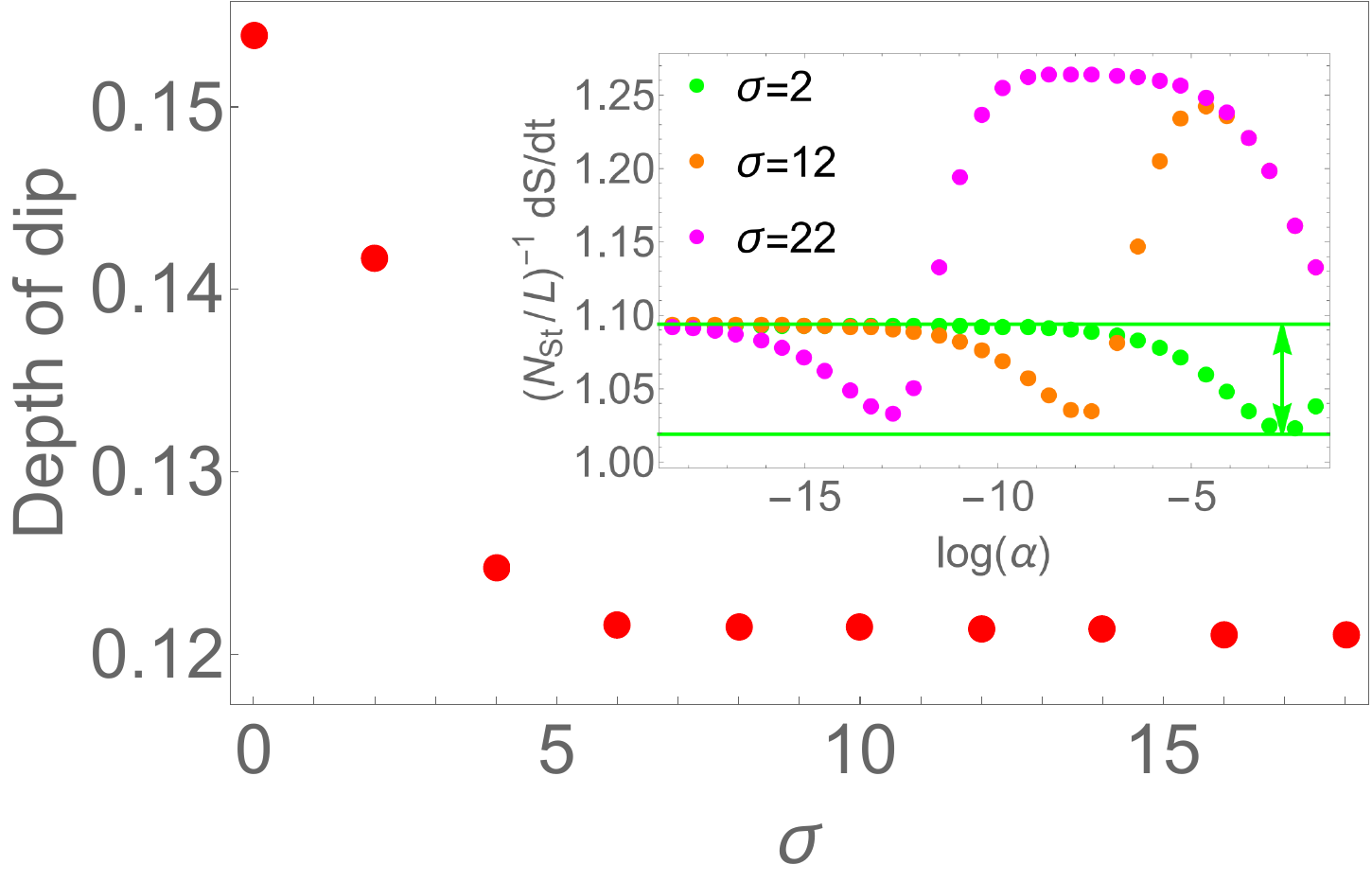}
\includegraphics[width=7.5cm]{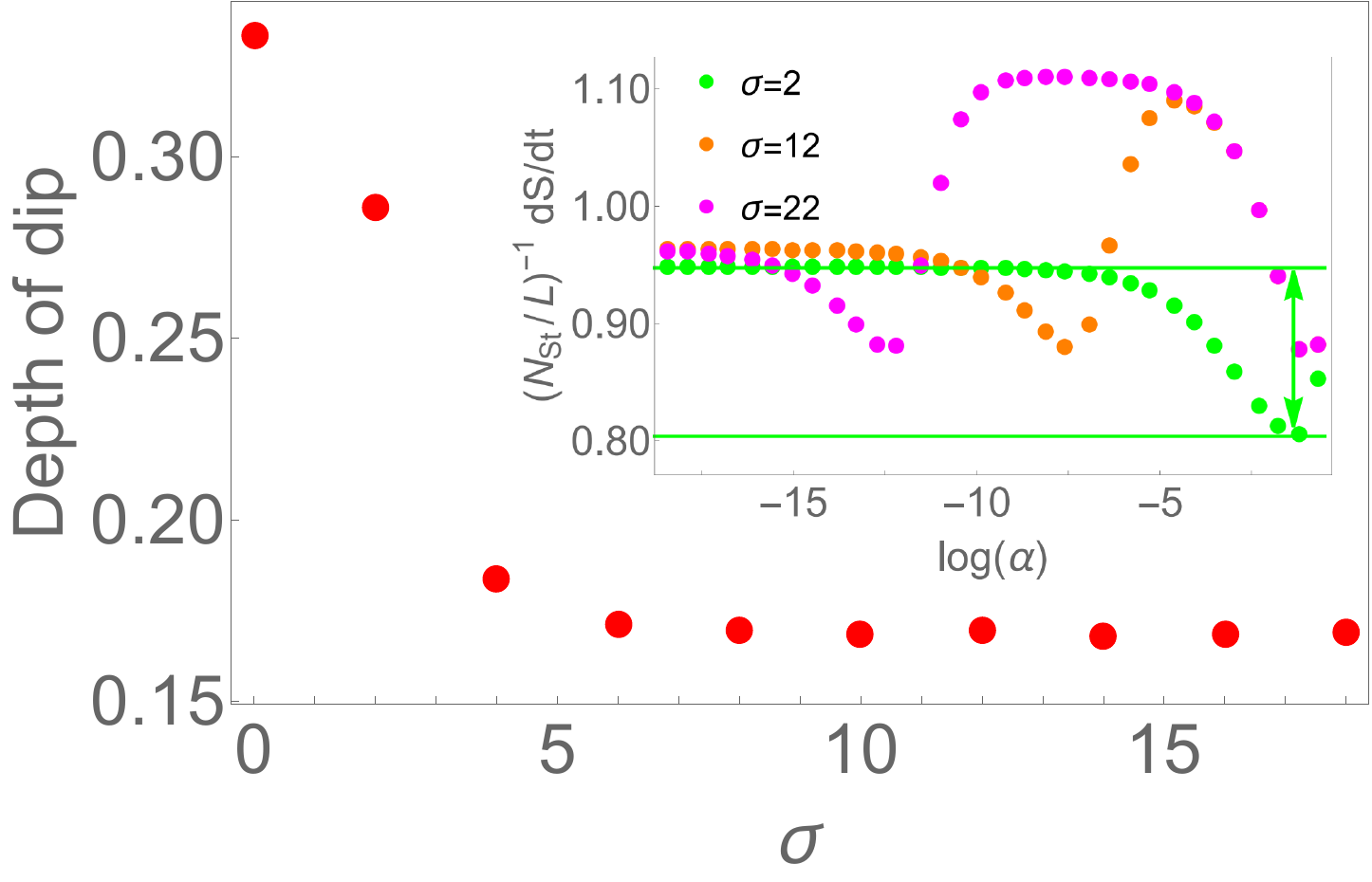}
\caption{\label{Depths} The combination $\left.\frac{L}{N_{\text{st}}} \frac{{\rm d}S}{{\rm d}t} \right|_{\alpha=0}-\underset{\alpha}{\text{min}}\left\{\frac{L}{N_{\text{st}}} \frac{{\rm d}S}{{\rm d}t}\right\}$ as a function of $\sigma$ for $K_{\rm B}(\theta)$ and $K_{\rm S}(\theta)$. 
This is a measure of the depletion of the normalised entropy production rate, defined as the difference between its value in the interacting regime (lower plateau in the inset) and its local minimum value (see the double arrowed segment in the inset). 
From both the main panels and the insets, we conclude that the depth of the minimum is larger for $\sigma \rightarrow 0$.}
    \end{center}
\end{figure}

We now consider again the figures in the second row of Fig.~\ref{SPerNFixedSigma} and ask the question: how does the shape and depth of the minimum change as functions of the resonance parameter?  
The answer to this question is helped by Fig.~\ref{Depths}. 
In the inset, we show several functions of the same type seen in Fig.~\ref{SPerNFixedSigma} for more values of $\sigma \in[2,22]$, 
while the main panels report the 
depth of the minimum (the precise definition is given in the figure's caption). 
There are two main observations.
First, the depth of the minimum stays constant for $\sigma \gtrsim 5$, as seen both in the insets and in the main panel.  
Second, the depth of the minimum and its shape start to change as $\sigma \rightarrow 0$  which is the limit where the lifetime of the unstable particle becomes infinite, namely, it becomes a virtual particle. 
In this limit, the depth  becomes more pronounced. 

It is in fact expected that the depth and the shape of the minimum should change for small $\sigma$. First of all, we observe that, from the point of view of the minimum's position, small $\sigma$ corresponds to $\alpha\ll 1$. We have previously noted that most functions depend on a combination of the variables $\sigma, \alpha$ which we have called $\kappa_i(\alpha,\sigma)$ with $i=\rm B,S$ and defined in (\ref{kappa2}). We also noted in (\ref{asym}) that $\kappa_i(\alpha,\sigma)=-\sigma-2\log \alpha$ when $\alpha\ll 1$. This is the reason why many of the functions in the insets of Fig.~\ref{Depths} look identical under translation when plotted against $\log\alpha$. However, this dependence on $\alpha$ no longer holds for values of $\alpha$ near 1, which is where we clearly see a larger minimum in the inset of the right panel of Fig. \ref{Depths} (the effect is more subtle on the left panel). Now the dependence in $\alpha$ is more involved, and the shape and depth of the minimum are changed. Minima for such large values of $\alpha$ correspond to $\sigma$ small too and so when looking at the red dots we see an increase in depth for $\sigma$ small.

In summary, our discussion and the numerics on the last two subsections demonstrate that the appearance of a local minimum and the formation of the unstable particle are features that can be naturally linked. 
However, the fact that this minimum is more pronounced precisely when the unstable particle becomes virtual is an intriguing feature that requires further study. 

\subsection{Changing the Resonance Parameter}

We reconsider the functions $K_{\text{B}}(\theta)$ and $K_{\text{S}}(\theta)$ defined in (\ref{freeB}) and (\ref{asi}) and vary the resonance parameter $\sigma$ or equivalently tune the mass of the unstable particle, while fixing the energy injected in the system by the quench. In this section, we will introduce a new energy scale, characterised by an effective inverse temperature at the free fermion point $\beta_{\rm FF}$ {\DXH according to
\begin{equation}
\frac{1}{L}\frac{\text{Tr} \left[  H_\text{FF}\, e^{-\beta_\text{FF}H_\text{FF}}\right]}{\text{Tr} \left[ e^{-\beta_\text{FF}H_\text{FF}}\right]}=\frac{\langle\Delta E\rangle}{L}\,,
\label{BetaFFDefinition}
\end{equation}
that is, via equating the injected energy density during the quench $\langle\Delta E\rangle/L$ and the energy density of the free fermion (FF) theory in a Gibbs ensemble.}
This scale provides an alternative measure of the quench magnitude and has the advantage of having a clearer physical interpretation than the parameter $\alpha$, while being closely related to it. This effective temperature is uniquely determined via the expectation value of the injected energy density after the quench. 
For simplicity, the forthcoming figures will show just two representative values for the effective inverse temperatures, but our findings are qualitatively general. More numerical data are presented in Table \ref{TableSigmasFixedEAllBosonSquared}.

\begin{figure}[H]
\begin{subfigure}{0.44\textwidth}
\centering
\includegraphics[width=\textwidth]{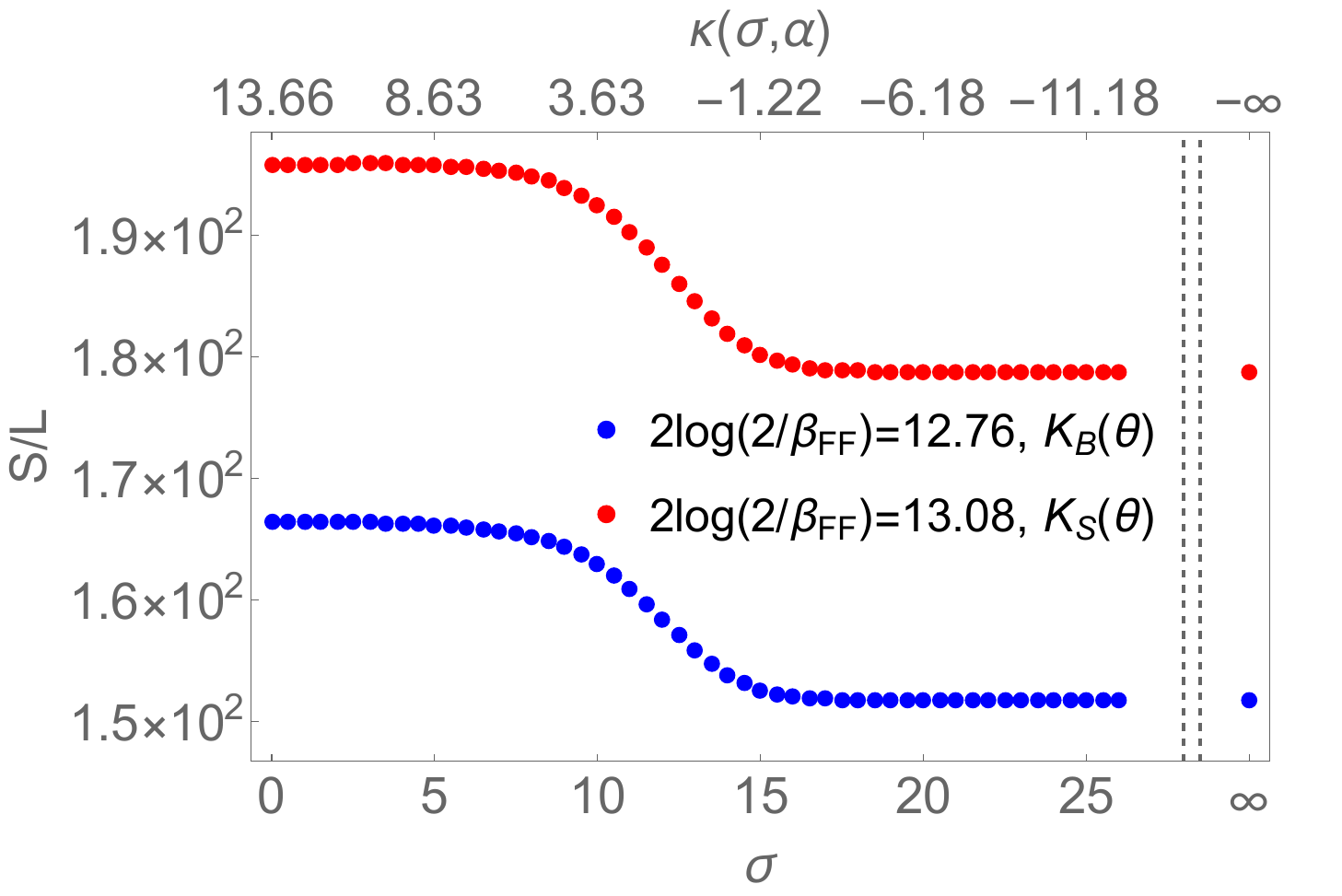}
\end{subfigure}\hfill
\begin{subfigure}{0.44\textwidth}
\centering
\includegraphics[width=\textwidth]{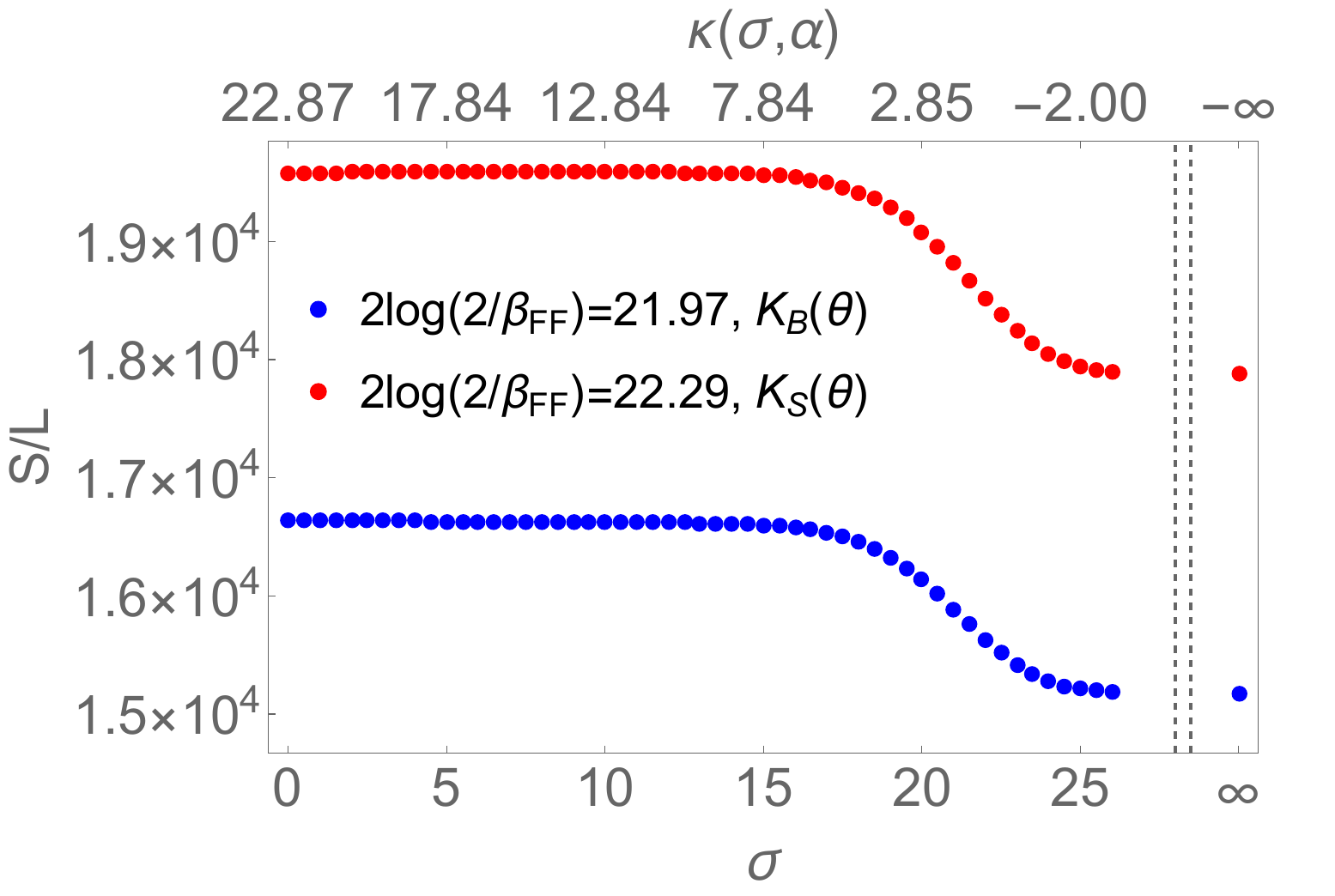}
\end{subfigure}
\begin{subfigure}{0.44\textwidth}
\centering
\includegraphics[width=\textwidth]{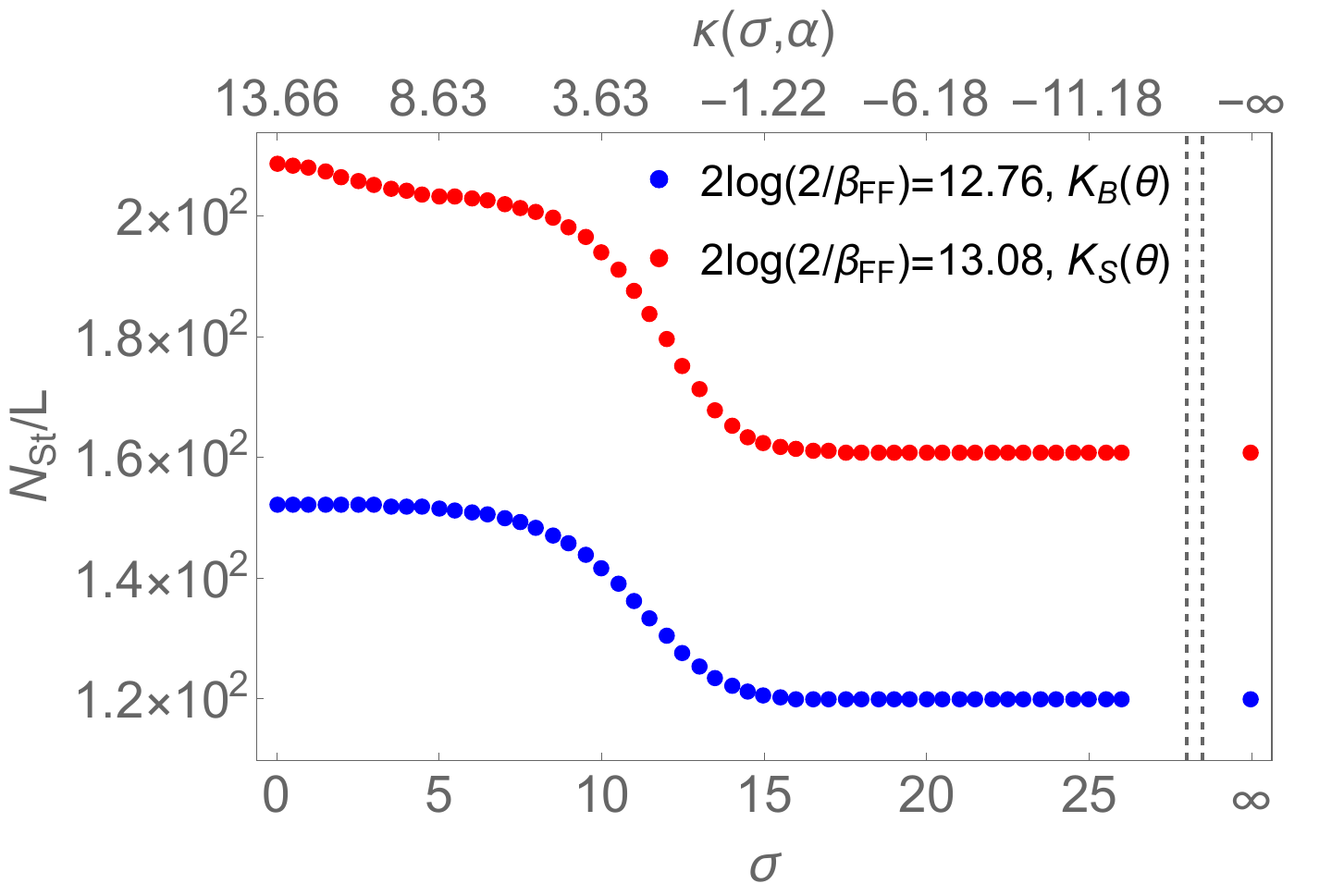}
\end{subfigure}\hfill
\begin{subfigure}{0.44\textwidth}
\centering
\includegraphics[width=\textwidth]{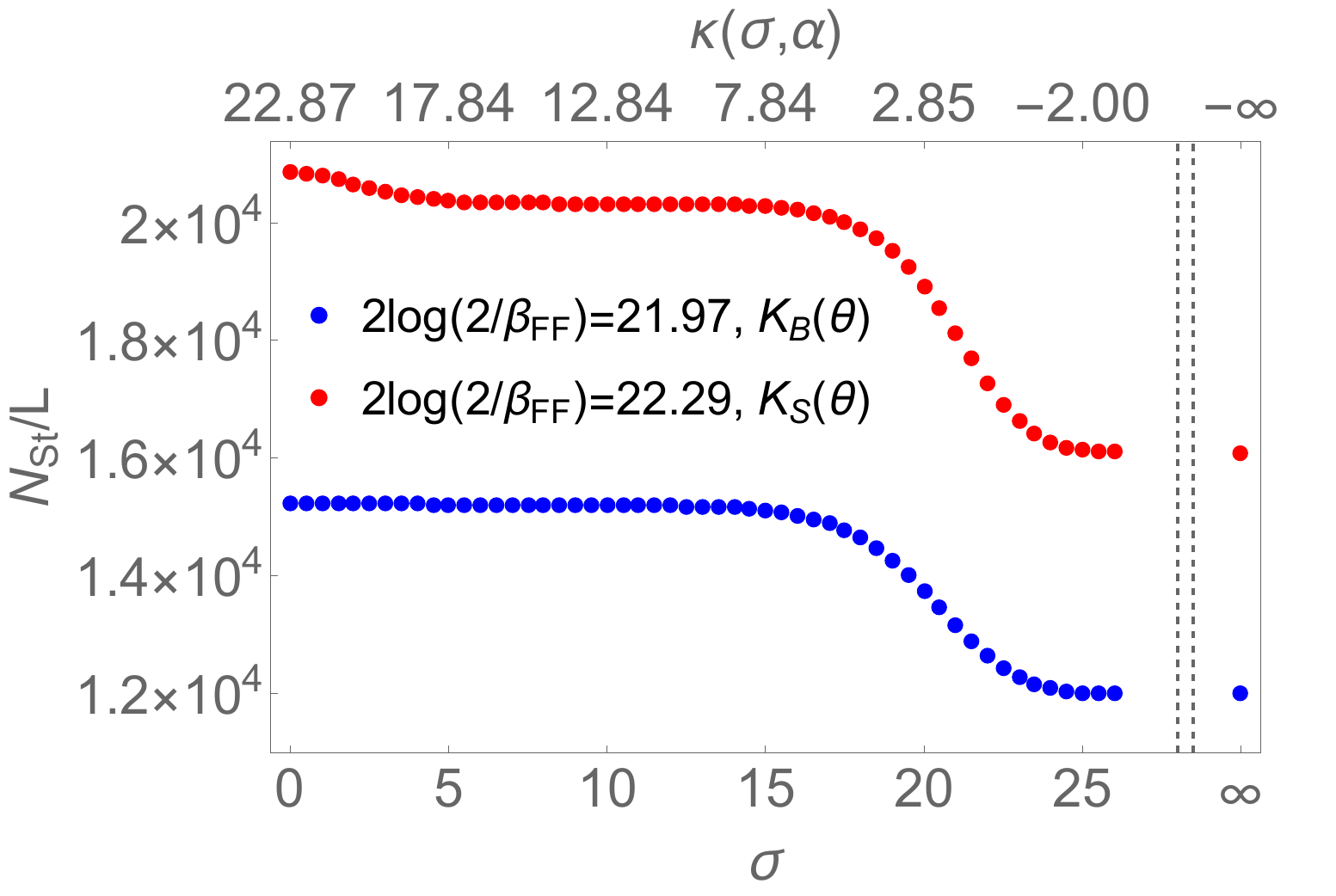}
\end{subfigure}
\begin{subfigure}{0.42\textwidth}
\centering
\includegraphics[width=\textwidth]{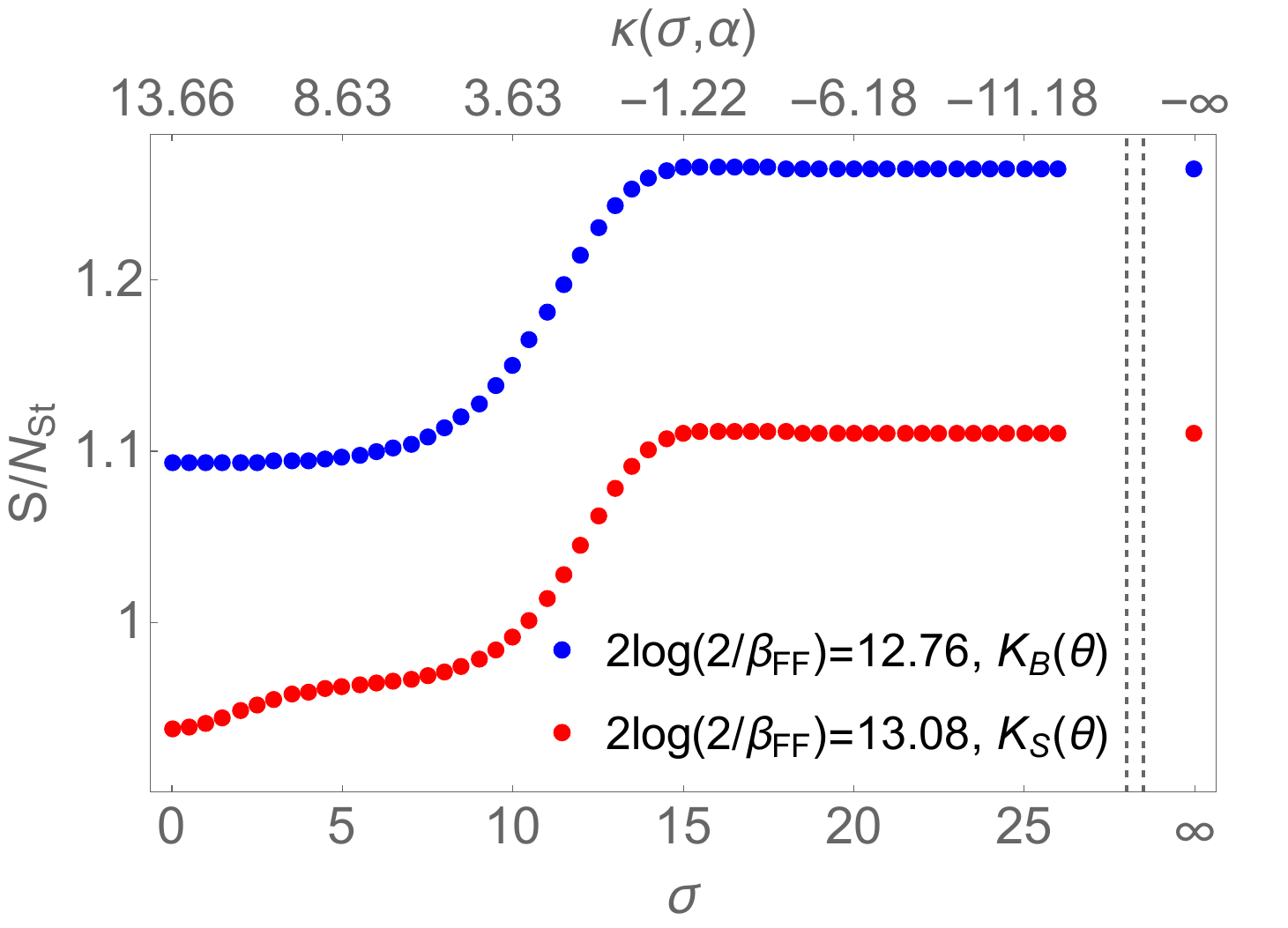}
\end{subfigure}\hfill\hspace{2mm}
\begin{subfigure}{0.42\textwidth}
\centering
\includegraphics[width=\textwidth]{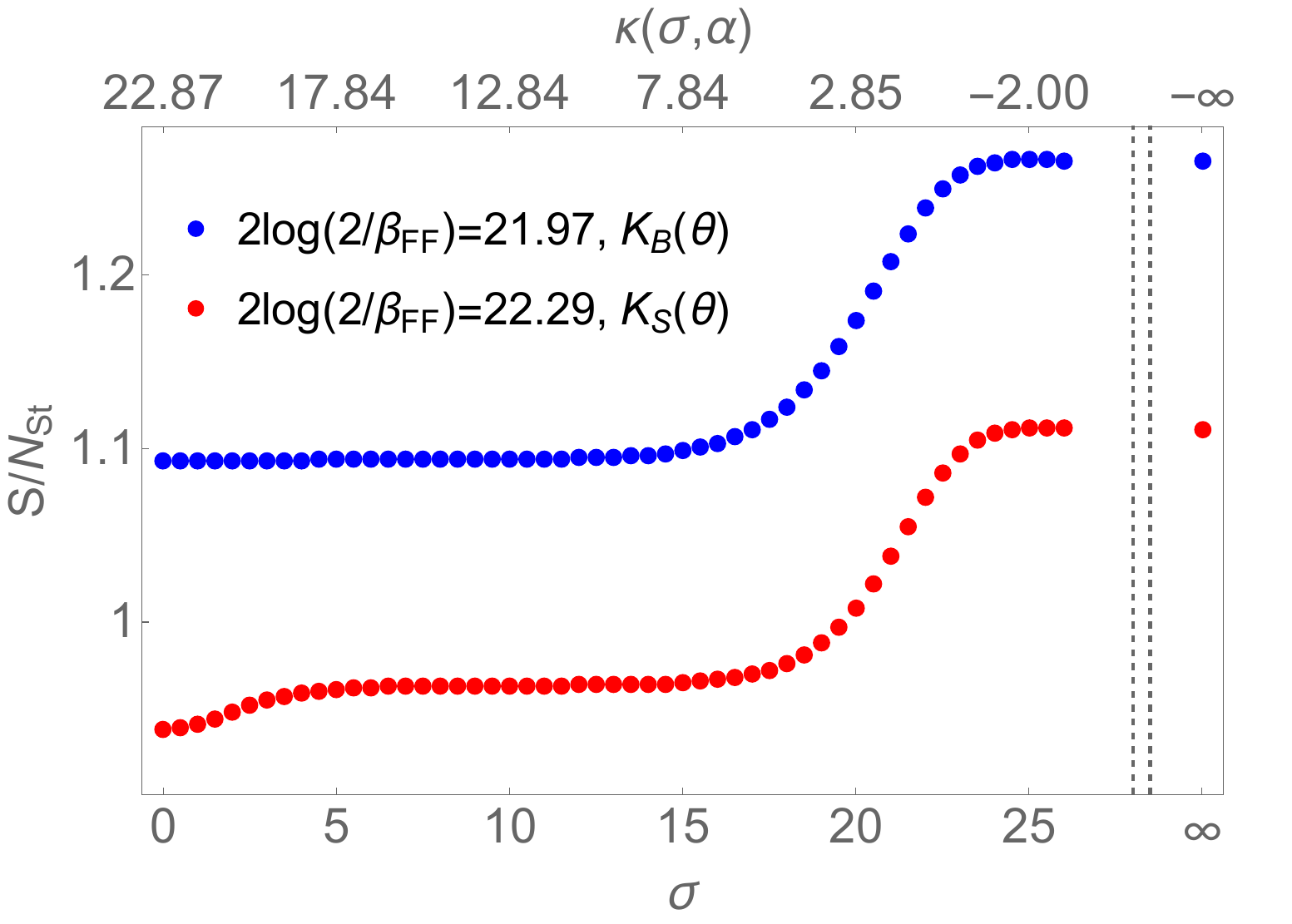}
\end{subfigure}
\caption{\label{SperLTwoKFixedE}
The entropy density $S/L$ (top), the particle density $N_{\text{st}}/L$ (middle) and their ratio (bottom), against the resonance parameter $\sigma$ for quenches with different $K$-functions and energy densities (left smaller energy, right larger energy). The injected energy density depends on the free fermion inverse temperature $\beta_{\text{FF}}$ and on the  $K$-functions ($K_{\rm B}(\theta)$  in blue, $K_{\rm S}(\theta)$ in red). 
}
\end{figure}

\subsubsection{Entropy and Particle Number Densities: Staircase Patterns}
We first briefly discuss the relatively simple behaviour of the steady state entropy $S/L$ and total particle number per unit length $N_{\text{st}}/L$.
Fig. \ref{SperLTwoKFixedE} reports the data as a function of $\sigma $ for fixed injected energy densities ($\beta_{\rm FF}$-s).
Both plotted quantities show two plateaux, a lower one at high $\sigma$-s when the unstable particles are too heavy to form, and a higher plateau at small $\sigma$-s when the unstable particles are present. 
This behaviour is reminiscent of other TBA quantities at equilibrium, such as the TBA scaling function \cite{ourtba,ourU}. 
The reason for the formation of two plateaux is exactly the same as in Fig. \ref{SPerNFixedSigma}, namely, the fact that for varying $\alpha$ ($\sigma$) and fixed $\sigma$ ($\alpha$), there is a transition between the interacting and  non-interacting regimes.

In Fig.~\ref{SperLTwoKFixedE} we can see very clearly that the presence of unstable particles naturally results into a higher stationary entropy density.
We also confirm the similarity of functions with different $\beta_{\rm FF}$:  
they are related to each other by a simple shift. Both these features are further explored in the next subsection. 

The presence of unstable particles is also linked to an increase in the stable particle number density $\rho(\theta)$, simply because creating unstable excitations requires a minimum energy threshold to be met and the higher the energy scale, the higher the population of stable particles too.  
For this reason, although $S/N_{\text{st}}$ exhibits also two plateaux, it has the relative height reversed with respect to $S/L$, just as observed in Fig.~\ref{SPerNFixedSigma} when varying $\alpha$.
\subsubsection{Universal Scaling of the Steady State Entropy}
\label{uni}

\begin{figure}[t]
\begin{center}
		\includegraphics[width=7.5cm]{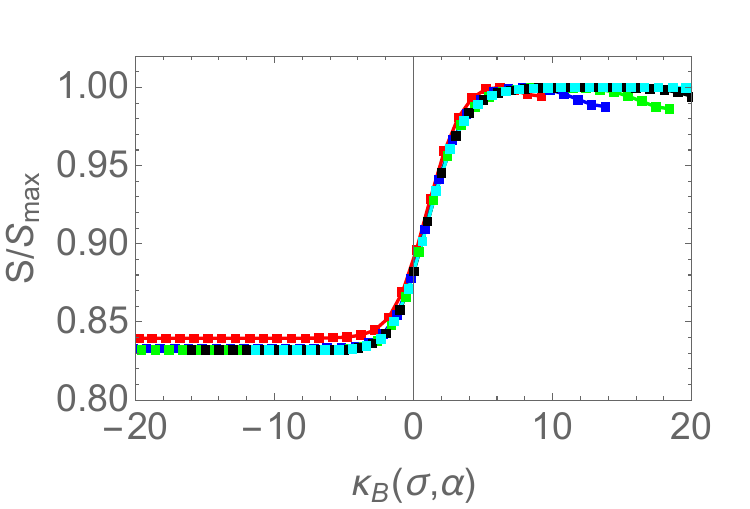}
		\includegraphics[width=7.5cm]{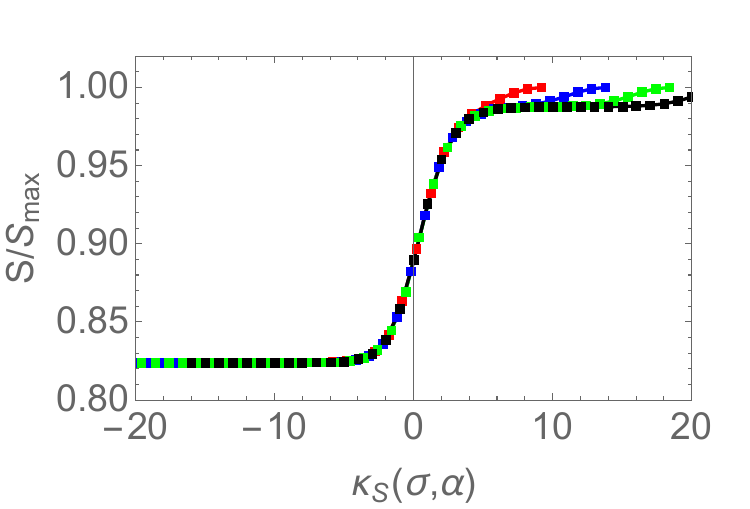}
\caption{\label{fig_exc3}The entropy normalised by its maximum value against the scale  $\kappa_i(\alpha,\sigma)$ defined in (\ref{kappa2}) for fixed $\alpha=10^{-k}$ and $k=2,3,4,5,6$ and varying $\sigma$ for  $K_{\text{B}}(\theta)$ (left) and $K_{\text{S}}(\theta)$ (right). 
We observe the collapse of curves, with the midpoint of the kink, located at $\kappa_{\rm S,B}(\alpha,\sigma)\approx 0$ signaling the threshold for the formation of the unstable particle. 
In both panels the lower plateau is at around $5/6=0.8333...$ which is the ratio of CFT central charges.}
    \end{center}
\end{figure}

A feature that we have now observed repeatedly for many functions is  that fixing $\alpha$ and varying $\sigma$ or viceversa give rise to figures which are qualitatively very similar. 
This similarity is not unexpected, but a consequence of the fact that all functions depend solely on the universal scale  $\kappa_i(\alpha,\sigma)$. 
This is very clearly illustrated in Fig. \ref{fig_exc3} for the function $S/L$ normalised by its local maximum value $S_{\rm{max}}/L$. 
Similar plots can be done for $N_{\rm st}/L$ and $S/N_{\rm st}$. 

The main feature of the Fig. \ref{fig_exc3} is the collapse of multiple curves along the kink that separates the two plateaux. This collapse occurs because we are plotting functions against $\kappa_{\rm B,S}(\sigma,\alpha)$. In all figures the value of $\sigma$ varies while $\alpha$ is fixed to different values corresponding to different symbols. 
Besides the double plateau structure, with a lower plateau for the free regime and a higher plateau for the interacting regime, we also see that after normalisation, the height of the lower plateau is once more well approximated by the value $5/6$, 
as predicted by CFT.

\subsubsection{Emergence of Two Plateaux and a Local Minimum}

\begin{figure}[H]
\begin{subfigure}{0.48\textwidth}
\centering
\includegraphics[width=\textwidth]{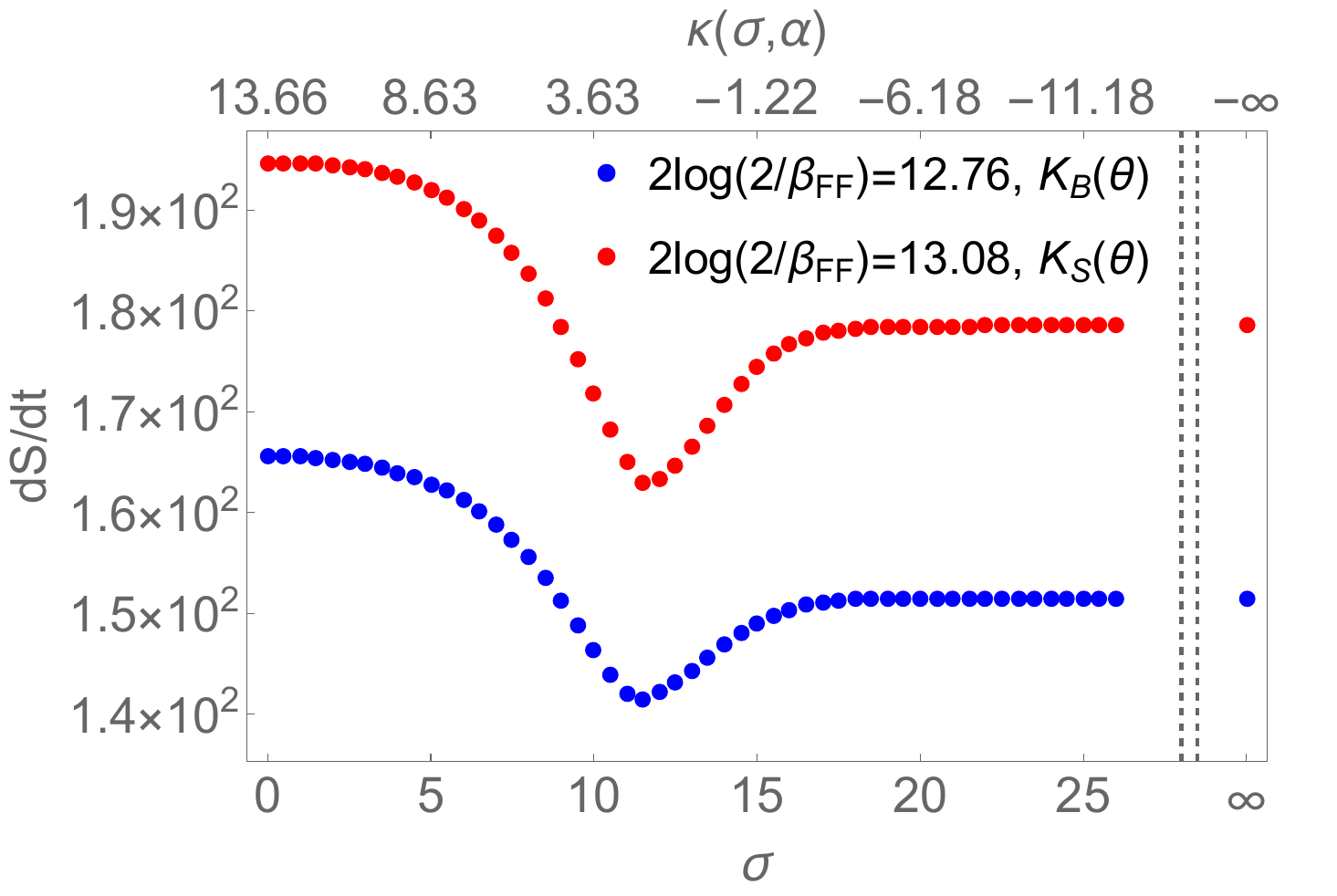}
\caption{\centering
$\frac{\text{d}S}{\text{d}t}$ for smaller energy density}
\vspace{.2cm}
\end{subfigure}\hfill
\begin{subfigure}{0.48\textwidth}
\centering
\includegraphics[width=\textwidth]{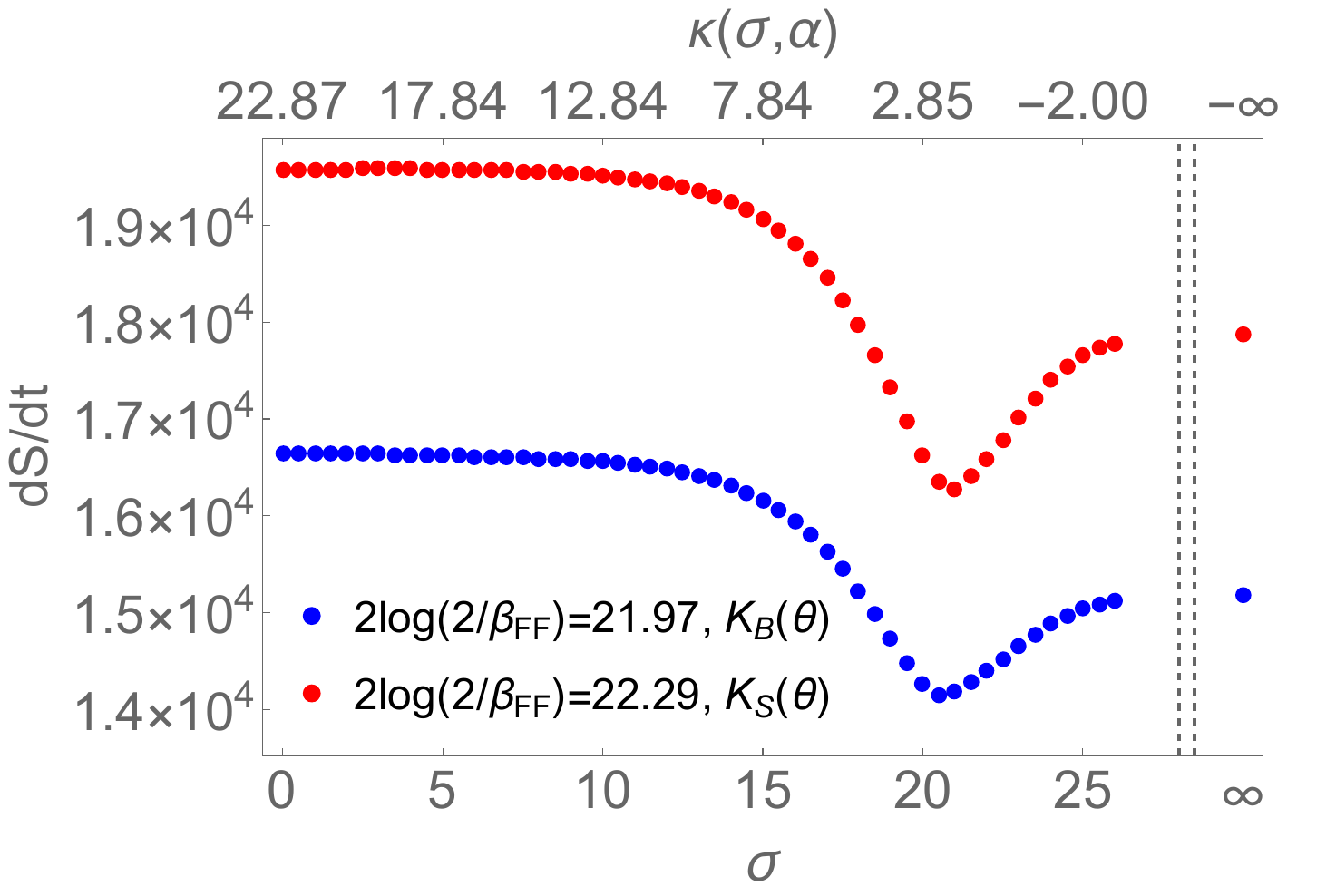}
\caption{\centering
$\frac{\text{d}S}{\text{d}t}$ for larger energy density}
\vspace{.2cm}
\end{subfigure}
\begin{subfigure}{0.46\textwidth}
\centering
\includegraphics[width=\textwidth]{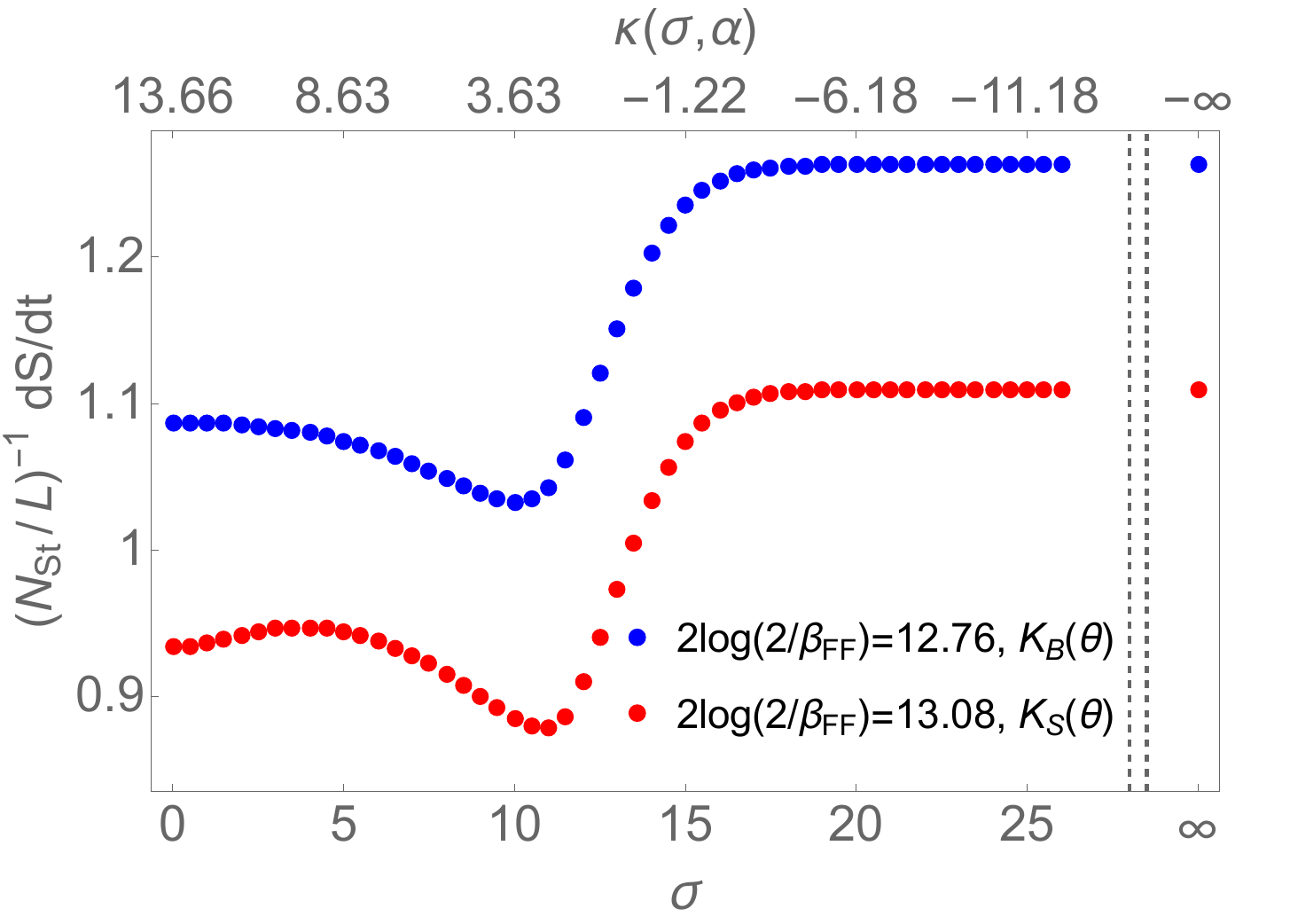}
\caption{\centering
$\frac{L}{N_{\text{st}}} \frac{\text{d}S}{\text{d}t}$ for smaller energy density}
\end{subfigure}\hfill
\begin{subfigure}{0.46\textwidth}
\centering
\includegraphics[width=\textwidth]{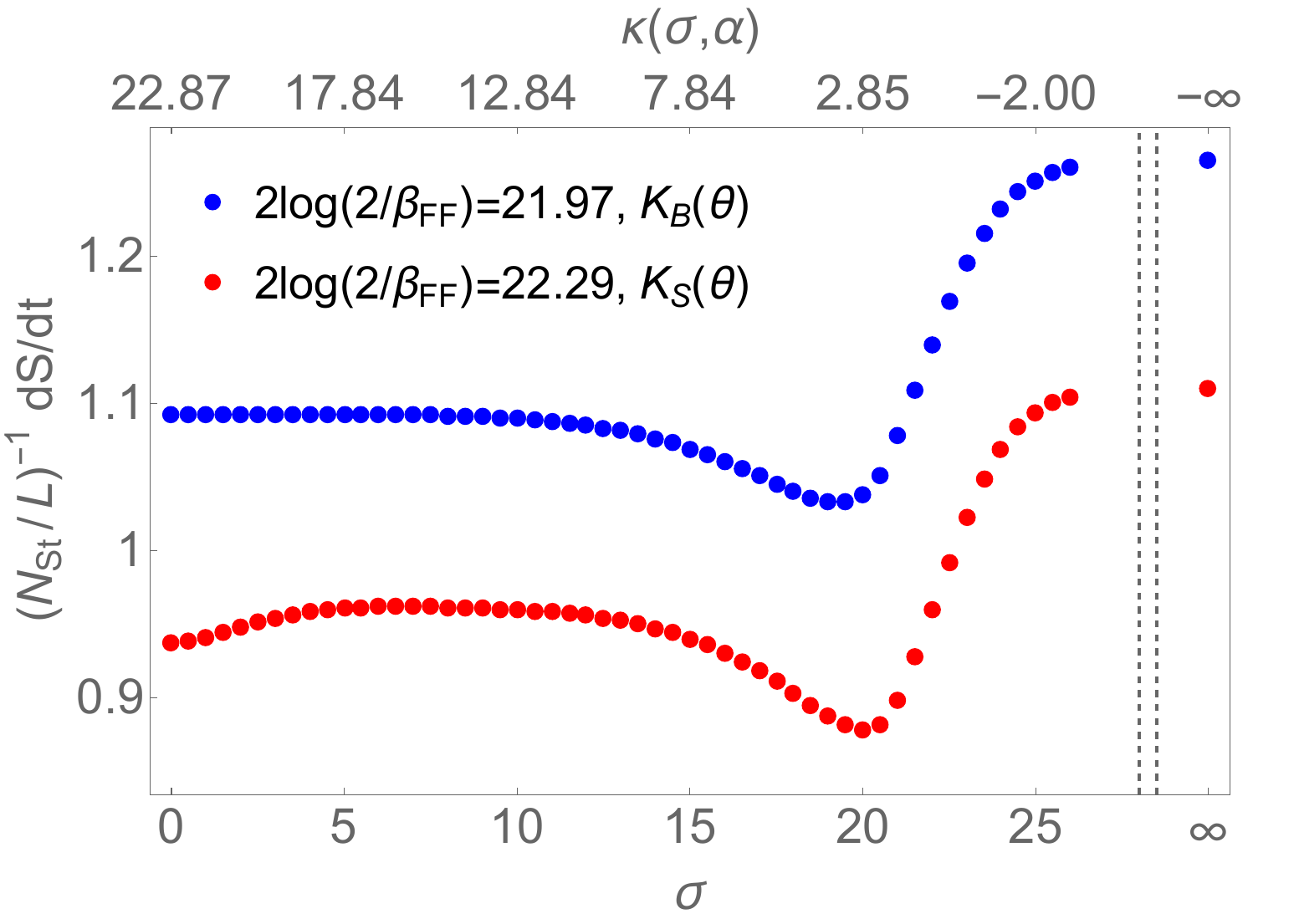}
\caption{\centering
$\frac{L}{N_{\text{st}}}\frac{\text{d}S}{\text{d}t}$ for larger energy density}
\end{subfigure}
\caption{
The entropy production rates d$S$/d$t$ without and with normalisation by $N_{\text{st}}/L$ against the resonance parameter $\sigma$.  We consider quenches with different $K$-functions ($K_{\rm B}(\theta)$  in blue, $K_{\rm S}(\theta)$ in red) and various fixed injected energy densities against the resonance parameter $\sigma$. The injected energy density corresponds to the same free fermion inverse temperatures as in Fig.~\ref{SperLTwoKFixedE}. The bottom and top labels show the values of $\sigma$ and of $\kappa(\sigma,\alpha)=\kappa_{\rm B}(\sigma,\alpha)\approx \kappa_{\rm S}(\sigma,\alpha)$, respectively, where the largest difference between $\kappa_{\rm B}$ and $\kappa_{\rm S}$ is $0.05$. A local minimum is seen at a value of $\sigma=\sigma_{\rm min}$ which is slightly below  $\sigma=2\log2/\beta_{\rm FF}$. The precise numbers are reported in Table \ref{TableSigmasFixedEAllBosonSquared}. }
\label{dSdtFixedE}
\end{figure}

Another indication of the presence of unstable excitations is observed when studying the entropy production rate. 
Both the production rate d$S$/d$t$ and $L/N_{\text{st}}\times$d$S$/d$t$ develop a local minimum at a specific  value of $\sigma$ as shown in Fig. \ref{dSdtFixedE}. We denote this value by $\sigma_{\text{min}}$ and we argue that it is related, once more, to the formation of the unstable particle. 
More precisely, the values of $\sigma_{\text{min}}$ can be compared to the scale $2\log2/\beta_{\text{FF}}$ which in turn characterises the energy available in the system. A 
comparison between $\sigma_{\text{min}}$ and
$2\log2/\beta_{\text{FF}}$ can be found in Table \ref{TableSigmasFixedEAllBosonSquared}.

We always find $\sigma_\text{min}<\log2/\beta_{\rm FF}$: the minimum occurs slightly below the mass of the unstable particle $M\approx \sqrt{2} m e^{\frac{\sigma}{2}}$. We should recall however, that  this approximation of the mass works only for large $\sigma$, thus explaining in part the difference (indeed, as $\sigma$ is increased, the relative difference between the values in the rows of the Table decreases). It is also worth pointing out that the separation between interacting and non-interacting regime is not sharp with respect to the energy scale (the transition between free fermion and interacting regime has a certain width, cf. Fig.~\ref{fig_exc3}). 
Comparing Fig.~\ref{SperLTwoKFixedE} with Fig.~\ref{dSdtFixedE}, we can see that the values $\sigma_{\rm{min}}$ (for d$S$/d$t$) approximately correspond to the top of the kink that connects the first and second plateaux in Fig.~\ref{SperLTwoKFixedE}, or to the end of the first plateau (for $L/N_{\text{st}}\times$d$S$/d$t$) whereas $2\log 2/\beta_{\rm FF}$ corresponds roughly to the midpoint of the kink. 



\begin{table*}
\begin{center}
\begin{tabular}{|c||c||c|c|c|c|c|}
  \hline
    \multirow{3}{*}{$K_{\rm B}(\theta)$} & $2\log 2/\beta_\text{FF} $ & $8.2$ & $12.8$ & $17.4$ & $22.0$ & $26.6$ \\ \cline{2-7} \cline{2-7}
        &   $\sigma_\text{min}$ (d$S$/d$t$)& 7.1 &  11.4 & 16.0  & 20.6 & 25.7 \\  \cline{2-7} 
         &  $\sigma_\text{min}$ ($N_{\text{st}}/L \times$d$S$/d$t$)& 5.4 & 10.1  & 14.7 & 19.3 & 23.9 \\  \hline\hline
     \multirow{3}{*}{$K_{\rm S}(\theta)$} & $2\log{2}/{\beta_\text{FF}}$ & $8.5$  &  $13.1$ & $17.7$ & $22.3$ & $27.0$ \\\cline{2-7}  \cline{2-7}
          &   $\sigma_\text{min}$ (d$S$/d$t$)& 7.3 &  11.7 & 16.3 & 20.9 & 25.5 \\ \cline{2-7}
           &  $\sigma_\text{min}$ ($N_{\text{st}}/L \times$d$S$/d$t$)& 6.2 & 10.8  & 15.4 & 20.0 & 24.6 \\ \hline
\end{tabular}
\caption
{$\sigma_\text{min}$ for the quench with $K_{\text{B}}(\theta)$ and $K_{\text{S}}(\theta)$.
}
\label{TableSigmasFixedEAllBosonSquared}
\end{center}
\end{table*}

\subsubsection{Depletion as a Function of Quench Magnitude}
\begin{figure}[H]
\begin{center}
 \includegraphics[width=7.5cm]{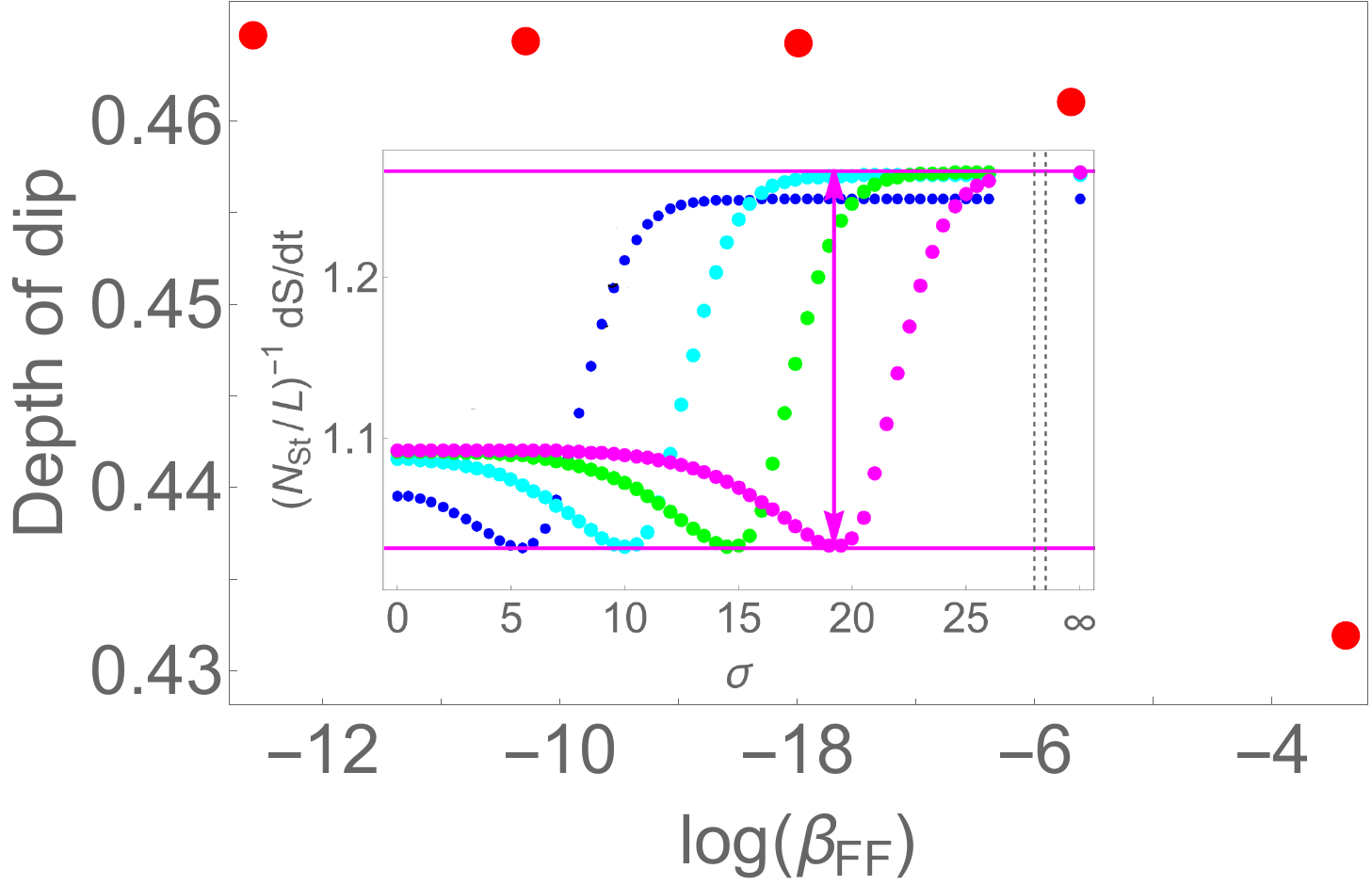}
 \includegraphics[width=7.5cm]{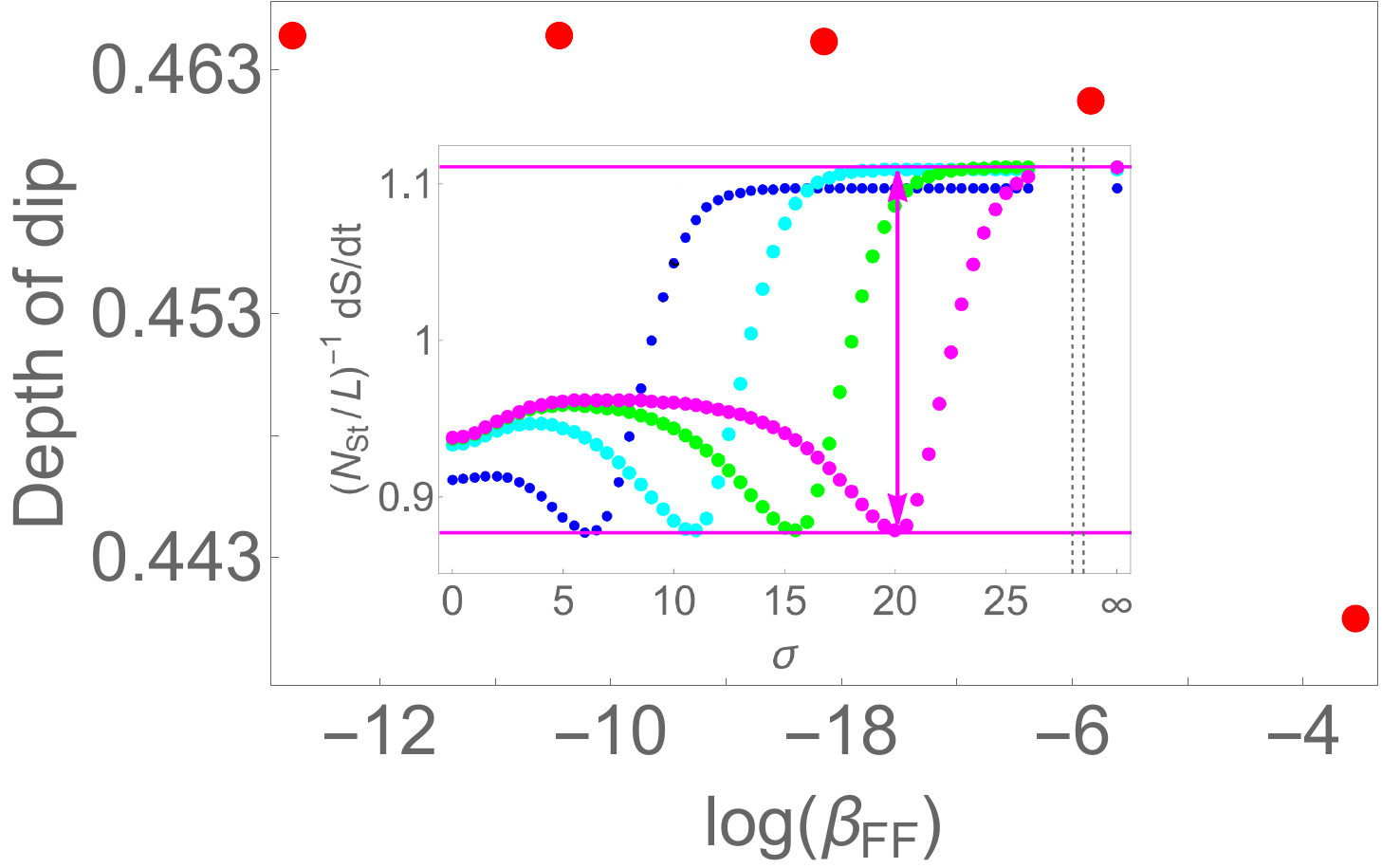}
\caption{\label{Depths2} 
		   The depletion of the normalised entropy production rate $\frac{L}{N_{\text{st}}}\left.\frac{\text{d}S}{\text{d}t}\right|_\text{FF}-\underset{\sigma}{\text{min}}\left\{\frac{L}{N_{\text{st}}}\frac{\text{d}S}{\text{d}t}\right\}$ is visualised as a function of $\beta_\text{FF}$ for $K_{\rm B}(\theta)$ and $K_{\rm S}(\theta)$ (left and right). The colours blue, cyan, green and magenta in the insets correspond to $-\log\beta_\text{FF}=3.38,5.69,7.99,10.29$ for  $K_{\rm B}(\theta)$ and $3.54,5.85,8.15,10.45$ for  $K_{\rm S}(\theta)$. 
		   }
    \end{center}
\end{figure}

We finally take a closer look at the local minimum of the function $(N_{\text{st}}/L)^{-1}\times\text{d}S/\text{d}t$, in particular at its depth as a function of $\sigma$. The insets of Fig.~\ref{Depths2} present several figures of the type seen in the last row of Fig.~\ref{dSdtFixedE} (up to a scaling that is explained in the caption). As $\beta_{\rm FF}$ is varied, there is hardly any change to the depth or shape of the minimum. However, if we consider further values of $\beta_\text{FF}$ and plot the depth of the minimum (red dots) we see that there is a change for small energies. 
There are two main properties worth highlighting:
(i) for $-\log {\beta_\text{FF}\approx -\log\alpha} \gg 1$, the depth of the minimum saturates to a maximum value which is roughly the same for both $K$-functions (right and left figures are very similar); (ii) the depth of the minimum and its shape start to change as 
{$\beta_\text{FF}\approx \alpha \rightarrow 1 $} which is the limit of no quench. In this case the depth of the minimum becomes slightly smaller.
The reasons for changes around $\alpha=1$ are the same as discussed in Subsection \ref{depli}. In addition, it is rather natural that the minimum should reduce as $\alpha$ approaches 1, since in the limit of no quench there should be no minimum either. 
\subsection{Spectral Densities and Effective Velocities}\label{SpectralDensities}

One of our main observations so far is that the entropy production rate is suppressed when the unstable particle starts to form. In this subsection, we argue that this suppression is explained at least in part by the slowdown of the stable particles that precedes the formation of unstable ones. The slowdown can be understood by studying the effective velocities of stable quasiparticles and various other spectral quantities, in particular the spectral entropy density $s(\theta)$ and the spectral particle density $\rho(\theta)$.

\begin{figure}[H]
\begin{subfigure}{0.48\textwidth}
\centering
\includegraphics[width=\textwidth]{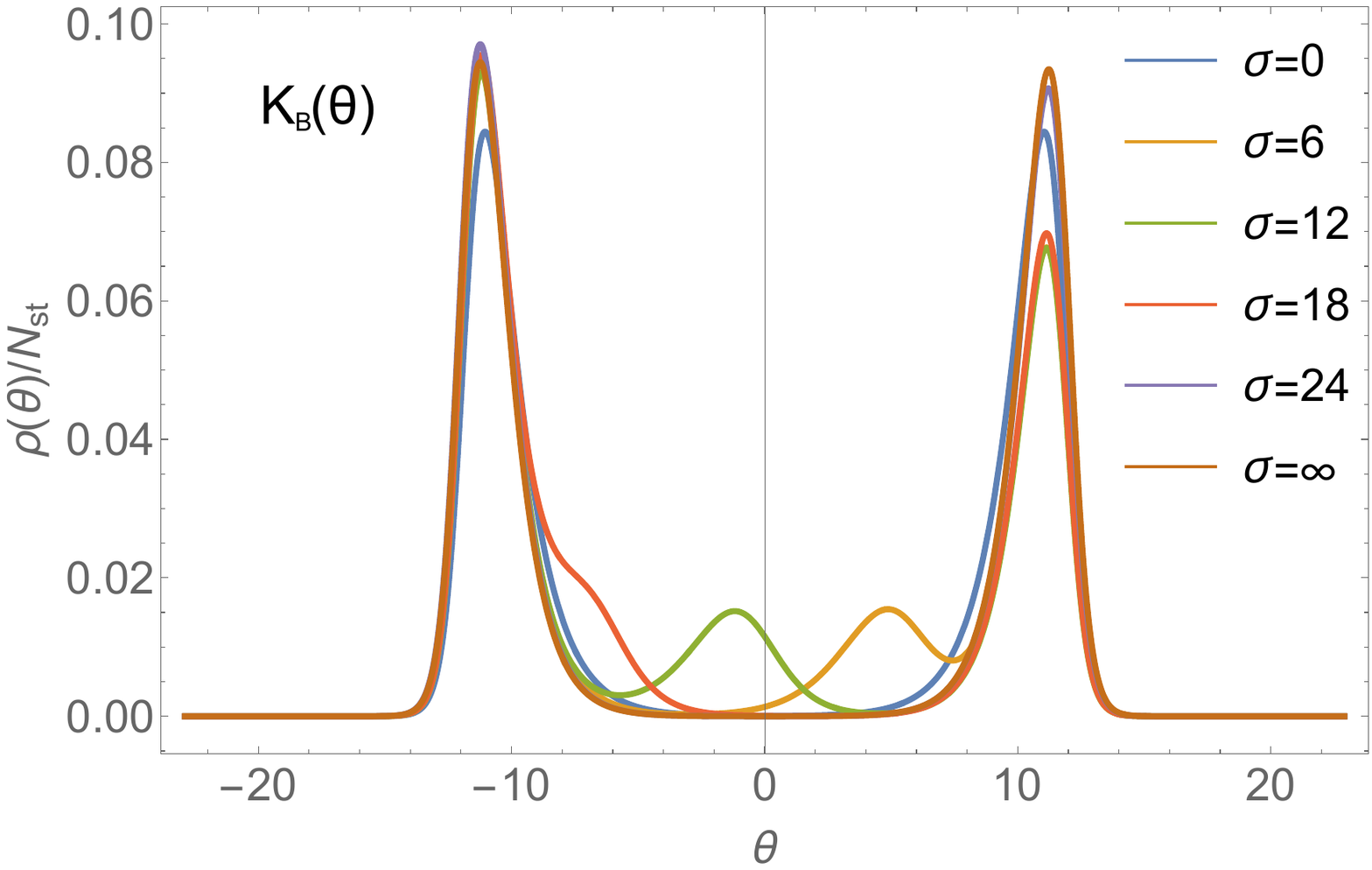}
\end{subfigure}\hfill
\begin{subfigure}{0.48\textwidth}
\centering
\includegraphics[width=\textwidth]{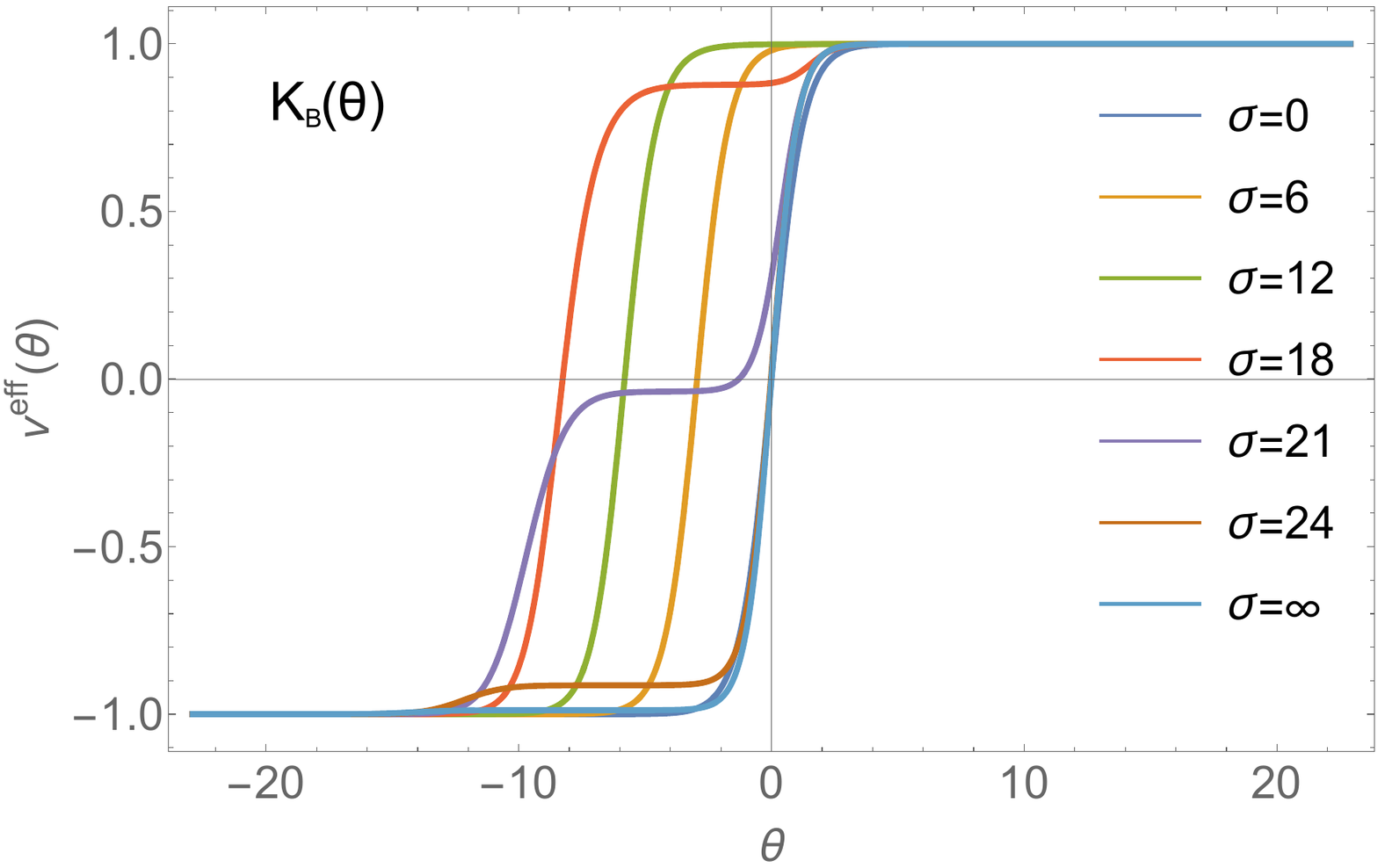}
\end{subfigure}
\caption{
The spectral particle density divided by the total number of stable particles $\rho(\theta)/N_\text{st}$ (left) and the effective velocity $v^\text{eff}(\theta)$ (right) for fixed energy density and various $\sigma$ parameters. 
We considered a quench with $K_{\rm B}(\theta)$ and
$\beta_{\rm{FF}}=3 \times10^{-5}$, i.e. $2\log2/\beta_{\rm FF}=22.215$. 
}
\label{SpectralDensitiesBosonFixedEVaryingSigmaFullRange}
\end{figure}

Let us consider, with the help of Fig. \ref{SpectralDensitiesBosonFixedEVaryingSigmaFullRange}, the behaviour of the spectral particle density $\rho(\theta)$ and the effective velocity $v^\text{eff}(\theta)$ in a typical situation. We focus on the quench function $K_{\rm B}(\theta)$, fix the injected energy to  $\beta_{\text{FF}}=3 \times10^{-5}$, that is $2\log2/\beta_{\rm FF}=22.215$, and vary the resonance parameter $\sigma$. The behaviour that we observe is very similar to that of the thermal case, analysed in Ref. \cite{ourU}.

In particular the spectral density $\rho(\theta)$ (and also its normalised version, in Fig.~\ref{SpectralDensitiesBosonFixedEVaryingSigmaFullRange}) exhibits either two or three peaks. The leftmost peak is referred to as the interaction peak, the rightmost one as the free fermion peak. If present, the additional peak in between is called the subsidiary peak. We observe that both at $\sigma=0$ and $\sigma=\infty$, the interaction and the free fermion peaks are mirror images of each other. These are the two limits where the parity symmetry of the model is restored at TBA level and correspond to either infinitely lived unstable excitations or no unstable excitations at all, respectively. Increasing $\sigma$ from zero, first the size of these two peaks changes, and then the subsidiary peak forms and emerges from the free fermion peak. Upon further increase of $\sigma$ the peak move towards the interaction peak and eventually merges with it for $\sigma=\infty$. 

The behaviour of the effective velocity can also be easily described. At $\sigma=0$ and $\sigma=\infty$,  $v^\text{eff}(\theta)$ is a parity-odd function. Increasing $\alpha$, this function undergoes first a shift towards the left and then develops an extra plateau for negative rapidity values. This plateau then moves down towards the value $-1$ restoring the parity-odd function at the free fermion point.

\begin{figure}[t]
\begin{subfigure}{0.48\textwidth}
\centering
\includegraphics[width=\textwidth]{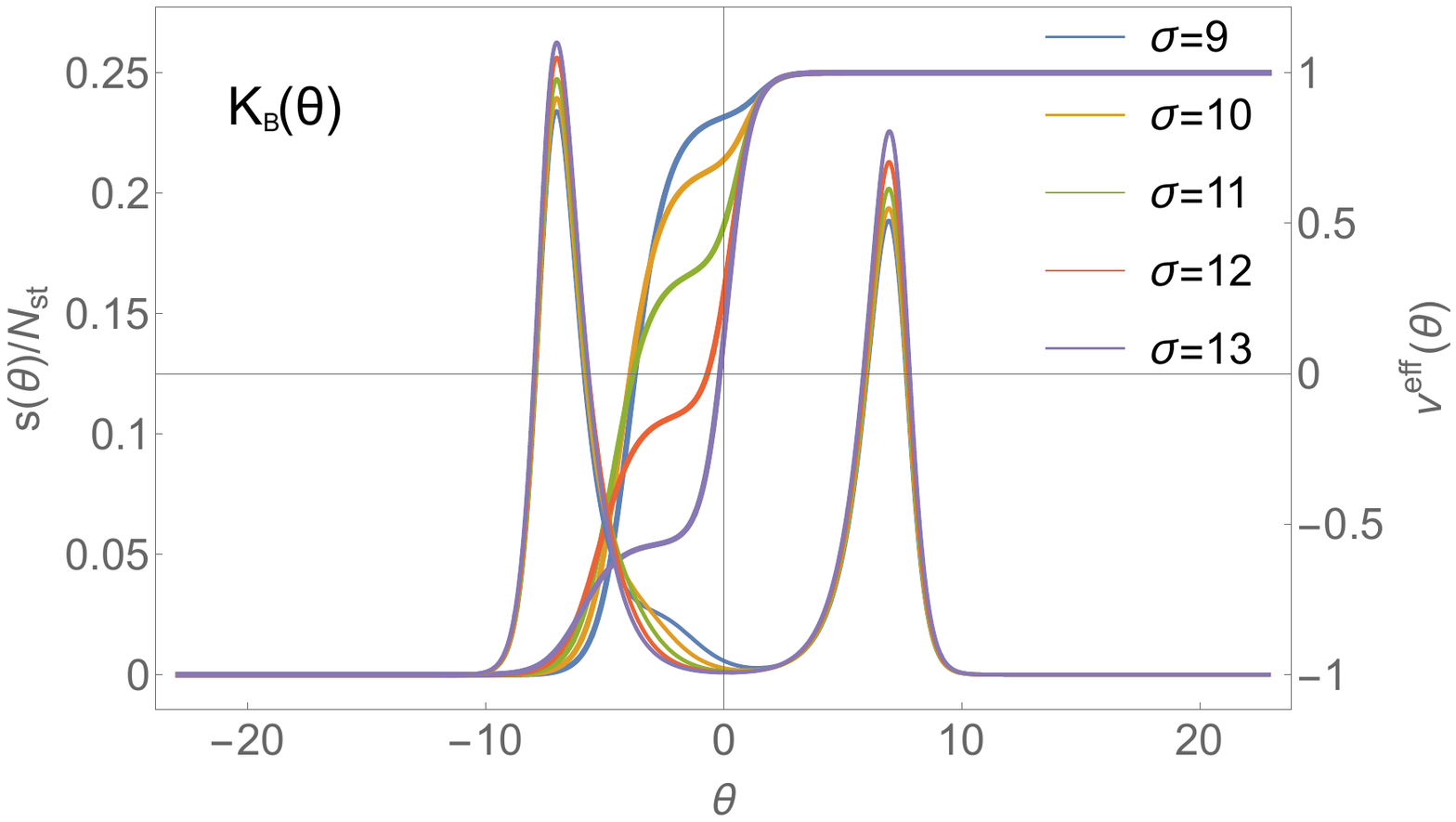}
\end{subfigure}\hfill
\begin{subfigure}{0.48\textwidth}
\centering
\includegraphics[width=\textwidth]{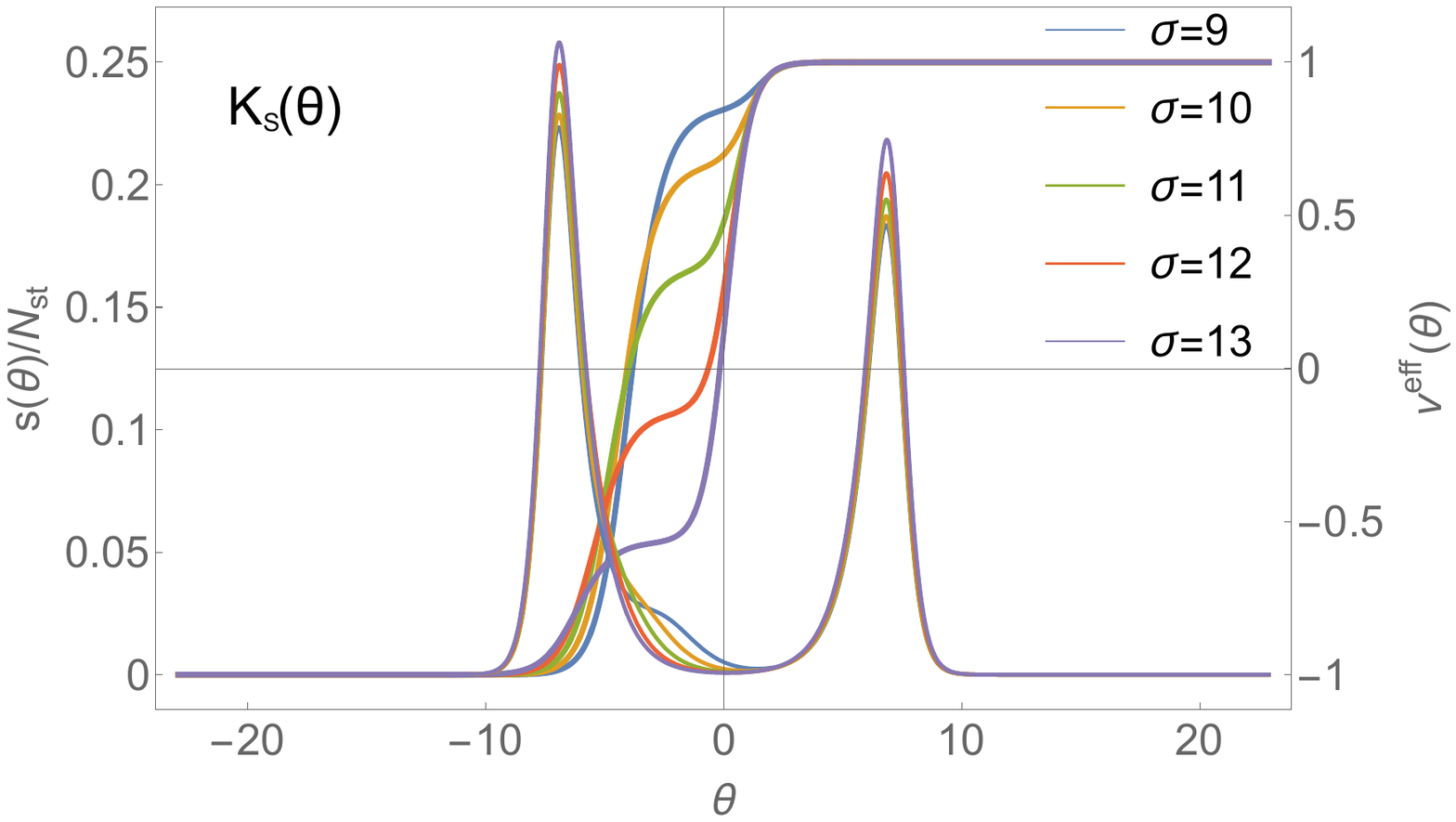}
\end{subfigure}\\

\begin{subfigure}{0.48\textwidth}
\centering
\includegraphics[width=\textwidth]{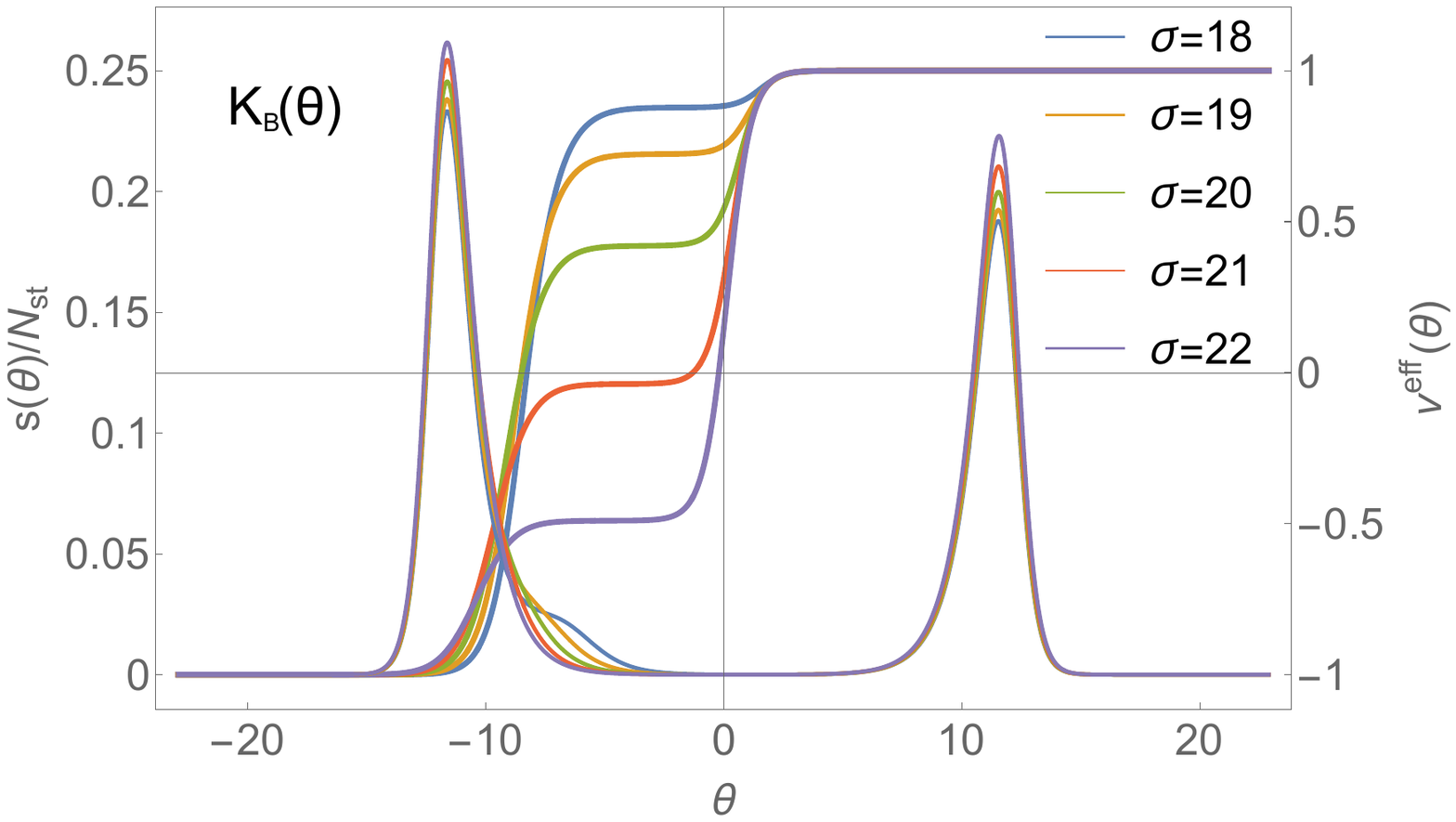}
\end{subfigure}\hfill
\begin{subfigure}{0.48\textwidth}
\centering
\includegraphics[width=\textwidth]{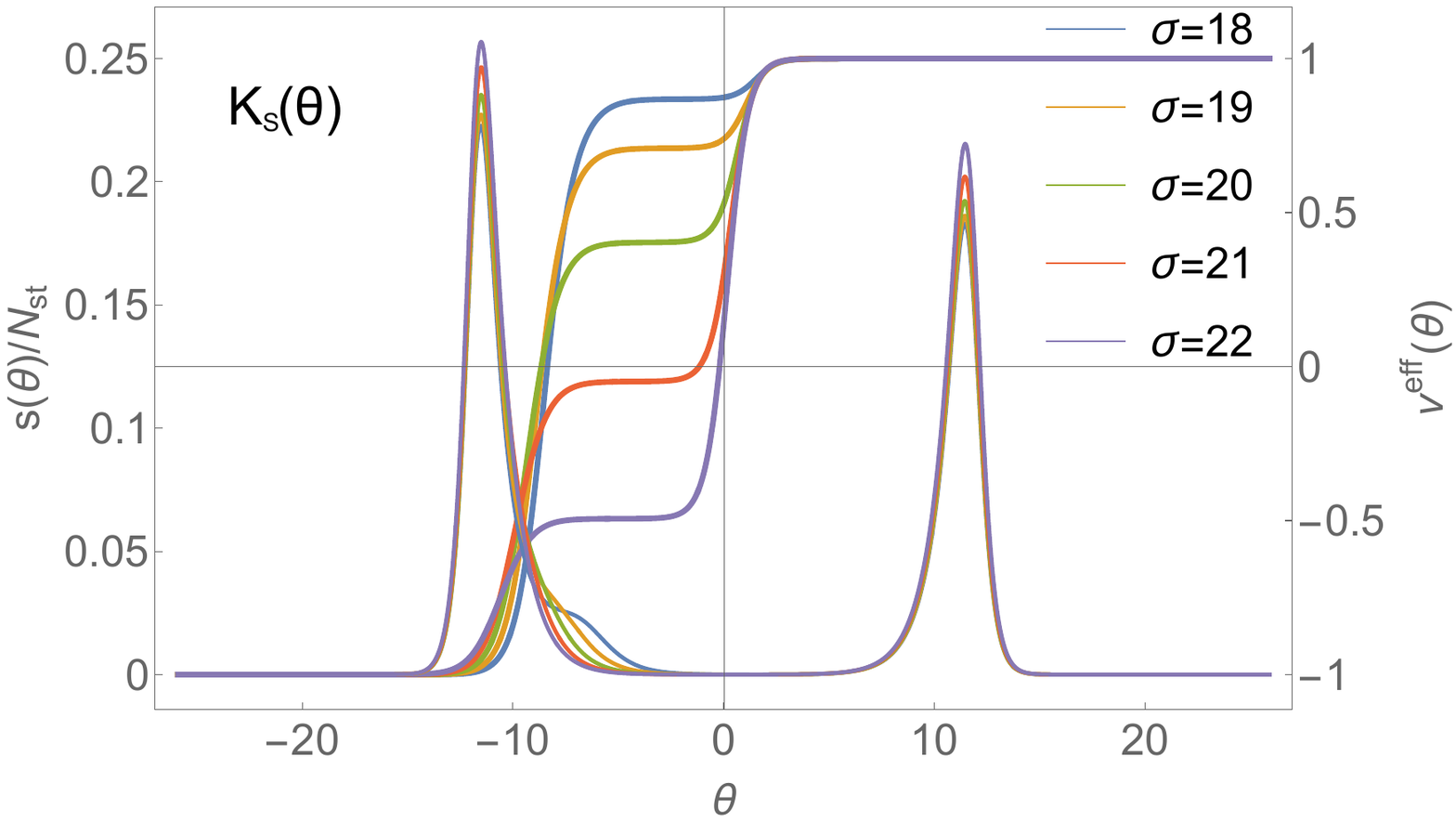}
\end{subfigure}
\caption{
The (normalised) spectral steady-state entropy density $s(\theta)/N_\text{st}\in[0,0.25]$ (left axis in each panel) and the effective velocity $v^\text{eff}(\theta)\in[-1,1]$  (right axis) for fixed injected energy densities and various $\sigma$ parameters near the local minimum of the production rates.
We have $2\log2/\beta_{\rm FF}=13.005$ for the left figures and $2\log2/\beta_{\rm FF}=22.215$ for the right figures.}
\label{SpectralDensitiesFixedE}
\end{figure}

We now turn back to the phenomenon of suppression in the entropy production rates and take a closer look at spectral quantities in the vicinity of the local minimum.
A first fundamental observation is that the effective velocity $v^\text{eff}(\theta)$ is zero (or at least much smaller than $1$) in a region where the spectral density of the stable particle has some support. 
This is visualised in Fig. \ref{SpectralDensitiesFixedE}, especially in Fig.~\ref{SpectralDensitiesFixedE} (d) for $\sigma=21$. Consequently, we can argue that as soon as interaction starts and unstable particles are formed, pairs of stable particles start to slow down and some of the energy goes into these particles which are more massive and so slower, slowing down entropy growth. 

In Fig.~\ref{SpectralDensitiesFixedE} we once more present the plots of the (normalised) spectral steady state entropy and effective velocities. We notice that the structure of the spectral entropy density is very similar to that of the spectral particle density and there is an overlap between the intermediate plateau of the effective velocity and the extra peak of the spectral entropy, which is associated with the formation of unstable particles.

Based on the behaviour of these  spectral quantities, we conclude that the main mechanism responsible for the decrease in the entropy production rates is that some stable particles are slowed down to form unstable ones. This is only possible due to the strong separation of scales that is enabled by the free parameter $\sigma$ and would not be observed for a similar quench in a theory with only stable bound states in which all masses are fixed and, typically, of the same order of magnitude.

\section{Conclusions and Outlook}
\label{conclus}

In this paper we have used the quench action approach in conjunction with the quasiparticle picture to study the stationary value of the thermodynamic entropy, its production rate and other related quantities. We have done so in a theory which has the peculiarity of including two stable excitations and one unstable quasiparticle. Following on from previous works \cite{ourU,nextU,nnextU} we have established the presence of unstable particles in the steady state resulting from equilibration after quite generic quantum quenches. We have found several precise signatures of the onset and presence of unstable particles, which we summarise below.

\medskip 
The starting point of our analysis has been the choice of quench in the context of the quasiparticle picture. Given that the model in question is non-trivially interacting, the construction of a squeezed coherent state and the associated $K$-function from first principles is rather difficult. Instead, we have chosen three known $K$-functions with standard features, depending on a quench parameter $\alpha$ and used those as our starting point. 

The quench protocols are the following: either keeping $\sigma$ fixed and varying $\alpha$ or \textit{keeping the injected energy fixed} and varying the resonance parameter  (this is not completely equivalent to keeping $\alpha$ fixed but nearly so). The main observations, irrespective of the choice of $K(\theta)$, are as follows.

\begin{itemize}
    \item{When the unstable particles are present, there is an increase in the final steady-state entropy density ($\frac{S}{L}$) and on the  number of stable particles ($\frac{N_\text{st}}{L}$).} 
    
    \item As functions of the resonance parameter $\sigma$, these functions as well as their ratio develop a staircase shape consisting of two steps and one connecting kink. The position of this kink is directly related to the amount of energy injected in the system. The midpoint of the kink corresponds to matching this energy with the mass of the unstable particle. 
    \item We recover therefore a familiar picture for this model, namely that the increase in degrees of freedom that occurs when the available energy is compatible with the formation of unstable particles leads to many quantities ``flowing" from their free fermion values to a new value. For the entropy, this new value in the interacting regime is higher, which confirms the intuition that increasing degrees of freedom leads to increased thermodynamic entropy. 
    \item Indeed the relative ratio of plateau height is given by the ratio of central charges in the deep UV of the two regimes seen in the theory: a non-interacting regime, described by two free fermions and an interacting regime described by a non-trivial CFT. The respective central charges are $c=1$ and $c=\frac{6}{5}=1.2$.
    \end{itemize}

The above trends are also observed  for the entropy production rates $(\text{d}S/\text{d}t)$  and related quantities, where $\text{d}S/\text{d}t$ is in analogy with $S/L$, and  $(N_{\text{st}}/L)^{-1}\times\text{d}S/\text{d}t$ with $S/N_\text{st}$. However, these exhibit an additional feature, that is a local minimum  which occurs when the unstable particles start to appear in the system. The behaviour/presence of this minimum is, nevertheless, more subtle and depends on the way the parameters change.

\begin{itemize}
    \item{When \textit{the injected energy density is kept fixed} and the resonance parameter is varied the minimum occurs both in $\frac{{\rm d}S}{{\rm d}t}$ and $\frac{L}{N_{\text{st}}}\frac{{\rm d}S}{\rm{d} t}$. In contrast, if the quench parameter $\alpha$ is fixed and we vary $\sigma$ a pronounced minimum occurs only in $\frac{L}{N_{\text{st}}}\frac{{\rm d}S}{\rm{d} t}$. } 
     \item{When $\sigma$ is kept fixed and the \textit{quench parameter $\alpha$} is varied the minimum occurs only in $\frac{L}{N_{\text{st}}}\frac{{\rm d}S}{\rm{d} t}$,  whereas $\frac{{\rm d}S}{{\rm d}t}$ shows a completely monotonic behaviour. It is indeed possible to show that $\frac{{\rm d}S}{\rm{d} t}\,\propto \, \alpha^{-1}$ for all quenches considered here, with a coefficient which is numerically different in the interacting and non-interacting regimes, with mutual ratio once more given by the ratio of central charges.} 
     \item Once the $K$ function and quantity of interest are chosen, the depth of the minimum is constant for a wide range of parameters, although it is slightly larger  when $\sigma$ is close to 0 (the unstable particle becomes a virtual particle) and slightly smaller when $\alpha$ is close to 1 (small quench). The precise mechanism that leads to greater depletion when the unstable particle is longer lived, eventually becoming a virtual particle, is not fully understood. 
     \item Dynamically, the minimum is associated to the slowdown of stable quasiparticles that occurs exactly when unstable particles start to form. This can be explained by considering the spectral density and effective velocity of the stable quasiparticles.  Namely one can observe that the spectral peak both in $\rho(\theta)$ and in $s(\theta)$ (spectral entropy density) broadens in a range of rapidities for which the effective velocities are close to zero. 
\end{itemize}

A distinct feature of this model is that the position and relative height of plateaux as well as the position and depth of the local minimum are universal with respect to a global parameter $\kappa_i(\sigma,\alpha)$ with $i=\rm B,S,0$. This parameter provides a natural RG scale, with negative values associated with the free regime, positive values associated with interaction and zero value corresponding to the threshold for the formation of unstable particles. The existence of such a scale is due to the presence of the free parameter $\sigma$, so it is a distinct consequence of the presence of unstable particles.

We close this paper by pointing out that the phenomenon of entropy growth depletion followed by sharp increase has been connected to the increase in degrees of freedom in the spectrum, e.g.~the formation of bound states, for different theories \cite{Gibbsparadox,LPT,PPT} and termed the ``dynamical manifestation of the Gibbs paradox". 
While there are many analogies between the 
findings of Refs. \cite{Gibbsparadox,LPT,PPT} and our results (most notably (i) the connection to counting degrees, 
(ii) the formation of the minimum related to a reduction in velocities of the binding particles), there are important differences related to the fact that our particles are unstable. 
In Refs. \cite{Gibbsparadox,LPT,PPT} it is pointed out that also virtual particles (i.e. stable excitations below the threshold of formation) should be responsible of a reduction of the entropy.
As observed earlier, it seems that in our case too virtual particles play an interesting role since depletion is maximised when the unstable particles become virtual. 

\medskip 
There are several further extensions of this work that could be carried out, in particular by considering other models of the same family where multiple unstable excitations of tunable masses are present. More fundamentally, it would be very interesting, even for the present model, to have a derivation of a $K$-function that we could associate to a physical quench such as a mass or resonance parameter quench. 
Finally, improved understanding of the entanglement dynamics in the presence of unstable bound states should also help in the study of the breakdown of confinement of elementary excitations and the onset of thermalisation \cite{confi,scb-21,mrw-17,cr-19,clsv-20,lsmc-21,Liu2018,tan2019,jkr-19,rjk-19}. 

\medskip

\noindent {\bf Acknowledgement:} We are grateful to G\'abor Tak\'acs for a very interesting discussion on the dynamical Gibbs paradox and its possible connection to some of our results. 
O.C.-A. is grateful to Vincenzo Alba and Benjamin Doyon for discussions and especially for first spotting that the relative height of the plateaux is related to the ratio of the central charges. O.C.-A. is also grateful to the organisers of the program on Randomness, Integrability and Universality, held at the Galileo Galilei Institute (Florence) April 19--June 3 (2022) and to the organisers of the conference ``Talking Integrability: Spins, Fields and Strings", held at Kavli Institute for Theoretical Physics (Santa Barbara) August 29--September 1 (2022) for financial support and for providing a great environment to complete some of this work. In the case of KITP, financial support was provided by the National Science Foundation under Grant No. NSF PHY-1748958. O. C.-A. gratefully acknowledges EPSRC's financial support under Small Grant
EP/W007045/1. P.C. and D.X.H acknowledge support from ERC under Consolidator grant number 771536 (NEMO).

\appendix 

\section{Some Analytical Derivations for Free Theories}
\label{AppA}
It is interesting to analytically derive some of the properties we have observed in the main text, such as the formation of plateaux. 
Our model allows us to consider some of these properties, at least in the limit when the theory is free, namely consisting just of two free fermions. 
In that case, the pseudoenergies are given by the driving term of the TBA, namely
\beq 
\varepsilon(\theta)=-\log|K(\theta)|^2\,,
\eeq 
and it is easy to show that the entropy per unit length becomes simply
\beq 
\frac{S}{L}=\frac{1}{2\pi}\int_{-\infty}^\infty d
\theta \frac{\cosh\theta}{1+|K(\theta)|^2}\left((1+|K(\theta)|^2)\log(1+|K(\theta)|^2)-|K(\theta)|^2\log|K(\theta)|^2\right)\,,
\label{sL}
\eeq 
We can now argue that the logarithm of this quantity should be linear in $\log\alpha$. In fact, this is most obvious for the quench with $K_{\rm S}(\theta)$, but the same sort of argument can be applied to the other cases. We can consider a very crude approximation of the function $|K_{\rm S}(\theta)|$, namely that it is essentially $1$ for $-\log\alpha \leq \theta \leq \log\alpha$ and zero otherwise. In this approximation, the function (\ref{sL}) becomes 
\beq 
\frac{S}{L}\approx \frac{\log 2}{2\pi} \int_{\log\alpha}^{-\log\alpha} \cosh\theta \, d\theta=\frac{1}{2\pi}\left(\frac{1}{\alpha}-\alpha\right) \log 2.
\label{consa}
\eeq 
Thus, for $\alpha\ll 1$ we have that
\beq 
\log\frac{S}{L}\approx \log({\log 2})-\log (2\pi) -\log\alpha - O(\alpha^2)\,.
\label{lscaling}
\eeq 
Although this is of course a very simple argument for free theories, we believe that a similar picture works for the interacting case too, which explains the linear scaling in $\log\alpha$ seen in the numerics. 
The piecewise constant approximation is less accurate for the other quenches, although they also develop a central plateau for $\alpha\ll 1$ (even $K_{\rm 0}(\theta)$ that presents two plateaux separated by a zero at $\theta=0$, which merge for small $\alpha$). 

For free fermions, this argument extends automatically to the entropy production rate, since this is given by the same integral with the extra factor $|v^{\rm eff}(\theta)|$. Furthermore, $v^{\rm eff}(\theta)=\tanh\theta$, can be approximated by $1$ inside the integral, giving exactly the same scaling (\ref{lscaling}).

For fixed $\alpha$ these results also imply that both $S/L$ and $\text{d}S/\text{d}t$  tend to saturate to the value (\ref{consa}) which scales as $1/\alpha$ for $\alpha\ll 1$. This is what we observed numerically as well. 
However, the $O(1)$ term $\log(\log2)-\log(2\pi)$ of Eq. (\ref{lscaling}) does not provide an accurate prediction even in the free fermion regime. This is due to the fact that the piecewise approximation of $K_i(\theta)$  approximation is very crude. In particular, the contributions from the decaying part of the $K$-functions are not as negligible as for typical TBA functions where the decay is often double exponential (like for $L$-functions). Here $K_{\rm S}(\theta)$ decays only exponentially and its logarithm only linearly in $\theta$. Thus approximating either function by just a box shape is a gross simplification which accurately predicts the $\alpha$ dependence but not the normalisation constants. 

\section{Results for $K_{\rm 0}(\theta)$}
\label{AppB}
\begin{figure}[H]
\begin{subfigure}{0.48\textwidth}
\centering
\includegraphics[width=\textwidth]{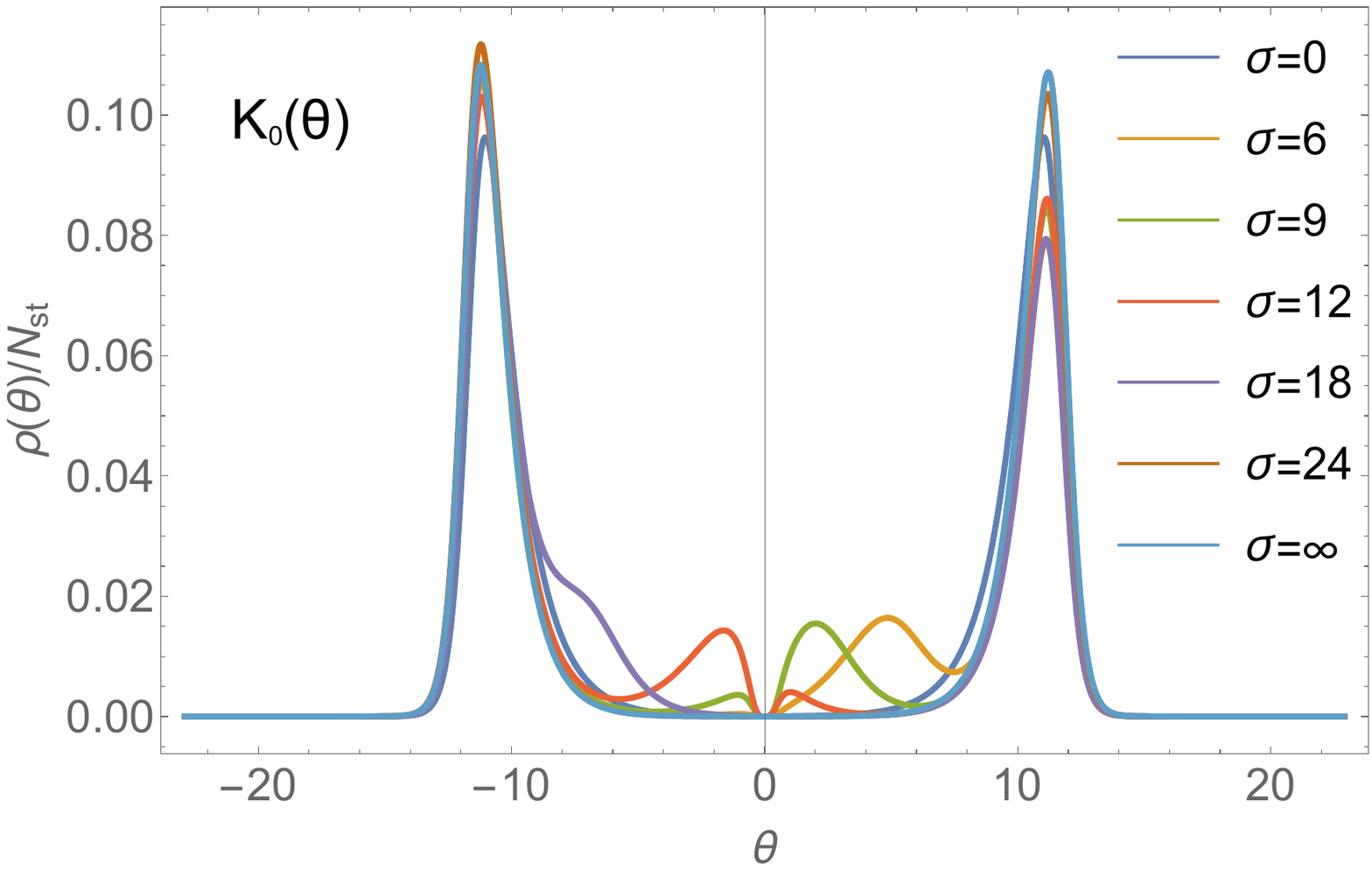}
\caption{\centering
$\rho(\theta)/N_\text{st}$ for $K_{\rm 0}(\theta)$ and various $\sigma$-s}
\end{subfigure}\hfill
\begin{subfigure}{0.48\textwidth}
\centering
\includegraphics[width=\textwidth]{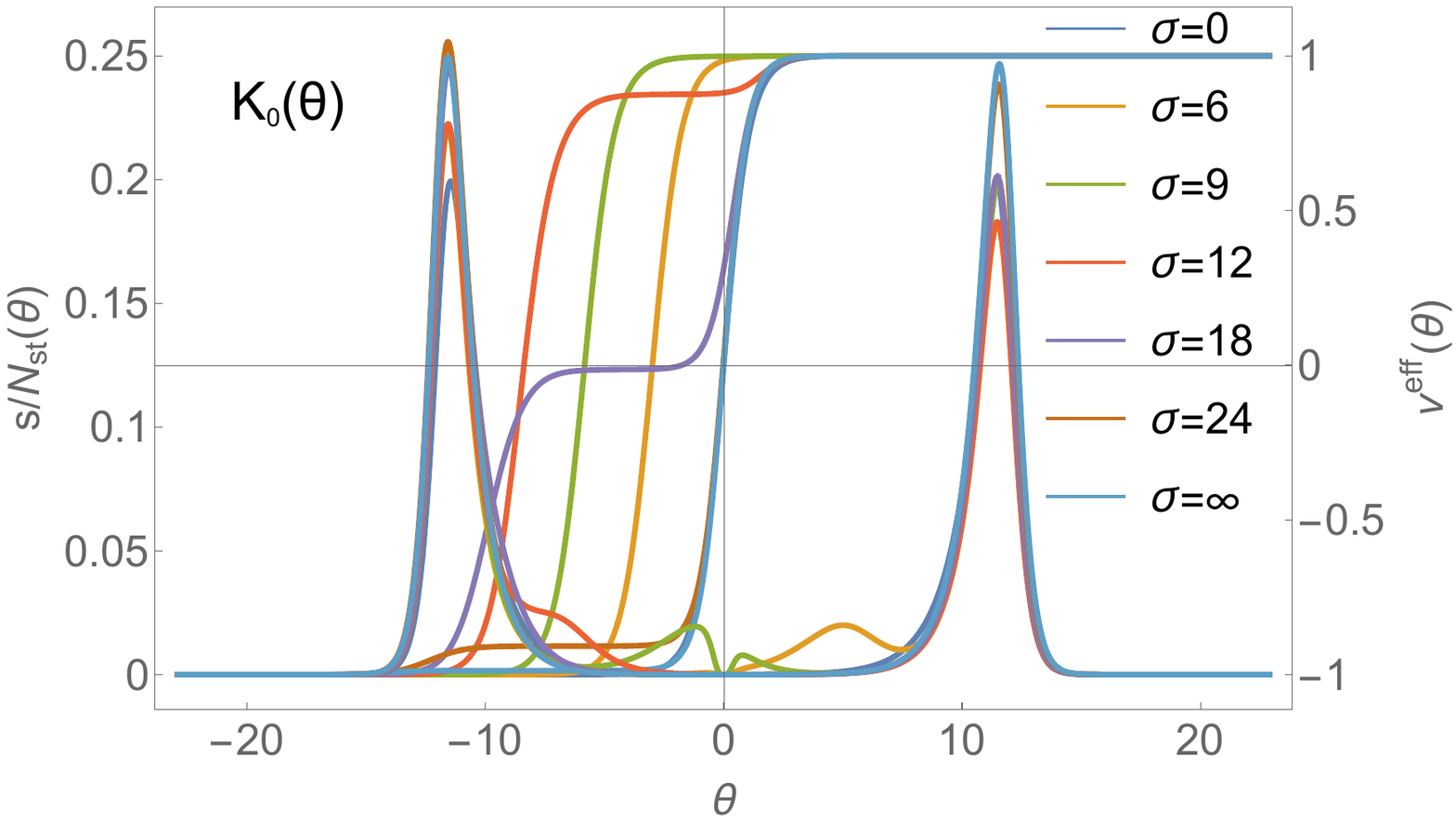}
\caption{\centering
$s(\theta)/N_\text{st}$ and $v^\text{eff}(\theta)$ for $K_\text{0}(\theta)$ and various $\sigma$-s}
\end{subfigure}
\caption{
The normalised spectral particle density divided by the total number of stable particles $\rho(\theta)/N_\text{st}$ (a) and the normalised spectral entropy density $s(\theta)/N_\text{st}$ with the effective velocity $v^\text{eff}(\theta)$ (b) for fixed energy density and various $\sigma$ parameters. We take $\beta_{\text{FF}}=2.8944\times10^{-5}$, that is $2\log2/\beta_{\rm FF}=22.28$. The spectral densities now have a zero at $\theta=0$ and the effective velocities develop a vanishing plateau around $\sigma=2\log2/\beta_{\rm FF}$. In inset (b) the left axis corresponds to $s(\theta)/N_\text{st}\in[0,0.25]$, and the right axis to $v^\text{eff}(\theta)\in[-1,1]$.}
\label{SpectralDensitiesCoshFixedEVaryingSigmaFullRange}
\end{figure}

In this Appendix we investigate the quench with the $K$-function $K_{\rm 0}(\theta)$. 
Although it shares many similarities with the other quenches, some  additional and pronounced features are present which are a direct consequence of the zero of the function $K_{\rm 0}(\theta)$ at $\theta=0$. 
Indeed, such zero of $K_{\rm 0}(\theta)$ imposes a constraint on the solution of the QA equations. More specifically, this zero at $\theta=0$ is inherited by both the spectral density of the particle number $\rho(\theta)$ and the Yang-Yang entropy density $S(\theta)/L$. 
Following the logic of Subsection \ref{SpectralDensities}, let us present first the behaviour of $\rho(\theta)/N_\text{st}$ accompanied with $v^\text{eff}(\theta)$ when the energy density is fixed ($\beta_{\text{FF}}=2.8944\times10^{-5}$) and $\sigma$ varies. The $\sigma$-dependence of these quantities is visualised in Fig. \ref{SpectralDensitiesCoshFixedEVaryingSigmaFullRange}. In fact, the main observation is that the subsidiary peak, when present, is forced to be zero at $\theta=0$, which results in a significant suppression of the spectral weight of this peak and in a notable fingerprint in the thermodynamic quantities, as we shall see. Otherwise the overall behaviour of $\rho(\theta)$ and $v^\text{eff}(\theta)$ are similar to those seen earlier.

Consequently, the observations we made for quenches with $K_{\rm B}(\theta)$ and $K_{\rm S}(\theta)$ still apply to the case of $K_{\rm 0}(\theta)$. 
That is, when the unstable particles are not present in the system after the quench, extended plateaux develop in $N_\text{st}/L$, $S/L$, $S/N_\text{st}$ as well as in d$S$/d$t$ and $(N_{\text{st}}/L)^{-1}\times\text{d}S/\text{d}t$ if the energy density is fixed. This plateau is displayed by the quantity $S/S_\text{max}$ also when instead of the injected energy density, the quench parameter $\alpha$ is fixed and $\sigma$ varies and the plateau is present in $S/N_\text{st}$ and $(N_{\text{st}}/L)^{-1}\times\text{d}S/\text{d}t$ as well if the resonance parameter is fixed and the unstable particles are still not present. In addition for a fixed resonance parameter, the other quantities $N_\text{st}/L$, $S/L$ and d$S$/d$t$ exhibit monotonic linear behaviour which can be explained as in the previous section, namely, it is due to the property
\beq 
\log C_{\rm 0}(\alpha) \approx -2\log\alpha \quad \mathrm{for}\quad \alpha\ll 1\,,
\eeq 
and the fact that, for fixed $\sigma$, all thermodynamic functions are function of this scale. See also the derivation in Appendix \ref{AppA}.
Additionally, the local minimum in d$S$/d$t$ and $(N_{\text{st}}/L)^{-1}\times\text{d}S/\text{d}t$ for fixed energy density and in $(N_{\text{st}}/L)^{-1}\times\text{d}S/\text{d}t$ also for fixed resonance parameter is also present in this case, for regimes when the unstable particles start to form. The mechanism is exactly the same as we discussed in the previous subsection \ref{SpectralDensities}. 

\begin{figure}[H]
\begin{subfigure}{0.495\textwidth}
\centering
\includegraphics[width=\textwidth]{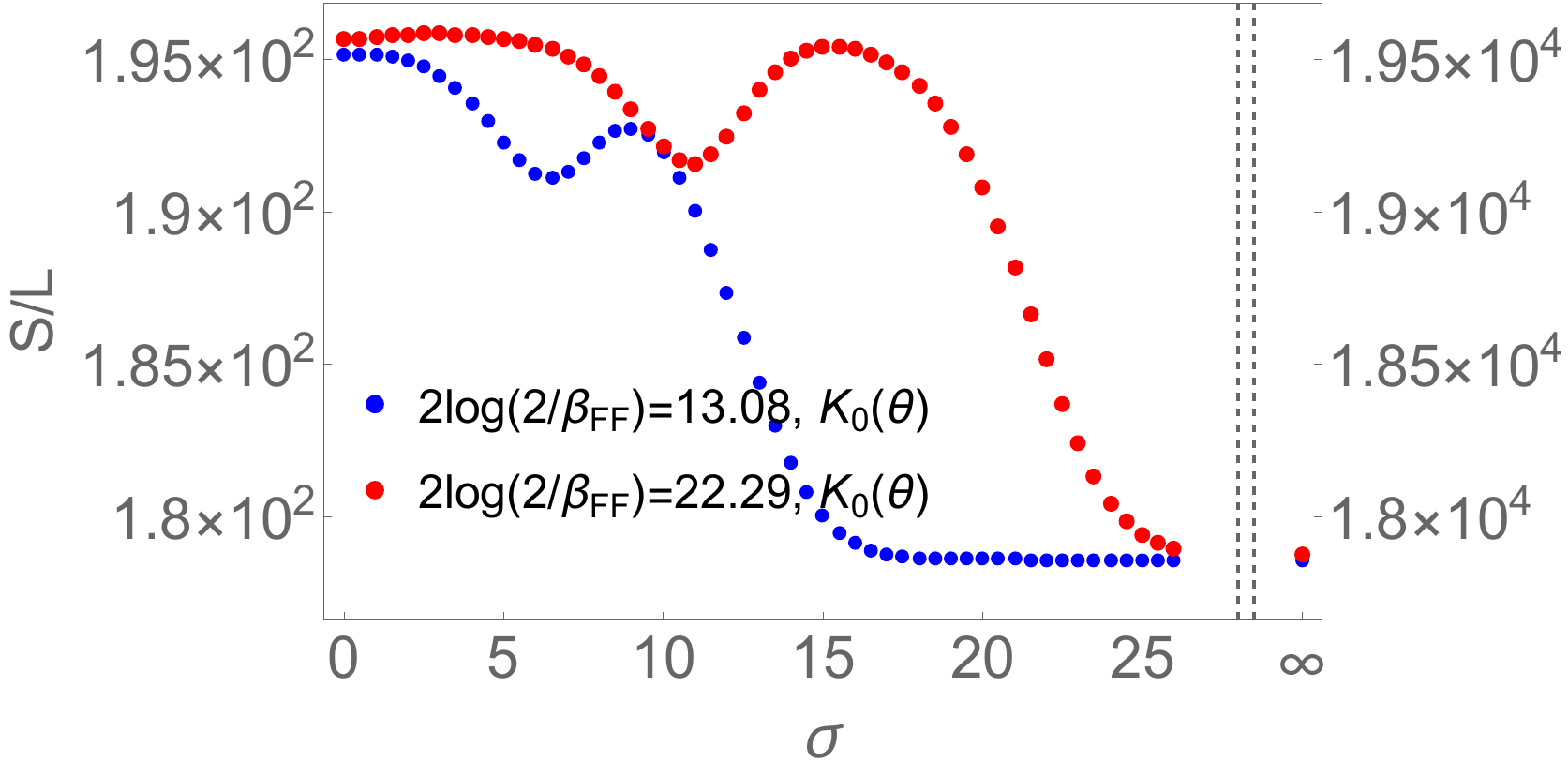}
\caption{\centering
$S/L$ for two energy densities}
\end{subfigure}\hfill
\begin{subfigure}{0.495\textwidth}
\centering
\includegraphics[width=\textwidth]{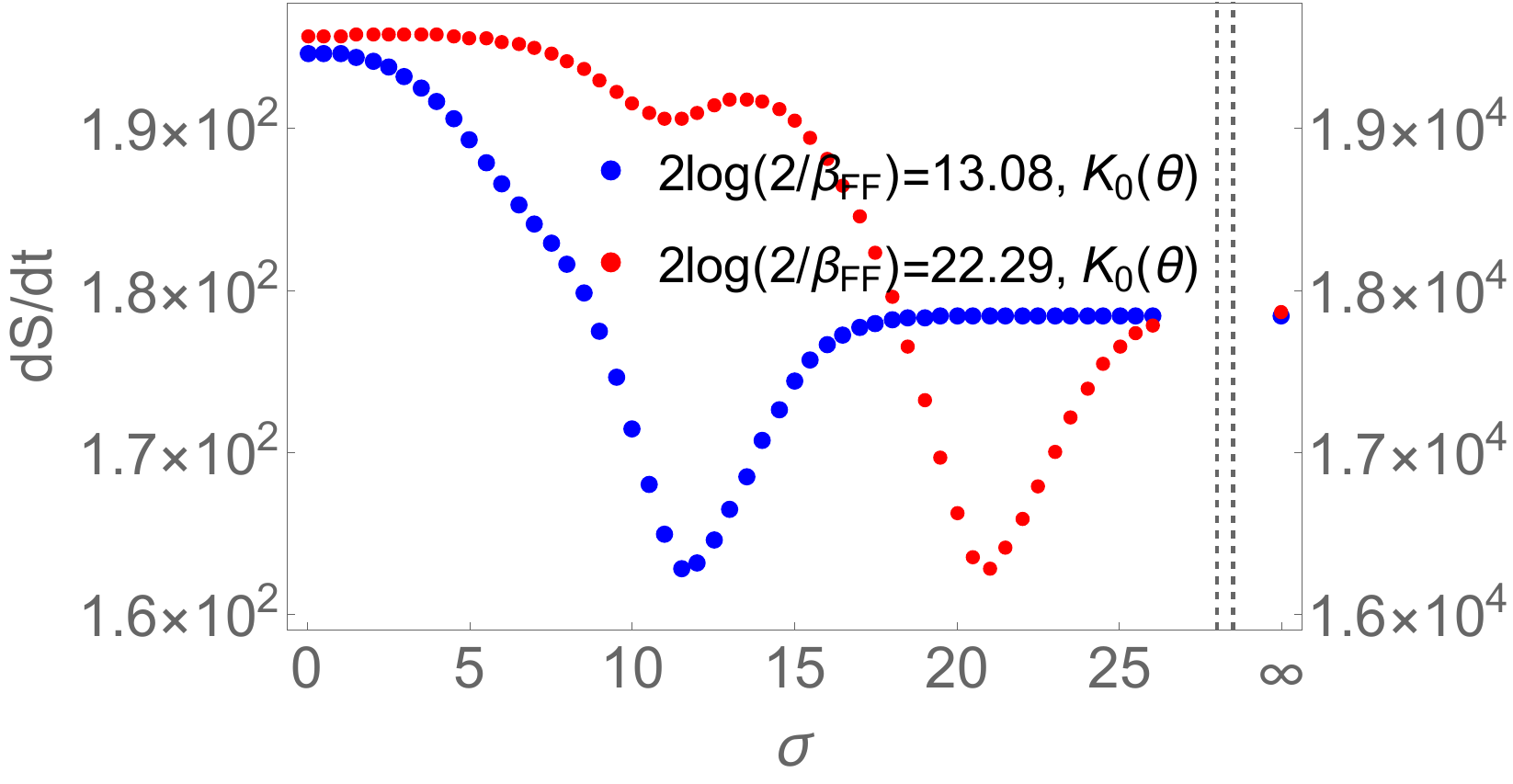}
\caption{\centering
$\frac{\text{d}S}{\text{d}t}$ for two energy densities}
\end{subfigure}
\caption{\label{SPerLAndProdRateK0FixedE}
The total entropy density $S/L$ (a) and its production rate $\text{d}S/\text{d}t$ (b) against $\sigma$ after quenches characterised by $K_0(\theta)$ and fixed injected energy density. The blue dots correspond to $2\log2/\beta_{\rm  FF}=13.08$ or $\beta_\text{FF}=2.8944\times10^{-3}$ (left axis) and the red dots to $2\log2/\beta_{\rm FF}=22.29$ or $\beta_\text{FF}=2.8944\times10^{-5}$ (right axis).}
\end{figure}

\begin{figure}[H]
\begin{subfigure}{0.48\textwidth}
\centering
\includegraphics[width=\textwidth]{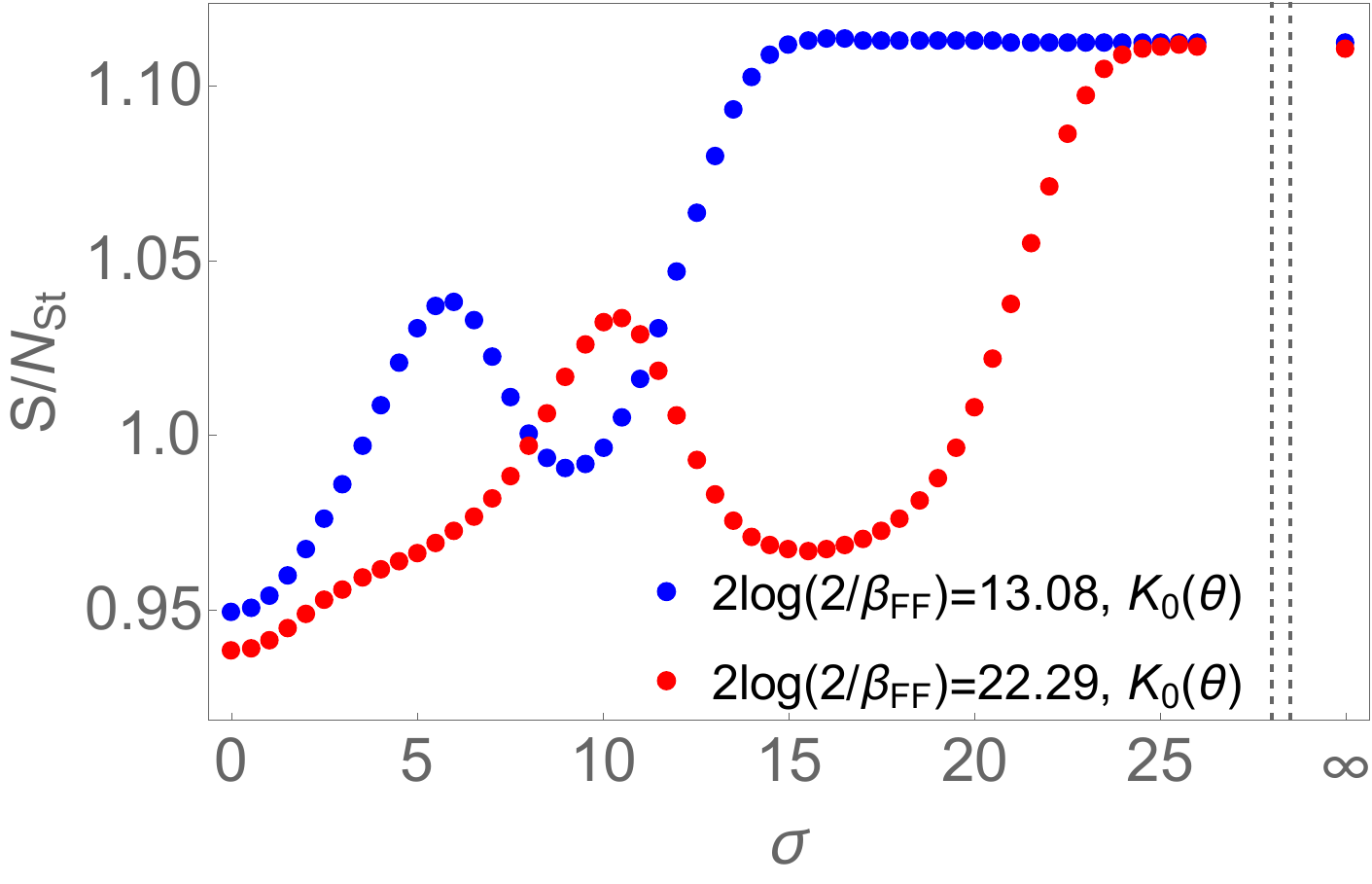}
\caption{\centering
$S/N_\text{st}$ for two energy densities}
\end{subfigure} 
\begin{subfigure}{0.48\textwidth}
\centering
\includegraphics[width=\textwidth]{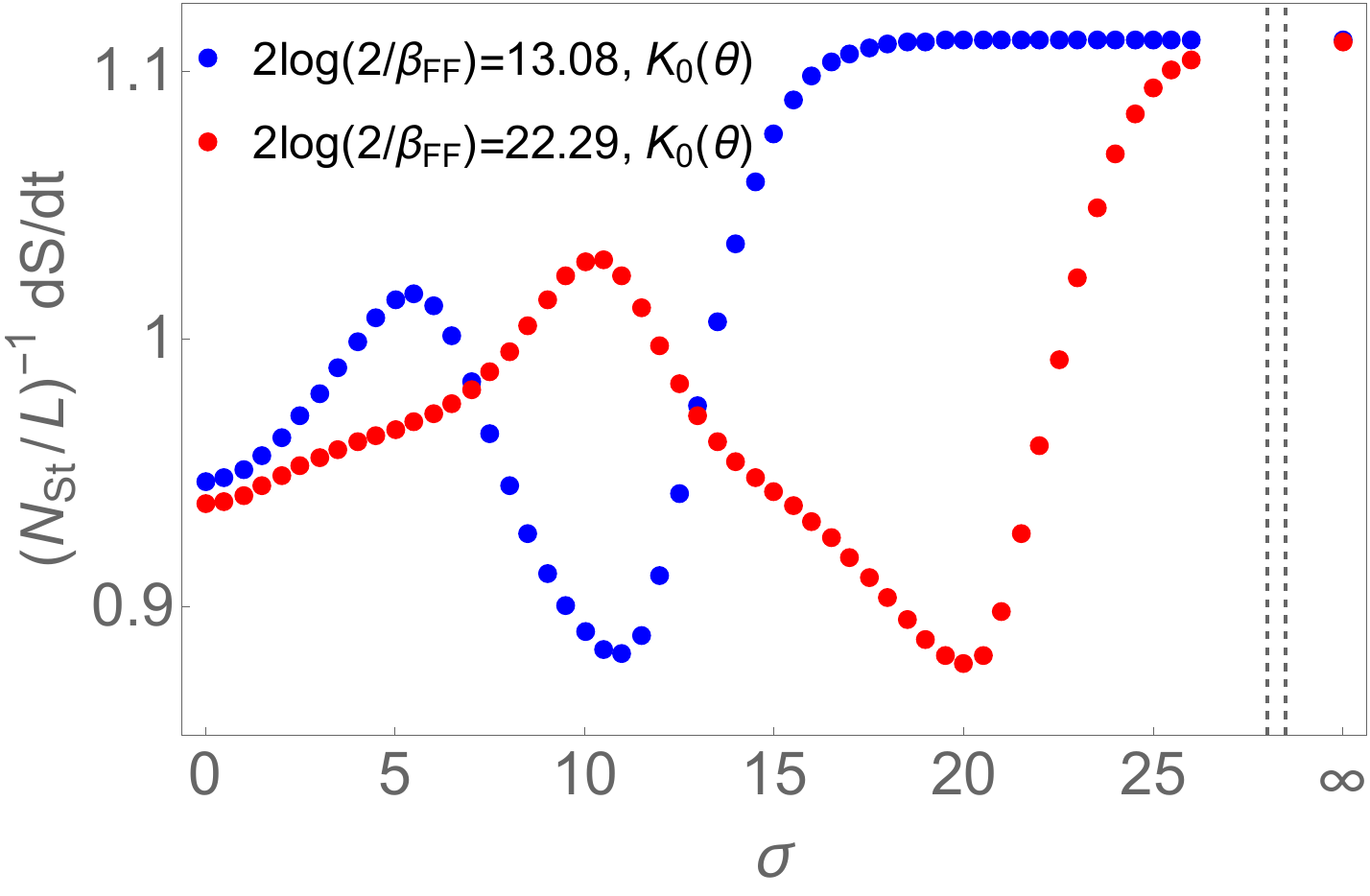}
\caption{\centering
$(N_\text{st}/L)^{-1}\frac{\text{d}S}{\text{d}t}$ for two energy densities}
\end{subfigure}
\caption{\label{SPerNAndProdRatesK0FixedE}
The total thermodynamic entropy density $S/N_{\text{st}}$ (a) with respect to the total particle number and the entropy production rate divided by the particle number density $(N_{\text{st}}/L)^{-1}\times\text{d}S/\text{d}t$ (b) against $\sigma$ after quenches characterised by $K_{\rm 0}(\theta)$ and fixed injected energy density. 
The blue dots correspond to $2\log2/\beta_{\rm  FF}=13.08$ or $\beta_\text{FF}=2.8944\times10^{-3}$ (left axis) and the red dots to $2\log2/\beta_{\rm FF}=22.29$ or $\beta_\text{FF}=2.8944\times10^{-5}$ (right axis).}
\end{figure}

\begin{figure}[H]
\begin{subfigure}{0.48\textwidth}
\centering
\includegraphics[width=\textwidth]{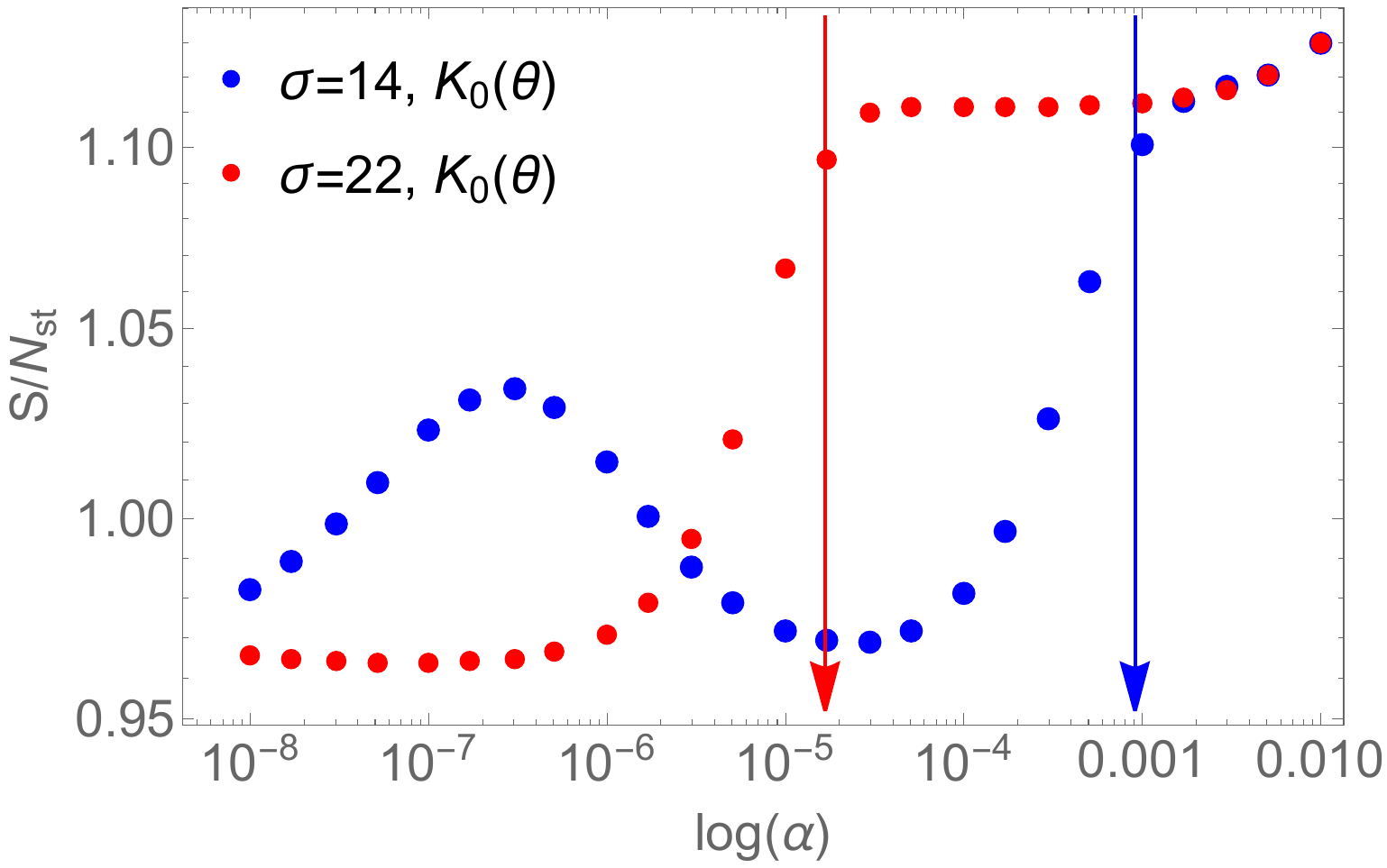}
\caption{\centering
$S/N_{\text{st}}$ for $\sigma=14$ and $\sigma=22$}
\end{subfigure}\hfill
\begin{subfigure}{0.48\textwidth}
\centering
\includegraphics[width=\textwidth]{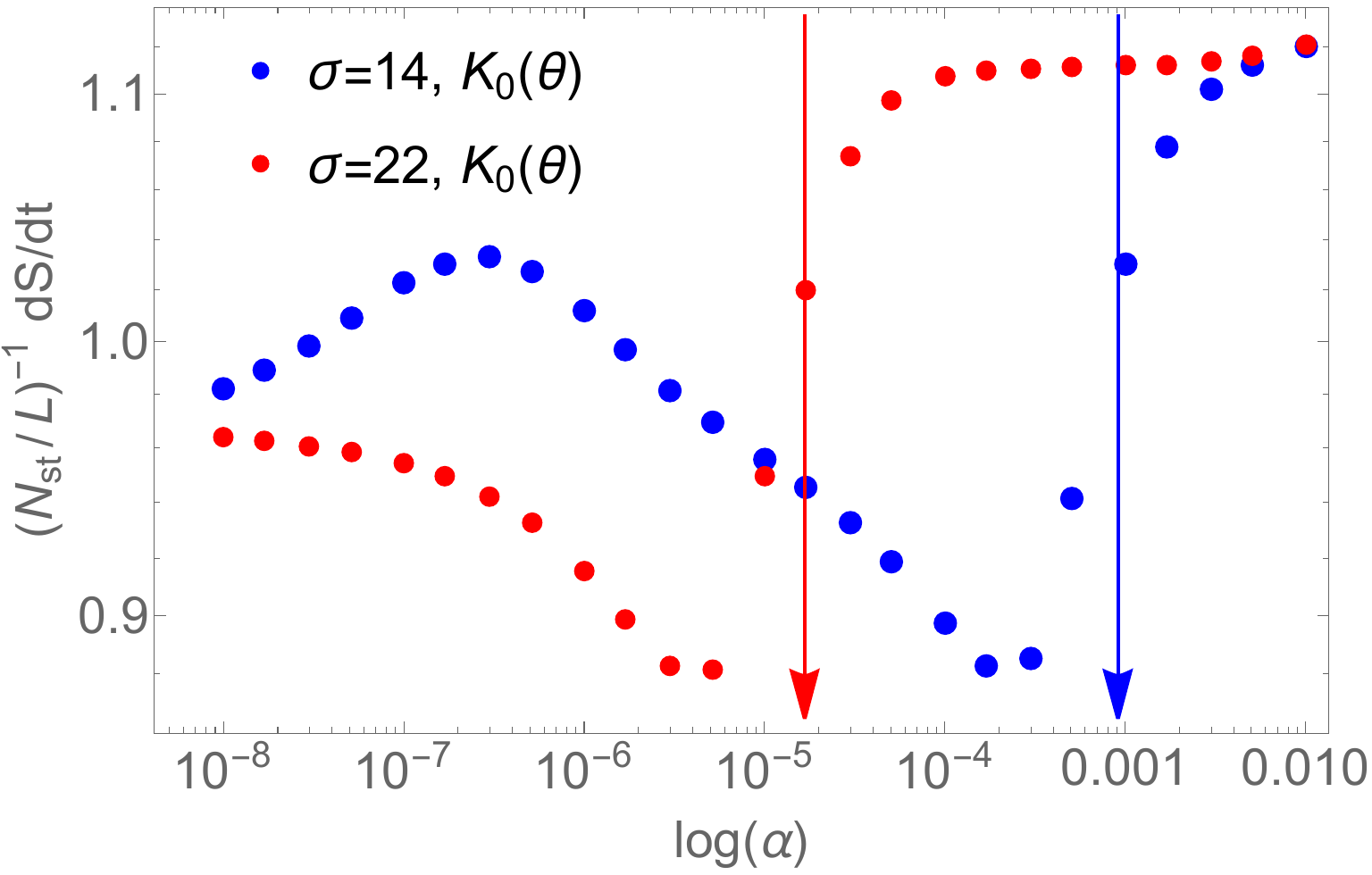}
\caption{\centering
$(N_{\text{st}}/L)^{-1}\frac{\text{d}S}{\text{d}t}$ for $\sigma=14$ and $\sigma=22$}
\end{subfigure}
\caption{
The total thermodynamic entropy density $S/N_{\text{st}}$ (a) with respect to the total particle number density and entropy production rates divided by the particle number density $(N_{\text{st}}/L)^{-1}\times\text{d}S/\text{d}t$ (b) against $\alpha$ after quenches characterised by $K_{\rm 0}(\theta)$ and fixed resonance parameters $\sigma$. The blue dots correspond to $\sigma=14$, and the red ones for $\sigma=22$. 
{\DXH The arrows show the particular $\alpha$ values at which the associated scale variable $\kappa(\sigma,\alpha)$ equals zero.}}
\label{SPerNdSdtFixedSigmaK0}
\end{figure}

There is however one main difference between the  $K_{\rm 0}(\theta)$ and other cases which we now discuss.  This occurs when the quench is such that the unstable particles are present in the post-quench system. The plateau for $\sigma$ small which we saw in previous cases with $K_{\rm B}(\theta)$ and $K_{\rm S}(\theta)$ (in the corresponding quench regime) is now modified by a pronounced local maximum or minimum depending on the specific quantity. In particular, $N_\text{st}/L$ and $S/L$ as well as d$S$/d$t$ develop a minimum for fixed energy densities. This is demonstrated by Fig. \ref{SPerLAndProdRateK0FixedE} via $S/L$ as well as d$S$/d$t$, and the minimum is naturally attributed to the zero and suppression in $\rho(\theta)$ and $s(\theta)/L$. The quantities divided by $N_\text{st}$ such as $S/N_\text{st}$ and $(N_{\text{st}}/L)^{-1}\times\text{d}S/\text{d}t$ instead show a maximum as one can see in Fig. \ref{SPerNAndProdRatesK0FixedE} together with Fig. \ref{SPerNdSdtFixedSigmaK0}. 
The two behaviours (i.e maximum/minimum) are of course correlated and result from the fact that if we consider quantities divided by $N_\text{st}$ the suppression, i.e., the minimum is slightly stronger in $N_\text{st}/L$ than in $S/L$.

\begin{thebibliography}{10}

\bibitem{quench1} P. Calabrese and J. Cardy, Time-dependence of correlation functions following a quantum quench, Phys. Rev. Lett. {\bf 96}, 136801 (2006).

\bibitem{quench2}  P. Calabrese and J. Cardy, Quantum quenches in extended systems, J. Stat. Mech. P06008 (2007).


\bibitem{pssv-11}
A.~Polkovnikov, K.~Sengupta, A.~Silva, and M.~Vengalattore,
{Colloquium: Nonequilibrium dynamics of closed interacting quantum systems}, 
{Rev. Mod. Phys. {\bf 83}, 863 (2011)}.

\bibitem{Eisert}
J.~Eisert, M.~Friesdorf, and C.~Gogolin,
\newblock {Quantum many-body systems out of equilibrium},
\newblock Nature Phys. {\bf 11}, 124 (2015).

\bibitem{CEM}
P.~Calabrese, H.~Essler, and G.~Mussardo~(ed.),
\newblock {Introduction to 'Quantum Integrability in Out-of-Equilibrium Systems'},
\newblock J. Stat. Mech. { 064001} (2016).

\bibitem{EF16}
F.~H.~L. Essler and M. Fagotti, 
\textit{Quench dynamics and relaxation in isolated integrable quantum spin chains}, {J. Stat. Mech.  064002 (2016)}.

\bibitem{EEDyn} P. Calabrese, Entanglement spreading in non-equilibrium integrable systems,  SciPost Phys. Lect. Notes {\bf 20} (2020).



\bibitem{kinoshita}
T.~Kinoshita, T.~Wenger, and D.~Weiss,
\newblock {A Quantum Newton's Cradle},
\newblock Nature {\bf 440}, 900 (2006).

\bibitem{Rigol}
M.~Rigol, V.~Dunjko, V.~Yurovsky, and M.~Olshanii,
\newblock Relaxation in a Completely Integrable Many-Body Quantum System: An Ab
  Initio Study of the Dynamics of the Highly Excited States of 1D Lattice
  Hard-Core Bosons,
\newblock Phys. Rev. Lett. {\bf 98}, 050405 (2007).



\bibitem{vr-16}
L. Vidmar and M. Rigol, { Generalized Gibbs Ensemble in Integrable Lattice Models},
{J. Stat. Mech.(2016) 064007}.


\bibitem{fcec-13}
M. Fagotti, M. Collura, F. H. L. Essler, and P. Calabrese,
{Relaxation after quantum quenches in the spin-1/2 Heisenberg XXZ chain},
{Phys. Rev. B {\bf 89}, 125101 (2014)} 

\bibitem{ilinardo}
E.~Ilievski, J.~De~Nardis, B.~Wouters, J.-S. Caux, F.~H.~L. Essler, and
  T.~Prosen,
\newblock Complete Generalized Gibbs Ensembles in an Interacting Theory,
\newblock Phys. Rev. Lett. {\bf 115}, 157201 (2015).

\bibitem{failure}
B.~Pozsgay, M.~Mesty\'an, M.~A. Werner, M.~Kormos, G.~Zar\'and, and
  G.~Tak\'acs,
\newblock Correlations after Quantum Quenches in the $XXZ$ Spin Chain: Failure
  of the Generalized Gibbs Ensemble,
\newblock Phys. Rev. Lett. {\bf 113}, 117203 (2014).

\bibitem{Prosen1}
M.~Mierzejewski, P.~Prelov\ifmmode~\check{s}\else \v{s}\fi{}ek, and T.~Prosen,
\newblock Breakdown of the Generalized Gibbs Ensemble for Current-Generating
  Quenches,
\newblock Phys. Rev. Lett. {\bf 113}, 020602 (2014).

\bibitem{Prosen2}
T.~Prosen,
\newblock {Quasilocal conservation laws in XXZ spin-1/2 chains: Open, periodic
  and twisted boundary conditions},
\newblock Nucl. Phys. B {\bf 886}, 1177 (2014).

\bibitem{Prosen3}
M.~Mierzejewski, P.~Prelov\ifmmode~\check{s}\else \v{s}\fi{}ek, and T.~Prosen,
\newblock Identifying Local and Quasilocal Conserved Quantities in Integrable
  Systems,
\newblock Phys. Rev. Lett. {\bf 114}, 140601 (2015).

\bibitem{Prosen4}
E.~Ilievski, M.~Medenjak, and T.~Prosen,
\newblock Quasilocal Conserved Operators in the Isotropic Heisenberg Spin-$1/2$
  Chain,
\newblock Phys. Rev. Lett. {\bf 115}, 120601 (2015).

\bibitem{doyon2017}
B.~Doyon,
\newblock Thermalization and Pseudolocality in Extended Quantum Systems,
\newblock Comm. Math. Phys. {\bf 351}, 155 (2017).





\bibitem{specialrec}  A.~Bastianello, B.~Bertini, B.~Doyon and R.~Vasseur~(ed.),  Emergent Hydrodynamics in Integrable Many-Body Systems, J. Stat. Mech. {014001}, (2022).



\bibitem{ourhydro}
O.~A. Castro-Alvaredo, B.~Doyon, and T.~Yoshimura,
\newblock {Emergent hydrodynamics in integrable quantum systems out of
  equilibrium},
\newblock Phys. Rev. X {\bf 6}, 041065 (2016).

\bibitem{theirhydro}
B.~Bertini, M.~Collura, J.~De~Nardis, and M.~Fagotti,
\newblock {Transport in Out-of-Equilibrium $XXZ$ Chains: Exact Profiles of
  Charges and Currents},
\newblock Phys. Rev. Lett.  {\bf 117}, 207201 (2016).

\bibitem{benreview}
B.~Doyon,
\newblock {Lecture notes on Generalised Hydrodynamics},  	SciPost Phys. Lect. Notes {\bf 18} (2020).

\bibitem{p-13}
B. Pozsgay, 
The dynamical free energy and the Loschmidt echo for a class of quantum quenches in the Heisenberg spin chain, J. Stat. Mech. {P10028} (2013).

\bibitem{IntegrableQuench} L. Piroli, B. Pozsgay and E. Vernier, What is an integrable quench?, {Nucl.
Phys.}  B \textbf{925}, 362 (2017).

\bibitem{ppv-18}
L. Piroli, B. Pozsgay, and E. Vernier, Non-analytic behavior of the Loschmidt echo in XXZ spin chains, Nucl. Phys.  B {\bf 933}, 454 (2018).


\bibitem{pvcp-18}
L. Piroli, E. Vernier, P. Calabrese, and B. Pozsgay, 
Integrable quenches in nested spin chains I: the exact steady states,
{J. Stat. Mech. {063103} (2019)}.

\bibitem{pvcp-18b}
L. Piroli, E. Vernier, P. Calabrese, and B. Pozsgay, 
Integrable quenches in nested spin chains II: the Quantum Transfer Matrix approach,
{J. Stat. Mech. {063104} (2019)}.

\bibitem{Cfab1} J.-S. Caux and F. H. L. Essler, Time evolution of local observables after quenching to an integrable model,  Phys. Rev. Lett. {\bf 110}, 257203 (2013).

\bibitem{QuenchActionReview} J-S. Caux, The Quench Action, J. Stat. Mech. {064006} (2016).

\bibitem{Ams1} B. Wouters, J. De Nardis, M. Brockmann, D. Fioretto, M. Rigol, and J.-S. Caux, Quenching the Anisotropic Heisenberg Chain: Exact Solution and Generalized Gibbs Ensemble Predictions, Phys. Rev. Lett. {\bf 113}, 117202 (2014).

\bibitem{Ams2} M. Brockmann, B. Wouters, D. Fioretto, J. De Nardis, R. Vlijm, and J.-S. Caux, 
Quench action approach for releasing the N\'eel state into the spin-1/2 XXZ chain, 
J. Stat. Mech. {P12009} (2014).

\bibitem{pvc-16}
L. Piroli, E. Vernier, and P. Calabrese,
{Exact steady states for quantum quenches in integrable Heisenberg spin chains}, 
{Phys. Rev. B {\bf 94}, 054313 (2016)}

\bibitem{pvcr-16}
L. Piroli, E. Vernier, P. Calabrese, and M. Rigol, { Correlations and diagonal entropy after quantum quenches in XXZ chains},
{Phys. Rev. B {\bf 95}, 054308 (2017)}


\bibitem{Hun2}  M. Mesty\'an, B. Pozsgay, G. Tak\'acs, and M. A. Werner, Quenching the XXZ spin chain: quench action approach versus generalized Gibbs ensemble, J. Stat. Mech. { P04001} (2015).



\bibitem{AC2} V. Alba and P. Calabrese, The quench action approach in finite integrable spin chains,  
J. Stat. Mech. {P043105} (2016).

\bibitem{mbpc-17}
M. Mesty\'an, B. Bertini, L. Piroli, and P. Calabrese, { Exact solution for the quench dynamics of a nested integrable system},
{J. Stat. Mech. (2017) 083103}.


\bibitem{NWB} J. De Nardis, B. Wouters, M. Brockmann, and J.-S. Caux, Solution for an interaction quench in the Lieb-Liniger Bose gas, Phys. Rev. A {\bf 89}, 033601 (2014).

\bibitem{PCE} L. Piroli, P. Calabrese, and F. H. L. Essler, Multiparticle Bound-State Formation following a Quantum Quench to the One-Dimensional Bose Gas with Attractive Interactions, Phys. Rev. Lett. {\bf 116}, 070408 (2016).

\bibitem{Bucci} L. Bucciantini, Stationary state after a quench to the Lieb-Liniger from rotating BECs,  J. Stat. Phys. {\bf 164}, 621 (2016).

\bibitem{pce-16}
L. Piroli, P. Calabrese, and F. H. L. Essler, {Quantum quenches to the attractive one-dimensional Bose gas: exact results},
{SciPost Phys. {\bf 1}, 001 (2016)}



\bibitem{NA1} J. De Nardis and J.-S. Caux, Analytical expression for a post-quench time evolution of the one-body density matrix of one-dimensional hard-core bosons, J. Stat. Mech. {P12012} (2014).

\bibitem{NA2} J. De Nardis, L. Piroli, and J.-S. Caux, Relaxation dynamics of local observables in integrable systems, J. Phys. A {\bf 48}, 43FT01 (2015).

\bibitem{NA3} R. Van Den Berg, B. Wouters, S. Eli\"ens, J. De Nardis, R. M. Konik, and J.-S. Caux, Separation of Timescales in a Quantum Newton's Cradle, Phys. Rev. Lett. {\bf 116}, 225302 (2016).

\bibitem{pc-17}
L. Piroli and P. Calabrese, { Exact dynamics following an interaction quench in a one-dimensional anyonic gas},
{Phys. Rev. A {\bf 96}, 023611 (2017)}.

\bibitem{bpc-18}
A. Bastianello, L. Piroli, and P. Calabrese, 
{Exact local correlations and full counting statistics for arbitrary states of the one-dimensional interacting Bose gas},
{Phys. Rev. Lett. {\bf 120}, 190601 (2018)}.



\bibitem{rbc-22}
C. Rylands, B. Bertini, and P. Calabrese, {Integrable quenches in the Hubbard model},
arXiv:2206.07985.

\bibitem{rcb-22b}
C. Rylands, B. Bertini, and P. Calabrese, { Exact Solution of the BEC-to-BCS Quench in One Dimension},
arXiv:2209.00956.

\bibitem{LMV} A. De Luca, G. Martelloni, and J. Viti, Stationary states in a free fermionic chain from the quench action method, Phys. Rev. A {\bf 91}, 021603(R).

\bibitem{Cfab2} B. Bertini, D. Schuricht, and F. H. L. Essler, Quantum quench in the sine-Gordon model, J. Stat. Mech. {P10035} (2014).


\bibitem{DS} E. Di Salvo and D. Schuricht, 
    Quantum quenches in the sinh-Gordon and Lieb-Liniger models,
2210.00316 (2022).


\bibitem{BPC} B. Bertini, L. Piroli, and P. Calabrese, Quantum quenches in the sinh-Gordon model: steady state and one point correlation
functions, J. Stat. Mech. { 063102}, (2016).

\bibitem{takahashi} M. Takahashi, {Thermodynamics of one-dimensional solvable models}, Cambridge University Press (1999).  

\bibitem{tba1}
A.~Zamolodchikov,
\newblock {Thermodynamic Bethe ansatz in relativistic models. Scaling three
  state Potts and Lee-Yang models},
\newblock Nucl. Phys. B {\bf 342}, 695 (1990).

\bibitem{tba2}
T.~R. Klassen and E.~Melzer,
\newblock {The Thermodynamics of purely elastic scattering theories and conformal perturbation theory},
\newblock Nucl. Phys. B {\bf 350}, 635 (1991).

\bibitem{Mossel}
J.~Mossel and J.-S. Caux,
\newblock {Generalized TBA and generalized Gibbs},
\newblock J. Phys. A {\bf 45}, 255001 (2012).

\bibitem{FM} D. Fioretto and G. Mussardo, Quantum Quenches in Integrable Field Theories, New J. Phys. {\bf 12}, 
055015 (2010).

\bibitem{SFM} S. Sotiriadis, D. Fioretto, and G. Mussardo, Zamolodchikov-Faddeev Algebra and Quantum Quenches in Integrable Field Theories, J. Stat. Mech. P02017 (2012). 

\bibitem{HST}
D. X. Horváth, S. Sotiriadis, and G. Takács, Initial states in integrable quantum field theory quenches from an integral equation hierarchy, 	Nucl. Phys.  B {\bf 902}, 508 (2016).

\bibitem{ES} D. Schuricht and F.H.L. Essler, Dynamics in the Ising field theory after a quantum quench,  J. Stat. Mech. {P04017} (2012).

\bibitem{TIsing}  O.A. Castro-Alvaredo, M. Lencs\'es, I.M.~ Sz\'ecs\'enyi and  J. Viti,  Entanglement Dynamics after a Quench in Ising Field Theory: A Branch Point Twist Field Approach, JHEP {\bf 2019}, 79 (2019). 


\bibitem{hsg}
C.~R. Fernandez-Pousa, M.~V. Gallas, T.~J. Hollowood, and J.~L. Miramontes,
\newblock {Solitonic integrable perturbations of parafermionic theories},
\newblock Nucl. Phys. B {\bf 499}, 673 (1997).

\bibitem{ntft}
C.~Fernandez-Pousa, M.~Gallas, T.~Hollowood, and J.~Miramontes,
\newblock {The symmetric space and homogeneous sine-Gordon theories},
\newblock Nucl. Phys. B {\bf 484}, 609 (1997).

\bibitem{FernandezPousa:1997iu}
C.~R. Fernandez-Pousa and J.~L. Miramontes,
\newblock {Semi-classical spectrum of the homogeneous sine-Gordon theories},
\newblock Nucl. Phys. B {\bf 518}, 745 (1998).

\bibitem{smatrix}
J.~L. Miramontes and C.~R. Fernandez-Pousa,
\newblock {Integrable quantum field theories with unstable particles},
\newblock Phys. Lett. B {\bf 472}, 392 (2000).

\bibitem{ourU} O.A. Castro-Alvaredo, C. De Fazio, B. Doyon and F. Ravanini, On the Hydrodynamics of Unstable Excitations, JHEP {\bf 2020}, 45 (2020).

\bibitem{nextU} O.A. Castro-Alvaredo, C. De Fazio, B. Doyon and A. Zi\'o\l{}kowska, Tails of Instability and Decay: a Hydrodynamic Perspective, SciPost Phys. {\bf 12}, 115 (2022).

\bibitem{nnextU} O.A. Castro-Alvaredo, C. De Fazio, B. Doyon and A. Zi\'o\l{}kowska, Generalised Hydrodynamics of Particle Creation and Decay,  JHEP {\bf 2022}, 35 (2022).

\bibitem{Gibbsparadox} M. Collura, M. Kormos and G. Tak\'acs, Dynamical manifestation of Gibbs paradox after a quantum quench, Phys. Rev. A {\bf 98}, 053610 (2018).

\bibitem{LPT} M. Lencs\'es, O. Pomponio and G. Tak\'acs, Relaxation and entropy generation after quenching quantum spin chains, SciPost Phys. {\bf 9}, 011 (2020). 

\bibitem{PPT} O.~Pomponio, L.~Pristy\'ak and G.~Tak\'acs, Quasi-particle spectrum and entanglement generation after a quench in the quantum Potts spin chain, J. Stat. Mech. {013104} (2019).

\bibitem{ourtba}
O.~A. Castro-Alvaredo, A.~Fring, C.~Korff, and J.~L. Miramontes,
\newblock {Thermodynamic Bethe ansatz of the homogeneous sine-Gordon models},
\newblock Nucl. Phys. B {\bf 575}, 535 (2000).

\bibitem{CastroAlvaredo:2002nv}
O.~A. Castro-Alvaredo, J.~Dreissig, and A.~Fring,
\newblock {Integrable scattering theories with unstable particles},
\newblock Eur. Phys. J. C {\bf 35}, 393 (2004).

\bibitem{Dorey:2004qc}
P.~Dorey and J.~Miramontes,
\newblock {Mass scales and crossover phenomena in the homogeneous sine-Gordon
  models},
\newblock Nucl. Phys. B {\bf 697}, 405 (2004).


\bibitem{KW}
M.~Karowski and P.~Weisz, Exact S matrices and form-factors in (1+1)-dimensional field
  theoretic models with soliton behavior, Nucl. Phys. B {\bf 139}, 455 (1978).

\bibitem{SmirnovBook}
F.~Smirnov, Form factors in completely integrable models of quantum field theory, Adv. Series in Math. Phys. {\bf 14}, World Scientific, Singapore
  (1992).


\bibitem{CastroAlvaredo:2000em}
O.~A. Castro-Alvaredo, A.~Fring, and C.~Korff,
\newblock {Form factors of the homogeneous sine-Gordon models},
\newblock Phys. Lett. B {\bf 484}, 167 (2000).

\bibitem{CastroAlvaredo:2000nk}
O.~A. Castro-Alvaredo and A.~Fring,
\newblock {Identifying the operator content, the homogeneous sine- Gordon
  models},
\newblock Nucl. Phys. B {\bf 604}, 367 (2001).

\bibitem{CastroAlvaredo:2000ag}
O.~A. Castro-Alvaredo and A.~Fring,
\newblock {Renormalization group flow with unstable particles},
\newblock Phys. Rev. D {\bf 63}, 021701 (2001).

\bibitem{CastroAlvaredo:2000nr}
O.~A. Castro-Alvaredo and A.~Fring,
\newblock {Decoupling the $SU(N)_2$-homogeneous sine-Gordon model},
\newblock Phys. Rev. D {\bf 64}, 085007 (2001).

\bibitem{MC1} Z. Bajnok, J. Balog, K. Ito, Y. Satoh and G.Z. T\'oth, On the mass-coupling relation of
multi-scale quantum integrable models, JHEP {\bf 06} (2016) 071.

\bibitem{MC2} Z. Bajnok, J. Balog, K. Ito, Y. Satoh and G.Z. T\'oth, Exact mass-coupling relation for the
homogeneous sine-Gordon model, Phys. Rev. Lett. {\bf 116} (2016) 181601.

\bibitem{WZNW1}  J.~Wess and B.~Zumino, Consequences of anomalous ward identities, Phys. Lett. B {\bf 37}, 95 (1971).

\bibitem{WZNW2} E.~Witten, Global aspects of current algebra, Nucl. Phys. B {\bf 223}, 422 (1983).

\bibitem{WZNW3} E.~Witten, Non-abelian bosonization in two dimensions, Comm. Math. Phys. {\bf 92}, 455 (1984).

\bibitem{WZNW4} S.P.~Novikov, Multivalued functions and functionals. An analogue of the Morse theory, Sov. Math. Dokl. {\bf 24}, 222 (1981).

\bibitem{WZNW5} S.P.~Novikov, The Hamiltonian formalism and a many-valued analogue of Morse theory, Russ. Math. Sur. {\bf 37}, 1 (1982).

\bibitem{ZA} A.B. Zamolodchikov and Al.B. Zamolodchikov, {Factorized $S$-matrices in two-dimensions as the exact solutions of certain relativistic quantum field models}, {Ann. Phys. {\bf 120},  253 (1979)}.

\bibitem{FA} L. D. Faddeev, {Quantum completely integrable models in field theory},{Cont. Math. Phys., {\bf 1C}, 107 (1980)}. 


\bibitem{Lechner1}{G. Lechner, Construction of Quantum Field Theories with Factorizing S-Matrices, Comm.
Math. Phys. \textbf{277},  821 (2008).}


\bibitem{Lechner2} G. Lechner, Algebraic constructive quantum
field theory: Integrable models and deformation techniques, arXiv:1503.03822 (2015).

\bibitem{Gutkin} E. Gutkin, Quantum nonlinear Schr\"odinger equation: two solutions, Phys. Rep. \textbf{167}, 1 (1988).



\bibitem{GZ} S. Ghoshal and A. Zamolodchikov, Boundary S-Matrix and Boundary State in Two-Dimensional Integrable Quantum Field Theory, Int. J. Mod. Phys. A {\bf 9}, 3841 (1994); Erratum-ibid. A {\bf 9} 4353 (1994).

{\DXH
\bibitem{SingularOverlaps}
 D.~X. Horváth, M. Kormos, and G. Takács,
 Overlap singularity and time evolution in integrable quantum field theory,
 JHEP {\bf 08} 170 (2018).
 }

\bibitem{ExcitedBetheStates} B. Pozsgay, Mean values of local operators in highly excited Bethe states, J. Stat. Mech.
{P01011} (2011).

\bibitem{IlievskiCauxQPPicture} E. Ilievski, E. Quinn and J.-S. Caux, From interacting particles to equilibrium statistical ensembles,
{Phys. Rev.} B \textbf{95},  115128 (2017).

\bibitem{QuasiAgain} V.~Alba and P.~Calabrese, Entanglement and thermodynamics after a quantum quench in integrable systems, PNAS {\bf 114}, 7947 (2017).

\bibitem{ac-17c}
V. Alba and P. Calabrese, { Entanglement dynamics after quantum quenches in generic integrable systems},
{SciPost Phys. {\bf 4}, 017 (2018)}.

\bibitem{ac-17a}
V. Alba and P. Calabrese, {Quench action and R\'enyi entropies in integrable systems}, {Phys. Rev. B {\bf 96}, 115421 (2017)}.

\bibitem{ac-17b}
V. Alba and P. Calabrese, { R\'enyi entropies after releasing the N\'eel state in the XXZ spin-chain}, 
{J. Stat. Mech. {113105} (2017)} 

\bibitem{mac-18}
M. Mesty\'an, V. Alba, and P. Calabrese,  {R\'enyi entropies of generic thermodynamic macrostates in integrable systems},
{J. Stat. Mech. {083104} (2018)}.

\bibitem{pvcc-22}
L. Piroli, E. Vernier, M. Collura, and P. Calabrese, { Thermodynamic symmetry resolved entanglement entropies in integrable systems},
J. Stat. Mech. (2022) 073102.

\bibitem{bka-22}
B. Bertini, K. Klobas, V. Alba, G, Lagnese, and P. Calabrese, { Growth of R\'enyi Entropies in Interacting Integrable Models and the Breakdown of the Quasiparticle Picture},
Phys. Rev. X {\bf 12}, 031016 (2022).


\bibitem{QuasiParticlePicture}
P.~Calabrese, and J. ~Cardy,
\newblock{Evolution of entanglement entropy in one-dimensional systems}, 
\newblock J. Stat. Mech. P04010 (2005).


\bibitem{BetheWaveFunction} {H. Bethe, Zur Theorie der Metalle. I. Eigenwerte und Eigenfunktionen der linearen Atomkette, {Zeit. Phys.} A \textbf{71}, 205 (1931).}

\bibitem{Luscher} M.~L\"uscher, Volume dependence of the energy spectrum in massive quantum field theories
I. Stable particle states, Commun. Math. Phys.\textbf{104}, 177 (1986).

\bibitem{KlassenM} T.R.~Klassen and E.~Melzer, On the relation between scattering amplitudes and finite size mass corrections in QFT, Nucl. Phys. B \textbf{362}, 329 (1991).

\bibitem{BajnokJanik} Z.~Bajnok and R.A.~Janik, Four-loop perturbative Konishi from strings and finite size effects for multiparticle states, Nucl. Phys. B
\textbf{807}, 625 (2009).

\bibitem{Hatsuda} Y.~Hatsuda and R.~Suzuki, Finite-Size Effects for Multi-Magnon States, {JHEP} \textbf{0809}
 025 (2008).

\bibitem{OnePointFunctions} M. Kormos and B. Pozsgay, One-point functions in massive integrable QFT with boundaries, JHEP
\textbf{2010} 112 (2010).

{\OCA \bibitem{108} B. Pozsgay and G. Takacs, Form factor expansion for thermal correlators, J. Stat. Mech. (2010) P11012.}

\bibitem{mck-22}
S. Murciano, P. Calabrese, and R. M. Konik, {Post-Quantum Quench Growth of Renyi Entropies in Low Dimensional Continuum Bosonic Systems},
{Phys. Rev. Lett. {\bf 129}, 106802 (2022)}.

\bibitem{ik-22}
P. Emonts and I. Kukuljan,
Reduced density matrix and entanglement in interacting quantum field theory with Hamiltonian truncation,
Phys. Rev. Res. {\bf 4}, 033039 (2022).

\bibitem{bfpc-18}
B. Bertini, M. Fagotti, L. Piroli, and P. Calabrese, { Entanglement evolution and generalised hydrodynamics: noninteracting systems},
{J. Phys. A {\bf 51}, 39LT01 (2018)}

\bibitem{abf-19}
V.~Alba, B.~Bertini, and M.~Fagotti, {Entanglement Spreading and Generalized Hydrodynamics}, 
{SciPost Phys. {\bf 7}, 005 (2019)}.  

\bibitem{a-19}
V. Alba, {Towards a generalized hydrodynamics description of Renyi entropies in integrable systems},
{Phys. Rev. B {\bf 99}, 045150 (2019)}. 


\bibitem{bc-18}
A. Bastianello and P. Calabrese, {Spreading of entanglement and correlations after a quench with intertwined quasiparticles},
{SciPost Phys. {\bf 5}, 033 (2018)}

\bibitem{btc-18}
B. Bertini, E. Tartaglia, and P. Calabrese,
{Entanglement and diagonal entropies after a quench with no pair structure}, {J. Stat. Mech. 063104 (2018).}

\bibitem{lcp-22}
G. Lagnese, P. Calabrese, and L. Piroli, {Entanglement dynamics of thermofield double states in integrable models},
{J. Phys. A {\bf 55}, 214003 (2022)}.


\bibitem{confi} M. Kormos, M. Collura, G. Takács, and P. Calabrese, Real time confinement following a quantum quench to a non-integrable model, Nature Phys. {\bf 13}, 246 (2017).

\bibitem{Liu2018}
F. Liu, R. Lundgren, P. Titum, G. Pagano, J. Zhang, C. Monroe, and A. V. Gorshkov,
{Confined Quasiparticle Dynamics in Long-Range Interacting Quantum Spin Chains},
{Phys. Rev. Lett. {\bf 122}, 150601 (2019)}.

\bibitem{jkr-19}
A. J. A. James, R. M. Konik, and N. J. Robinson,
{Nonthermal States Arising from Confinement in One and Two Dimensions},
{Phys. Rev. Lett. {\bf 122}, 130603 (2019)}.

\bibitem{rjk-19}
N. J. Robinson, A. J. A. James, and R. M. Konik,
{Signatures of rare states and thermalization in a theory with confinement},
{Phys. Rev. B {\bf 99}, 195108 (2019)}.

\bibitem{tan2019}
W. L. Tan, P. Becker, F. Liu, G. Pagano, K. S. Collins, A. De, L. Feng, H. B. Kaplan, A. Kyprianidis, R. Lundgren, W. Morong, S. Whitsitt, A. V. Gorshkov, and C. Monroe,
{Observation of Domain Wall Confinement and Dynamics in a Quantum Simulator},
{Nature Phys. {\bf 17}, 742 (2021)}.

{ \DXH \bibitem{DelfinoOscillations}
G. Delfino and J. Viti,
{On the theory of quantum quenches in near-critical systems}, 

\bibitem{clsv-20}
O.~A. Castro-Alvaredo, M. Lencs{\'e}s, I.~M.  Sz{\'e}cs{\'e}nyi, and J. Viti,
{Entanglement oscillations near a quantum critical point}, 
{Phys. Rev. Lett. {\bf 124}, 230601 (2020)}.

{ \DXH \bibitem{ch-sG-REOscillations}
O.~A. Castro-Alvaredo and D. ~X. Horv{\'a}th,
{Branch Point Twist Field Form Factors in the sine-Gordon Model I: Breather Fusion and Entanglement Dynamics}, 
{SciPost Phys. {\bf 10}, 132 (2021)}.}



\bibitem{scb-21}
S. Scopa, P. Calabrese, and A. Bastianello, {Entanglement dynamics in confining spin chains},
{Phys. Rev. B {\bf 105}, 125413 (2022)}.

\bibitem{mrw-17}
R. C. Myers, M. Rozali, and B. Way, {Holographic quenches in a confined phase},
{J. Phys. A {\bf 50}, 494002 (2017)}.

\bibitem{cr-19}
A. Cortes Cubero and N. J. Robinson, {Lack of thermalization in $(1+1)$-d quantum chromodynamics at large $N _c$},
{J. Stat. Mech. (2019) 123101}.


\bibitem{lsmc-21}
G. Lagnese, F. M. Surace, M. Kormos, and P. Calabrese, {False vacuum decay in quantum spin chains},
{Phys. Rev. B {\bf 104}, L201106 (2021)}.

\end{thebibliography}
\end{document}